\documentclass[a4paper,12pt,oneside]{article}
\usepackage[utf8]{inputenc}
\usepackage{lmodern}
\usepackage[ngerman, english]{babel}
\usepackage{amsmath}
\usepackage{amssymb}
\usepackage{amsthm}
\usepackage{mathtools}
\usepackage{microtype}
\usepackage{amsopn}
\usepackage{varwidth} 
\usepackage{xcolor,colortbl}
\usepackage{tabu}
\usepackage{comment}
\usepackage{enumerate}
\usepackage{array}
\usepackage{csquotes}
\usepackage{tikz}
\usepackage{tikz-qtree}
\usetikzlibrary{trees,positioning,shapes}
\usepackage{pgfplots}
\usepackage{algpseudocode}
\usepackage{algorithm}
\usepackage{nicefrac}
\usepackage{cmap}
\usepackage{multicol}
\usepackage[left=2.75cm, right=2.75cm, top=3cm, bottom=3cm]{geometry}
\usepackage{listings}
\usepackage[section]{placeins}
\usepackage{subcaption}
\usepackage{stmaryrd}
\usepackage{pdfpages}
\usepackage{hyperref}

\usepackage[capitalise,nameinlink]{cleveref}
\usepackage[retainorgcmds]{IEEEtrantools}

\pgfplotsset{compat=1.11}
\newlength\Origarrayrulewidth

\theoremstyle{plain}
\newtheorem{theorem}{Theorem}

\newtheorem{lemma}[theorem]{Lemma}
\newtheorem{proposition}[theorem]{Proposition}
\newtheorem{Theorem}[theorem]{Theorem}
\newtheorem{corollary}[theorem]{Corollary}

\theoremstyle{definition}
\newtheorem{definition}[theorem]{Definition}

\newtheorem{remenv}[theorem]{Remark}
\newtheorem{example}[theorem]{Example}

\newtheorem{optimization}[theorem]{Optimization}

\newenvironment{remark}
  {\pushQED{\qed}\remenv}
  {\popQED\endremenv}

\numberwithin{equation}{section}
\numberwithin{theorem}{section}

\newcommand{\bbN}{\mathbb{N}}

\newcommand{\bbR}{\mathbb{R}}
\newcommand{\bbS}{\mathbb{S}}

\newcommand{\bbZ}{\mathbb{Z}}

\newcommand{\calB}{\mathcal{B}}

\newcommand{\calF}{\mathcal{F}}
\newcommand{\calG}{\mathcal{G}}

\newcommand{\calI}{\mathcal{I}}

\newcommand{\calL}{\mathcal{L}}
\newcommand{\calM}{\mathcal{M}}

\newcommand{\calO}{\mathcal{O}}

\newcommand{\calS}{\mathcal{S}}

\newcommand{\calV}{\mathcal{V}}

\newcommand{\quot}[1]{\enquote{#1}}
\newcommand{\equalDef}{\coloneqq}

\newcommand{\equivDef}{:\Leftrightarrow}

\DeclareMathOperator{\id}{id}

\DeclareMathOperator{\conv}{conv}
\DeclareMathOperator{\aff}{aff}
\DeclareMathOperator{\polytopeVert}{vert}
\DeclareMathOperator{\opFacets}{facets}
\DeclareMathOperator{\opSection}{section}

\newcommand{\natOne}{\bbN_{\geq 1}}
\newcommand{\natZero}{\bbN_0}

\newcommand{\reals}{\bbR}

\newcommand{\compLeq}[1]{\calO\left( #1 \right)}

\DeclareMathOperator{\intdiv}{div}
\DeclareMathOperator{\Aut}{Aut}

\newcommand{\assign}{\coloneqq}

\DeclareTextFontCommand{\defEmph}{\bfseries}

\allowdisplaybreaks

\begin{document}

\newcommand{\childIndices}{\calB}
\newcommand{\rootv}{r}
\newcommand{\vertices}{\calV}
\newcommand{\parent}{P}
\newcommand{\child}{C}
\newcommand{\idx}{I}
\newcommand{\level}{\ell}
\newcommand{\facets}{\calF}
\newcommand{\pface}{F^p}
\newcommand{\cface}{F^c}
\newcommand{\states}{\calS}
\newcommand{\state}{S}
\newcommand{\istate}{G}
\newcommand{\pstate}{S^p}
\newcommand{\cstate}{S^c}
\newcommand{\nfunc}{N}
\newcommand{\ofunc}{\Omega}
\newcommand{\ndef}{{\perp}} 
\newcommand{\faceperm}{\pi}
\newcommand{\depth}{k}
\newcommand{\neighborDepth}{D_T}
\newcommand{\neighborDepthWithoutTree}{D}
\newcommand{\neighborRuntime}{t_{\textsc{Neighbor}}}
\newcommand{\neighborRuntimeBound}{\hat{t}_{\textsc{Neighbor}}}
\newcommand{\vRuntime}{\hat{t}}
\newcommand{\rootState}{s_\rootv}
\newcommand{\rootPoints}{Q^{(r)}}
\newcommand{\transitionMat}{M}
\newcommand{\convHull}[1]{\conv(#1)}
\newcommand{\facetsOf}[1]{\opFacets(#1)}
\newcommand{\facetIndices}[2]{\mathrm{indices}(#1, #2)}
\newcommand{\facetFromIndices}[2]{\mathrm{facet}(#1, #2)}
\newcommand{\polytopeFacetIndices}[1]{\mathrm{indexFacets}(#1)}
\newcommand{\matrixCol}[2]{#1_{\cdot, #2}}
\newcommand{\oneVec}[1]{\mathbf{1}_{#1}}
\newcommand{\bStateSystem}{\bbS}
\newcommand{\vertexIso}{\varphi^V}
\newcommand{\stateIso}{\varphi^S}
\newcommand{\facetIso}{\varphi^F}
\newcommand{\affAut}{\Aut_{\mathrm{aff}}}

\newcommand{\vertexFace}[2]{(#1)_{#2}}
\newcommand{\polytopeSection}[2]{\opSection(#1, #2)}

\newcommand{\sfcpp}{\texttt{sfcpp}}

\includepdf{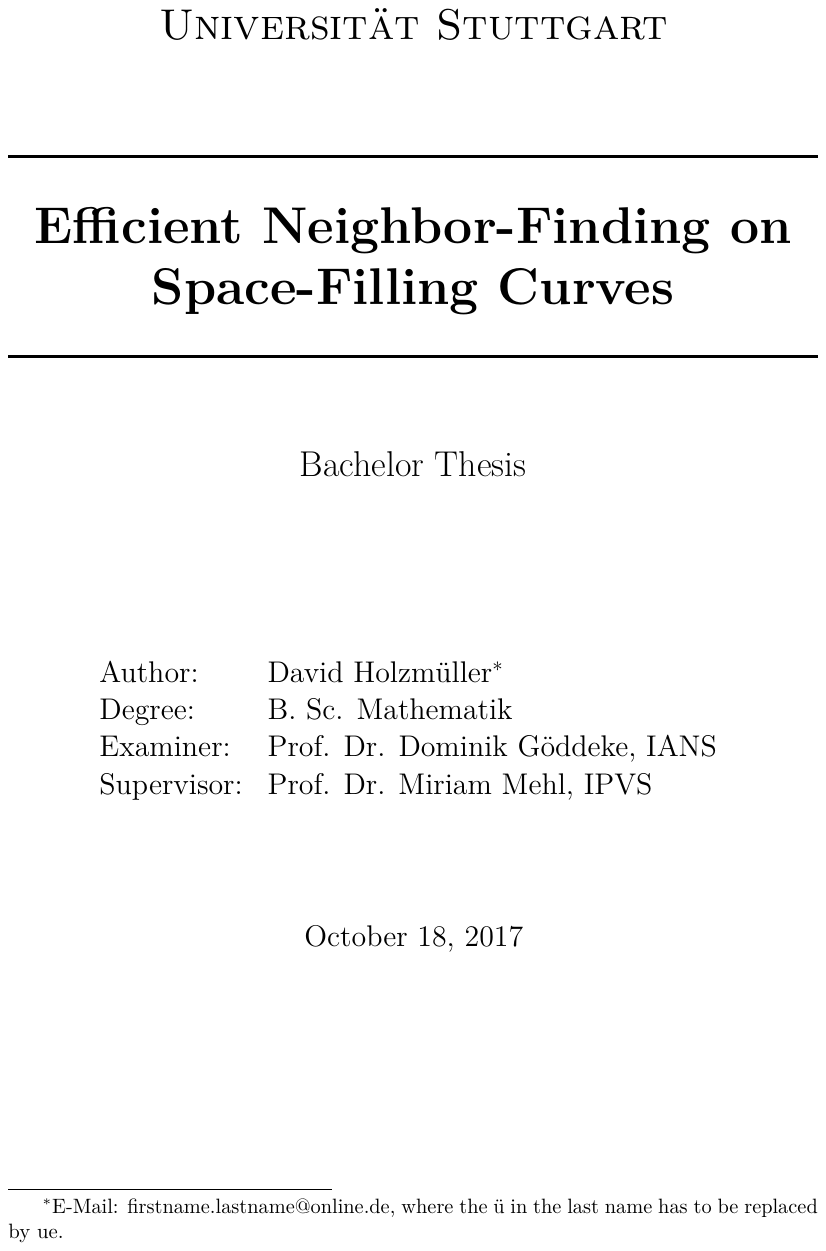}

\cleardoublepage

\begin{abstract}
Space-filling curves (SFC, also known as FASS-curves) are a useful tool in scientific computing and other areas of computer science to sequentialize multidimensional grids in a cache-efficient and parallelization-friendly way for storage in an array. Many algorithms, for example grid-based numerical PDE solvers, have to access all neighbor cells of each grid cell during a grid traversal. While the array indices of neighbors can be stored in a cell, they still have to be computed for initialization or when the grid is adaptively refined. A fast neighbor-finding algorithm can thus significantly improve the runtime of computations on multidimensional grids.

In this thesis, we show how neighbors on many regular grids ordered by space-filling curves can be found in an average-case time complexity of $\compLeq{1}$. In general, this assumes that the local orientation (i.e.\ a variable of a describing grammar) of the SFC inside the grid cell is known in advance, which can be efficiently realized during traversals. Supported SFCs include Hilbert, Peano and Sierpinski curves in arbitrary dimensions. We assume that integer arithmetic operations can be performed in $\compLeq{1}$, i.e.\ independent of the size of the integer. We do not deal with the case of adaptively refined grids here. However, it appears that a generalization of the algorithm to suitable adaptive grids is possible. 

To formulate the neighbor-finding algorithm and prove its correctness and runtime properties, a modeling framework is introduced. This framework extends the idea of vertex-labeling to a description using grammars and matrices. With the \sfcpp\ library, we provide a C++ implementation to render SFCs generated by such models and automatically compute all lookup tables needed for the neighbor-finding algorithm. Furthermore, optimized neighbor-finding implementations for various SFCs are included for which we provide runtime measurements.
\end{abstract}

\clearpage

\tableofcontents
\cleardoublepage

\section{Introduction}

Many algorithms, especially in scientific computing, operate on data stored in multidimensional grids. Often, these grids are stored in a one-dimensional array, using a \emph{sequential order} on the grid cells that defines which array index corresponds to which grid cell. For example, matrices are usually stored row-major (i.e.\ row-by-row) or column-major (i.e.\ column-by-column). Yet, such a naive sequential order may be suboptimal when certain geometric operations are performed on the grid. Space-filling curves (SFC), originally known as a topological curiosity, yield sequential orders on grids that are more cache-efficient and parallelization-friendly \cite{bader2012}.

For example, solving a partial differential equation on a grid numerically often involves combining values of a grid cell with values of its geometrical neighbors. In a sequential order induced by a SFC, a much higher percentage of pairs of geometrical neighbors have array indices that are \quot{close}, which means that geometrical neighbors are often loaded into cache memory together. Due to the access speed difference between RAM and cache memory, this locality property can have a significant impact on the algorithm's overall efficiency. Another consequence of the better locality of SFCs is that when splitting the array in $p$ similarly-sized partitions, a high percentage of pairs of geometrical neighbors lie in the same part. If these $p$ parts of the array are processed in $p$ parallel threads, each pair of neighboring cells lying in different parts of the array leads to communication between the corresponding threads or increases the number of cells that are stored in multiple threads. This is another reason why SFCs can help to reduce computational effort \cite{bader2012}.

SFC-induced sequential orders are usually recursively defined, starting with a single-cell grid and in each recursion step splitting each cell into $b \geq 2$ subcells and arranging them in a certain order. A disadvantage of such a SFC-induced sequential order over simpler sequential orders is therefore that the algorithms needed to deal with them are often more complicated. It has to be taken care that the reduction of data access and transfer time described above is greater than the extra runtime introduced by more complicated algorithms. %
In particular, this thesis deals with the problem of finding the array index of a geometrical neighbor cell and related problems.

\subsection{Related Work}

An obvious way to find neighbors in a regular grid is to convert an array index to a coordinate vector, changing one coordinate to obtain the neighbor's coordinate vector, and convert the latter back into an array index. Many methods for these conversions have been suggested. The book by Bader \cite{bader2012} provides a good overview over such algorithms. Bartholdi and Goldsman \cite{bartholdi2001} introduced the technique of vertex-labeling. This technique is similar to the modeling approach introduced in this thesis. Moreover, it can be used to convert between array indices and coordinate vectors for a broad range of SFCs. Bartholdi and Goldsman also proposed a neighbor-finding algorithm for various SFCs based on vertex-labeling that uses a coordinate vector of an interior point of each edge of a cell. All neighbor-finding algorithms of this type have a time complexity equal to that of the corresponding conversion algorithm, meaning that they will scale linearly with the refinement level of the curve and thus logarithmically with the number of grid points (assuming that the grid is regular, i.e.\ equally refined everywhere).

For the popular Morton order, Schrack \cite{schrack1992} published a neighbor-finding algorithm with runtime $\compLeq{1}$, i.e.\ independent of the refinement level of the grid, assuming constant-time arithmetic integer operations. Aizawa and Tanaka \cite{aizawa2009} investigated the Morton order on adaptive 2D grids. They suggested to store level differences to neighbors inside the grid cells and presented an adaption of Schrack's algorithm to find \quot{location codes} of neighbors of equal or bigger size. If the used data structure efficiently allows to access data based on its location code, then such neighbors can be accessed efficiently.

In stack-based approaches (see Chapter 14 in Bader \cite{bader2012}), data is placed at the vertices of the grid cells, i.e.\ the grid points, and stored in stacks. Neighboring vertices are automatically found when a cell containing both neighbors is visited. Weinzierl and Mehl \cite{weinzierl2011} presented a stack-based framework for the Peano curve. This approach does not work for all space-filling curves. For example, the Hilbert curves in dimensions $d \geq 3$ are not suited for such a traversal \cite{bader2012}. In addition, it can be difficult to embed an existing simulation software in such a framework.

\subsection{Contribution}

In this thesis, we present an algorithm to find neighbors in space-filling curves with an average-case time complexity of $\compLeq{1}$ under certain assumptions:
\begin{itemize}
\item The space-filling curve is generated by a certain recursive pattern that satisfies some regularity conditions. These conditions are precisely given in \Cref{sec:modeling}. Supported SFCs include Morton, Hilbert, Peano, Sierpinski and variants thereof, some of which are shown in \Cref{sec:overview:otherSFC}.
\item Arithmetic operations on (non-negative) integers can be performed in $\compLeq{1}$ time, i.e.\ independent of the number of bits of the integer. For SFCs like Morton, Hilbert and Sierpinski, where $b$ (the number of subcells that a cell is split into in the recursive construction) is a power of two, only the following arithmetic operations are needed: 
\begin{itemize}
\item Increment, decrement and comparison on integers not greater than the level of the given grid cell. Even if no constant-time integer arithmetic is assumed, this only needs $\compLeq{\log(\text{level})}$ time.
\item Bit operations that operate on a constant number of bits.
\end{itemize}
\item The grid is regular. This usually means that every grid cell has the same size. %
We assume that an extension of the algorithm may also work on adaptive grids.
\item The \quot{state}, i.e.\ the local pattern of the curve inside the given grid cell, is known. During a traversal of the grid, this state can be computed very efficiently with $\compLeq{1}$ overhead per grid cell. If random access is intended, the state can be computed in $\compLeq{l} = \compLeq{\log n}$, where $l$ is the tree level of the grid cell in the tree and $n$ is the total number of grid cells. This state can then be used to find all neighbors and their neighbors, if necessary. %
For SFCs like Hilbert, Peano and Sierpinski, the neighbors still can be found in unknown order if the state of the grid cell is not known. This is explained in \Cref{sec:algorithms:symmetry}.
\end{itemize}

In this thesis, we develop a rigorous framework for modeling space-filling curves based on vertex-labeling \cite{bartholdi2001}. This framework is used to give a general formulation of the neighbor-finding algorithm and prove its correctness and runtime. Furthermore, it is employed to precisely formulate properties of SFCs, for example conditions for the algorithm to operate correctly or conditions needed to perform certain optimizations in its implementation. We intend to make all aspects of this modeling algorithmically computable. Together with this thesis, we provide a library called \sfcpp ,\footnote{\href{https://github.com/dholzmueller/sfcpp}{https://github.com/dholzmueller/sfcpp}} where some these aspects have been implemented, especially:
\begin{itemize}
\item neighbor-finding algorithms for the Peano curve in arbitrary dimensions, the Hilbert curve in 2D and 3D and the Sierpinski curve in 2D,
\item models of many SFCs,
\item \LaTeX\ rendering (visualization) of SFCs specified by such a model,
\item generation of lookup tables needed for the algorithms given here, and
\item checking of some properties of a specified model.
\end{itemize}

A particular version of the neighbor-finding algorithm for the Peano curve, an implementation and visualization code has been developed as a term paper.\footnote{David Holzmüller: Raumfüllende Kurven, 2016.} This thesis generalizes the algorithm, includes a further optimized version of the implementation into a bigger library and partially replaces the visualization code with a more general implementation.

\subsection{Outline}

In \Cref{sec:overview:hilbert}, we will give an overview over the modeling framework, using the 2D Hilbert curve as an example. In \Cref{sec:overview:algorithm}, the idea behind the neighbor-finding algorithm is presented using some examples. \Cref{sec:overview:otherSFC} presents other curves to motivate the introduction of a general model. 

\Cref{sec:modeling:treesMatricesStates} then introduces formal definitions for models and related objects such as trees. \Cref{sec:modeling:algNeighbors} establishes a definition of neighborship on trees using lookup tables. This definition is used later to formulate the neighbor-finding algorithm. \Cref{sec:modeling:geometricNeighbors} formally introduces a geometric definition of neighborship for geometric models and proves that this definition is, under certain regularity conditions, equivalent to the neighborship definition using lookup tables. \Cref{sec:modeling:verification} then shows how to verify these regularity conditions algorithmically. \Cref{sec:modeling:isomorphisms} introduces isomorphisms between trees.

A general formulation of our neighbor-finding algorithm is given in \Cref{sec:algorithms:generalNeighbors} and its correctness and runtime are proven. \Cref{sec:algorithms:other} shows how to use the neighbor-finding algorithm in a traversal. Based on tree isomorphisms, a framework for computing states and coordinates is given. In some cases, calling this algorithm with a wrong state still yields the correct set of neighbors, just in permuted order. \Cref{sec:algorithms:symmetry} gives formal criteria to verify these cases.

\Cref{sec:optimizations} shows optimizations that can be employed to make the algorithms faster in practice. Some of these optimizations work for all SFCs while some only apply to special curves.

\Cref{sec:implementation} presents implementations of algorithms, mainly those implemented in the \sfcpp\ library. It also provides runtime measurements for different algorithms that compute neighbors or states of grid cells.

Finally, \Cref{sec:conclusion} points out open research questions.

\cleardoublepage

\section{Overview and Motivating Examples} \label{sec:overview}

In this section, we want to give an overview over many concepts and algorithms that are introduced formally in the following sections.
First, we examine the two-dimensional Hilbert curve as a popular example of a space-filling curve. After discussing various related mathematical tools, we explain the ideas behind the neighbor-finding algorithm presented in \Cref{sec:algorithms:generalNeighbors}. Finally, we present more examples of space-filling curves, focusing on aspects that are important for the design of algorithms that should be applicable to a large class of SFCs. The choice of presented SFCs and their presentation is inspired by Bader \cite{bader2012}. %

\subsection{Modeling the Hilbert Curve} \label{sec:overview:hilbert}

The definition of the term \quot{space-filling curve} is not uniform across literature (\cite{bader2012}, \cite{bartholdi2001}). A space-filling curve (SFC) can be defined as a continuous function $f: I \subseteq \bbR \to \bbR^d, d \geq 2$, for which $f(I)$ has positive Jordan content. In 1890, Peano showed the first example of such a space-filling curve \cite{bader2012}.
SFCs relevant for grid sequentialization are usually specified as the uniform limit of a sequence of piecewise affine curves, where each curve in the sequence is a recursive refinement of the previous curve in the sequence. From an algorithmic perspective, the \quot{finite approximations} in this sequence are the main object of study. In the following, they will also be referred to as space-filling curves, although they do not satisfy the definition above.

\newcommand{\curveScale}{0.85}
\begin{figure}[htb]
\centering
\subcaptionbox{Level $0$ \label{fig:hilbertConstruction:l0}}{
\includegraphics[scale=\curveScale]{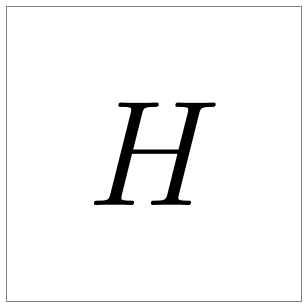}
}
\subcaptionbox{Level $1$ \label{fig:hilbertConstruction:l1}}{
\includegraphics[scale=\curveScale]{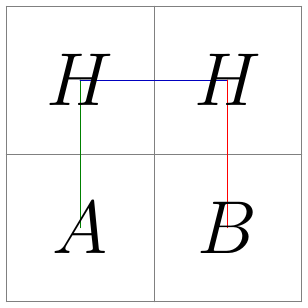}
}
\subcaptionbox{Level $2$ \label{fig:hilbertConstruction:l2}}{
\includegraphics[scale=\curveScale]{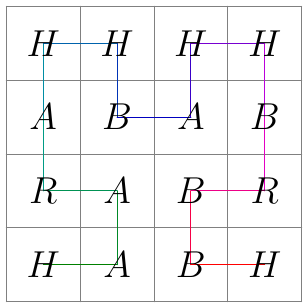}
}
\caption{Construction of the 2D Hilbert curve.} \label{fig:hilbertConstruction}
\end{figure}

\begin{figure}[htb]
\centering
\includegraphics[scale=0.8]{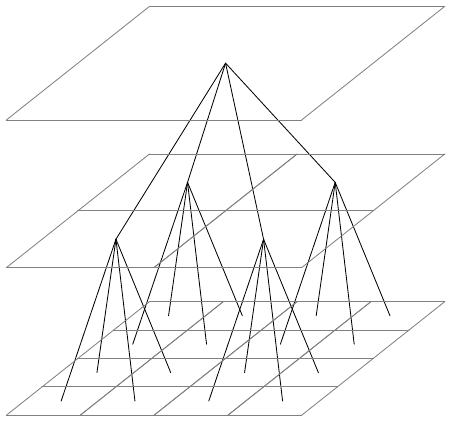}
\caption{Tree representation of the squares from the 2D Hilbert curve.} \label{fig:hilbertSquareTree}
\end{figure}

\Cref{fig:hilbertConstruction} shows the first three levels of the 2D Hilbert curve. From one level to the next level, each square is subdivided into four equal subsquares. These subsquares are then put into a certain order depending on the state $s \in \{H, A, B, R\}$ of the square. We can interpret these squares on different levels as nodes (vertices) of a tree: The square $\rootPoints$ at level 0 is the root of the tree. The subsquares $\child(Q, j), j \in \childIndices \equalDef \{0, 1, 2, 3\}$, of a square $Q$, which are located in the next level, form the four children of a square. Conversely, the square $Q$ is called the parent of its four children. Each square except the root $\rootPoints$ has a parent. The tree structure of the squares is also visualized in \Cref{fig:hilbertSquareTree}.

A tree consisting of $d$-dimensional hypercubes, where each hypercube ist partitioned into $k^d$ equal subcubes by its children, is called $k^d$-tree. For the 2D Hilbert curve, we have $k = 2$ and $d = 2$ and each square is partitioned into $b \equalDef k^d = 2^2 = 4$ equal subsquares. A $2^2$-tree is also called Quadtree \cite{bader2012}. 

\begin{figure}[htb]
\centering
\subcaptionbox{Level $0$ \label{fig:hilbertOrderingDecimal:l0}}{
\includegraphics[scale=\curveScale]{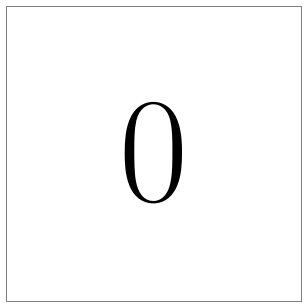}
}
\subcaptionbox{Level $1$ \label{fig:hilbertOrderingDecimal:l1}}{
\includegraphics[scale=\curveScale]{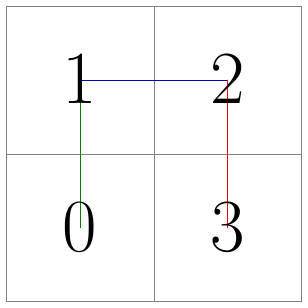}
}
\subcaptionbox{Level $2$ \label{fig:hilbertOrderingDecimal:l2}}{
\includegraphics[scale=\curveScale]{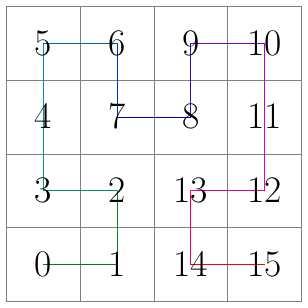}
}
\caption{Cell enumeration in the 2D Hilbert curve in base 10.} \label{fig:hilbertOrderingDecimal}
\end{figure}

To store all squares of a given level sequentially in memory, we take the order generated by the space-filling curve, see \Cref{fig:hilbertOrderingDecimal}.

\begin{figure}[htb]
\centering
\subcaptionbox{Level $0$ \label{fig:hilbertOrderingBaseFour:l0}}{
\includegraphics[scale=\curveScale]{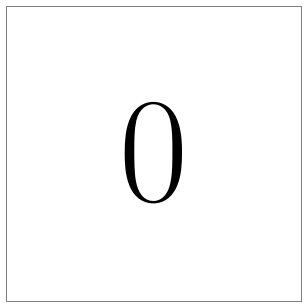}
}
\subcaptionbox{Level $1$ \label{fig:hilbertOrderingBaseFour:l1}}{
\includegraphics[scale=\curveScale]{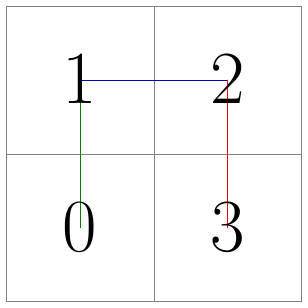}
}
\subcaptionbox{Level $2$ \label{fig:hilbertOrderingBaseFour:l2}}{
\includegraphics[scale=\curveScale]{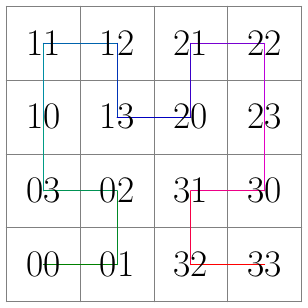}
}
\caption{Cell enumeration in the 2D Hilbert curve in base 4.} \label{fig:hilbertOrderingBaseFour}
\end{figure}

\Cref{fig:hilbertOrderingBaseFour} shows the numbers in base $b = 4$ instead of base 10. This exposes a simple pattern: The base-$4$ digits of a square in the tree consist of the base-$4$ digits of its parent square and a digit corresponding to its position inside its parent square.

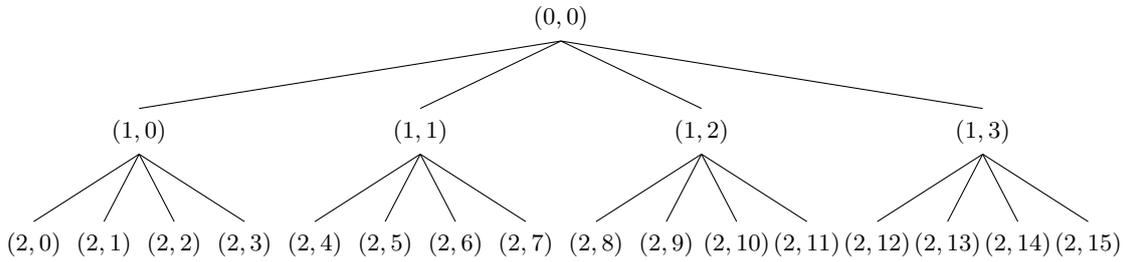
\begin{figure}[htb]
\centering
\begin{tikzpicture}[font={\fontsize{9pt}{12}\selectfont}, level/.style={sibling distance=37mm/((#1)*(#1))}]
\node (a) {$(0, 0)$}
	child{node (b) {$(1, 0)$}
		child{node (c) {$(2, 0)$} %
		}
		child{node (c) {$(2, 1)$}
		}
		child{node (c) {$(2, 2)$}
		}
		child{node (c) {$(2, 3)$}
		}
	}
	child{node (b) {$(1, 1)$}
		child{node (c) {$(2, 4)$}
		}
		child{node (c) {$(2, 5)$}
		}
		child{node (c) {$(2, 6)$}
		}
		child{node (c) {$(2, 7)$}
		}
	}
	child{node (b) {$(1, 2)$}
		child{node (c) {$(2, 8)$}
		}
		child{node (c) {$(2, 9)$}
		}
		child{node (c) {$(2, 10)$}
		}
		child{node (c) {$(2, 11)$}
		}
	}
	child{node (b) {$(1, 3)$}
		child{node (c) {$(2, 12)$}
		}
		child{node (c) {$(2, 13)$}
		}
		child{node (c) {$(2, 14)$}
		}
		child{node (c) {$(2, 15)$}
		}
	};
\end{tikzpicture}
\caption{Levels 0, 1, 2 of the level-position $b$-index-tree as defined in \Cref{def:levelPositionTree}.} \label{fig:levelPositionTree}
\end{figure}

The tree representation using squares is useful to draw SFCs. However, when using SFCs to create efficient algorithms, representing a square using coordinates is computationally slow. Moreover, depending on the application, a square might not be given by its coordinates. Instead, we may represent a square in the tree by a tuple $(l, j)$. Here, $l \in \natZero$ is the level of the square and $j \in \{0, \hdots, b^l - 1\}$ is its position along the curve. %
The first three levels of the corresponding tree (which will be called level-position $b$-index-tree later) are shown in \Cref{fig:levelPositionTree}. %

The base-$4$ pattern for the position of a node described above can be directly turned into arithmetic formulas: Consider a node $v = (l, j)$.
\begin{itemize} 
\item The parent $\parent(v)$ of $v$ is given by $P(v) = (l-1, j \intdiv 4)$, where $j \intdiv 4 \equalDef \lfloor j/4 \rfloor$ is the integer part of $j/4$. The operation $j \mapsto j \intdiv 4$ eliminates the last digit of $j$ in its base-$4$ representation.
\item The $i$-th child $\child(v, i)$ is given by $\child(v, i) \equalDef (l+1, 4j + i)$, where $j \mapsto 4j + i$ inserts the digit $i$ at the end of the base-$4$ representation of $j$.
\item The index $\idx(v)$ of $v$ inside its parent is given by $\idx(v) = j \bmod 4$, i.e.\ the last digit of the base-$4$ representation of $j$.
\end{itemize}

Now that we know how to work with levels and positions of nodes in the 2D Hilbert curve, we can turn to the states of the nodes. As noted above, the state $\state(v)$ of a node $v$ determines the geometric arrangement of the children of the node. For all SFCs examined here, the state $\state(\child(v, i))$ of the $i$-th child of $v$ is uniquely determined by $i$ and by the state of $v$. That is, we can specify a function $\cstate$ such that $\state(\child(v, i)) = \cstate(\state(v), i)$. For the 2D Hilbert curve, this function is specified in \Cref{table:hilbertChildState}.

\begin{table}[htb]
\centering
\caption{Values of the function $\cstate$ for the 2D Hilbert curve.} \label{table:hilbertChildState}
\vspace{0.1cm}
\begin{tabular}{c|cccc}
$\cstate(s, j)$ & $j = 0$ & $j = 1$ & $j = 2$ & $j = 3$ \\
\hline
$s = H$ & $A$ & $H$ & $H$ & $B$ \\
$s = A$ & $H$ & $A$ & $A$ & $R$ \\
$s = R$ & $B$ & $R$ & $R$ & $A$ \\
$s = B$ & $R$ & $B$ & $B$ & $H$
\end{tabular}
\end{table}

\begin{figure}[htb]
\centering
\subcaptionbox{Level $1$ \label{fig:hilbertConstruction2:l1}}{
\includegraphics[scale=\curveScale]{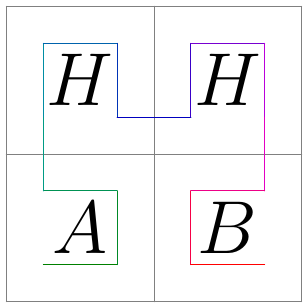}
}
\subcaptionbox{Level $2$ \label{fig:hilbertConstruction2:l2}}{
\includegraphics[scale=\curveScale]{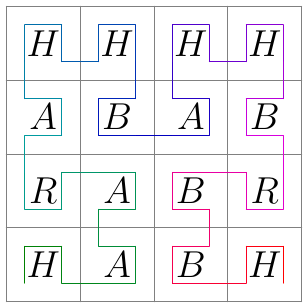}
}
\subcaptionbox{Level $3$ \label{fig:hilbertConstruction2:l3}}{
\includegraphics[scale=\curveScale]{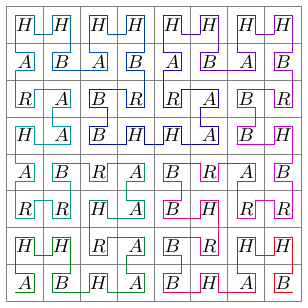}
}
\caption{Construction of the 2D Hilbert curve. The curve is drawn one level finer than the grid.} \label{fig:hilbertConstruction2}
\end{figure}

All of the entries except for $s = R$ can be extracted from \Cref{fig:hilbertConstruction}. The state $s = R$ occurs in level 2 of the curve for the first time. Thus, we need level 3 to read off %
the states of the children of a state-$R$ square. \Cref{fig:hilbertConstruction2} shows the construction up to level 3 using a finer Hilbert curve. Note that the function $\cstate$ may also be expressed as a grammar with production rules $H \to AHHB, A \to HAAR, R \to BRRA$ and $B \to RBBH$. We will use the function-based approach since it is better suited for this thesis.

Using the function $\cstate$, we can compute the state $\state(v) = \cstate(\state(\parent(v)), \idx(v))$ of a node $v$ only using the state $\state(\parent(v))$ of its parent and the index $\idx(v)$ of $v$ inside its parent. This can be efficiently implemented using a lookup table. Can we also compute the state of a parent from the state of its child? For the 2D Hilbert curve, the answer is yes. The function $s \mapsto \cstate(s, i)$ is invertible for all $i$ since all columns of \Cref{table:hilbertChildState} contain each state exactly once. Thus, we can find an inverse function $s \mapsto \pstate(s, i)$ for all $i$. This means that $\state(\parent(v)) = \pstate(\state(v), \idx(v))$, since
\begin{IEEEeqnarray*}{+rCl+x*}
\state(v) & = & \state(\child(\parent(v), \idx(v))) = \cstate(\state(\parent(v)), \idx(v))~.
\end{IEEEeqnarray*}
Here, we used the fact that a node $v$ is by definition the $\idx(v)$-th child of its own parent. 

\begin{remark} \label{remark:hilbert2DGroup}
In the case of the 2D Hilbert curve, the function $\pstate$ is equal to the function $\cstate$. From an algebraic perspective, the functions $\sigma_j: \states \to \states, s \mapsto \cstate(s, j)$ are invertible and thus permutations on the set $\states = \{H, A, B, R\}$ of states. For the 2D Hilbert curve, each $\sigma_j$ is equal to its own inverse, which is why we find that $\pstate = \cstate$. The permutation subgroup $\langle \sigma_j \mid j \in \{0, 1, 2, 3\}\rangle$ spanned by these permutations is isomorphic to the Klein four-group $\bbZ_2 \times \bbZ_2$. If we examine the Hilbert curve more closely, this is no surprise: The curves in squares of different states can be seen as mirrored versions of each other, where the mirroring corresponds to an orthogonal transformation given by one of the four matrices
\begin{IEEEeqnarray*}{+rCl+x*}
\begin{pmatrix}
1 & 0 \\ 0 & 1
\end{pmatrix}, \begin{pmatrix}
0 & 1 \\ 1 & 0
\end{pmatrix}, \begin{pmatrix}
-1 & 0 \\ 0 & -1
\end{pmatrix}, \begin{pmatrix}
0 & -1 \\ -1 & 0
\end{pmatrix}~.
\end{IEEEeqnarray*}
A change of basis to a new basis $b_1 = (1, 1)^\top$, $b_2 = (1, -1)^\top$ yields the matrices 
\begin{IEEEeqnarray*}{+rCl+x*}
\begin{pmatrix}
1 & 0 \\ 0 & 1
\end{pmatrix}, \begin{pmatrix}
1 & 0 \\ 0 & -1
\end{pmatrix}, \begin{pmatrix}
-1 & 0 \\ 0 & -1
\end{pmatrix}, \begin{pmatrix}
-1 & 0 \\ 0 & 1
\end{pmatrix}~,
\end{IEEEeqnarray*}
which clearly form a subgroup of $\mathrm{O}_2(\bbR)$ isomorphic to the Klein four-group.

In an implementation, this suggests to encode the states in a way such that this group operation can be applied efficiently. For example, one might use integers with bitwise XOR as the group operation.
\end{remark}

\subsection{Neighbor-Finding Algorithm} \label{sec:overview:algorithm}

In this section, we will outline the ideas behind our neighbor-finding algorithm along the example of the Hilbert curve. In a nutshell, the neighbor-finding algorithm takes the shortest path to the neighbor in the corresponding tree. It ascends $k$ levels and then descends $k$ levels, where $k$ is as small as possible and depends on the tree node. An investigation of the average size of $k$ with a standard geometric series argument yields that for all SFCs presented here, this average is bounded independent of the level of the nodes considered.

In \Cref{sec:overview:hilbert}, we have seen how to efficiently deal with levels, positions and states. Yet, there is one modeling decision remaining: When modeling general space-filling curves, one may or may not include the state as part of a node. In the modeling below, we decided to include the state into the node. Thus, we will from now on deal with nodes of the form $(l, j, s)$ instead of $(l, j)$. A disadvantage of this modeling is that for specifying a node, one already has to know its state. However, given the level and the position of a node inside the curve, the state can be computed as shown in \Cref{sec:algorithms:other}. It can also be efficiently tracked during traversals. Furthermore, our neighbor-finding algorithm assumes that the state of a node is known. The following examples show an application of our algorithm and different cases that have to be handled. A complete presentation of \Cref{alg:generalTrees} will be given in \Cref{sec:algorithms:generalNeighbors}.

\begin{example} \label{ex:neighbors:1}
Let $v$ be the node at level 2 with position 1 in the curve. In \Cref{fig:hilbertConstruction2}, we can see that its state is $A$, since the second square on the curve at level $2$ (i.e.\ the square with position 1) has state $A$. Thus, $v = (2, 1, A)$. Now assume that we want to find the upper neighbor. The algorithm proceeds as follows:
\begin{enumerate}[(1)]
\item Check if the level of $v$ is zero, in which case there would be no neighbor. This is not the case.
\item Compute the parent $p_v = (2-1, 1 \intdiv 4, \pstate(A, 1 \bmod 4)) = (1, 0, A)$ of $v$. Also, compute the index $j_v = 1 \bmod 4 = 1$ of $v$.
\item Every node with state $A$ in the Hilbert curve \quot{looks essentially the same},\footnote{This corresponds to the pre-regularity condition (P1) from \Cref{def:preregular}.} so we can consult a suitable lookup table to find out whether in a node of state $A = \state(p_v)$, the child with index $\idx(v) = 1$ has an upper neighbor inside $p_v$. Indeed, the lookup table will yield that the child with index 2 is an upper neighbor of $v$.
\item Compute the child with index 2: $w \equalDef \child(p_v, 2) = (1 + 1, 4 \cdot 0 + 2, \cstate(A, 2)) = (2, 2, A)$.
\item Return $w$ as the upper neighbor of $v$.
\end{enumerate}

Finding the left neighbor $(2, 0, H)$ of $v$ is completely analogous.
\end{example}

\begin{example} \label{ex:neighbors:2}
Now suppose that instead of the upper neighbor of $v = (2, 1, A)$, we are interested in the right neighbor of $v$. This time, the algorithm does the following:
\begin{enumerate}[(1)]
\item The level of $v$ is not zero, continue.
\item Compute the parent $p_v = (1, 0, A)$ of $v$.
\item Ask the lookup table if the child with index $1$ inside a node with state $A$ has a right neighbor. This time, the answer is no.
\item If there is a right neighbor of $v$, its parent must be a right neighbor of $p_v$.\footnote{This corresponds to the regularity conditions (R2) and (R3) of \Cref{def:geometric:regular}.} Apply the algorithm recursively to find a right neighbor of $p_v$. The recursive call will proceed similarly to \Cref{ex:neighbors:1} and yield the neighbor $p_w = (1, 3, B)$.
\item In the Hilbert curve, whenever a node of state $B$ is a right neighbor of a node of state $A$, the geometric arrangement of these two \quot{looks essentially the same}.\footnote{This corresponds to the regularity condition (R1) of \Cref{def:geometric:regular}.} Thus, we can consult a second lookup table using $\idx(v) = 1, \state(p_v) = A, \state(p_w) = B$ and the facet \quot{right} to find the index of the right neighbor of $v$ inside $w$. The lookup table will yield the index $i = 2$ as desired.
\item Now, compute the neighbor $w = \child(p_w, i) = (1 + 1, 4 \cdot 3 + 2, \cstate(B, 2)) = (2, 14, B)$ of $v$.
\item Return $w$ as the right neighbor of $v$.
\end{enumerate}
\end{example}

\begin{example} \label{ex:neighbors:3}
Finally, suppose that we are seeking the lower neighbor of $v = (2, 1, A)$. The algorithm proceeds as follows:
\begin{enumerate}[(1)]
\item The level of $v$ is not zero, continue.
\item Compute the parent $p_v = (1, 0, A)$ of $v$.
\item The first lookup table yields that there is no lower neighbor of $v$ inside $p_v$.
\item Call the algorithm recursively to find a lower neighbor of $p_v$:
\begin{enumerate}[(i)]
\item The level of $p_v$ is not zero, continue.
\item Compute the parent $p_{p_v} = (0, 0, H)$ of $p_v$.
\item The first lookup table yields that there is no lower neighbor of $v$ inside $p_v$.
\item Call the algorithm recursively to find a lower neighbor of $p_{p_v}$:
\begin{enumerate}[1.]
\item The level of $p_{p_v}$ is zero. Thus, $p_{p_v}$ cannot have a neighbor. Return a value indicating that there is no lower neighbor of $p_{p_v}$.
\end{enumerate}
\item Since there is no lower neighbor of $p_{p_v}$, return a value indicating that there is no lower neighbor of $p_v$.
\end{enumerate}
\item Since there is no neighbor of $p_v$, return a value indicating that there is no lower neighbor of $v$.
\end{enumerate}
\end{example}

In these three examples, it becomes evident that the runtime of the algorithm depends on how far one has to walk up and down the tree to find the neighbor. This concept is later introduced as the neighbor-depth of $v$. The upper-neighbor-depth and the left-neighbor-depth of $v$ in these examples is 1, since there is only one call of the algorithm in \Cref{ex:neighbors:1}. The right-neighbor-depth of $v$ in these examples is 2, since the algorithm needs to call itself recursively once. The lower-neighbor-depth of $v$ is 3, since the algorithm calls itself recursively twice. The neighbor-depth allows to characterize the runtime of the algorithm. In \Cref{thm:generalTrees:correct}, we will show that the runtime of the algorithm is in $\compLeq{\text{neighbor-depth}}$ (assuming constant-time arithmetic operations). Using a standard geometric series argument, we will then show that the average neighbor-depth at level $l$ is $\compLeq{1}$ for all \quot{normal} SFCs. 

\begin{remark} \label{remark:hilbert2DSymmetry}
In \Cref{remark:hilbert2DGroup}, it was mentioned that in the 2D Hilbert curve, nodes with states $A, B$ or $R$ are merely mirrored versions of nodes of state $H$. If the algorithm is given the wrong state of a node, it will behave as if the whole curve was mirrored. For example, given the node $v = (2, 1, H)$ instead of $(2, 1, A)$, requesting the upper neighbor of $v$ will return the right neighbor of $(2, 1, A)$ (also with an incorrect state), since the upper side in the coordinate system of a state-$H$-node corresponds to the right side in the coordinate system of a state-$A$-node. A formal criterion for this behavior is given in \Cref{sec:algorithms:symmetry}.
\end{remark}

\subsection{Other Space-Filling Curve Models} \label{sec:overview:otherSFC}

In this section, we want to take a look at some other curves with a focus on aspects that are different to the 2D Hilbert curve. The general space-filling curve model from \Cref{sec:modeling} and the formulation of the algorithm are motivated by many of these aspects presented here. Furthermore, special properties of certain SFCs presented here allow for special optimizations of the algorithm.

\begin{figure}[htb]
\centering
\subcaptionbox{Level $0$ \label{fig:peanoConstruction:l0}}{
\includegraphics[scale=\curveScale]{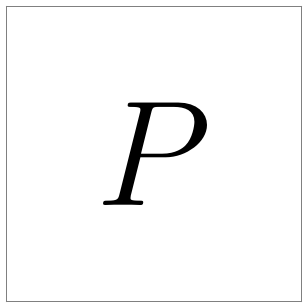}
}
\subcaptionbox{Level $1$ \label{fig:peanoConstruction:l1}}{
\includegraphics[scale=\curveScale]{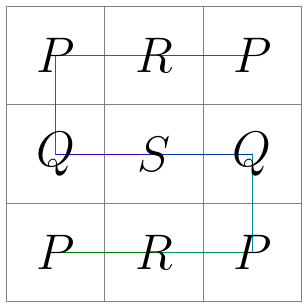}
}
\subcaptionbox{Level $2$ \label{fig:peanoConstruction:l2}}{
\includegraphics[scale=\curveScale]{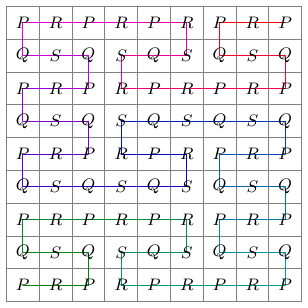}
}
\caption{Construction of the 2D Peano curve.} \label{fig:peanoConstruction}
\end{figure}

\Cref{fig:peanoConstruction} shows the construction of the 2D Peano curve. This is the first example of a SFC given by Peano in 1890 \cite{bader2012}. Note that while the Hilbert curve uses a $2^2$-tree, this Peano curve uses a $3^2$-tree, i.e.\ $k = 3$. There are also versions of the Peano curve for $k = 2m + 1, m \in \natOne$. The Peano curve satisfies most of the properties that the Hilbert curve has. In contrast to the Hilbert curve, it also satisfies the \emph{palindrome property}: At a boundary of two neighboring squares, the neighboring children are ordered inversely to each other \cite{bader2012}. For example, the node $w \equalDef (1, 1, R)$ is the right neighbor of $v \equalDef (1, 0, P)$ and the children $(2, 9, R), (2, 14, S), (2, 15, R)$ of $w$ are the right neighbors of the children $(2, 8, P), (2, 3, Q), (2, 2, P)$ of $v$. Here, the palindrome property implies that the indices of these child neighbors add up to $3^2 - 1 = 9 - 1 = 8$: Indeed, we find that $0 + 8 = 5 + 3 = 6 + 2 = 8$. To see that this does not work for the Hilbert curve, consider e.g.\ the upper neighbor $w = (1, 1, H)$ of $v = (1, 0, A)$ in \Cref{fig:hilbertConstruction2}.

A disadvantage of the 2D Peano curve (and other Peano curves) is that the branching factor $b = 3^2$ is not a power of two. This means that the arithmetic operations used for computing parents and children of nodes like division by $b$, multiplication with $b$ or computing modulo $b$ cannot be represented using bit-operations, so they are potentially slower. Another disadvantage is that because of the bigger branching factor $b$, adaptive refinement can be controlled less precisely.

\begin{figure}[htb]
\centering
\includegraphics[scale=0.8]{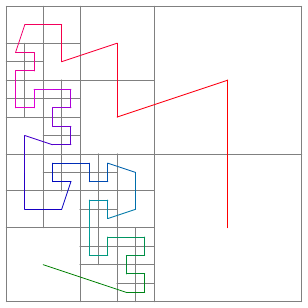}
\caption{The 2D Hilbert curve on an adaptive grid.} \label{fig:hilbert2DAdaptive}
\end{figure}

\Cref{fig:hilbert2DAdaptive} shows the 2D Hilbert curve on an adaptively refined grid, i.e.\ where not all of the nodes have the same level. In an adaptive grid, the question of specifying a neighbor or even finding these neighbors becomes more difficult. These questions are outside of the scope of this thesis. However, we think that it is possible to generalize the results presented here for adaptive grids, if the grid cells are efficiently accessible via their node encoding $v = (l, j, s)$. %

\begin{figure}[htb]
\centering
\subcaptionbox{Level $0$ \label{fig:mortonConstruction:l0}}{
\includegraphics[scale=\curveScale]{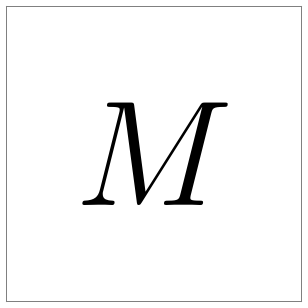}
}
\subcaptionbox{Level $1$ \label{fig:mortonConstruction:l1}}{
\includegraphics[scale=\curveScale]{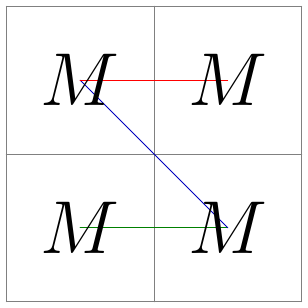}
}
\subcaptionbox{Level $2$ \label{fig:mortonConstruction:l2}}{
\includegraphics[scale=\curveScale]{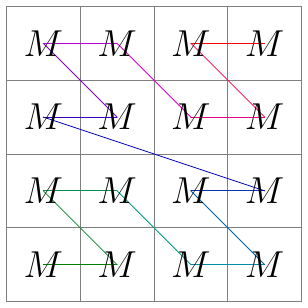}
}
\caption{Construction of the 2D Morton curve.} \label{fig:mortonConstruction}
\end{figure}

A particularly simple curve is the Morton curve shown in \Cref{fig:mortonConstruction}, also known as Morton code, Lebesgue curve, Morton order or Z-order curve. %
It is built from only one pattern and is popular due to the simplicity and efficiency of the algorithms that are used to deal with it. For example, there is a $\compLeq{1}$ neighbor-finding algorithm with a small constant by Schrack \cite{schrack1992}. Its drawback is that it is not a SFC in the sense that it does not converge to a continuous curve. This means that it has worse locality properties than the other curves presented here \cite{bader2012}. While the algorithms shown in this paper can be applied to the Morton curve, already existing algorithms perform better on it.

\begin{figure}[htb]
\centering
\subcaptionbox{Level $0$ \label{fig:sierpinskiConstruction:l0}}{
\includegraphics[scale=\curveScale]{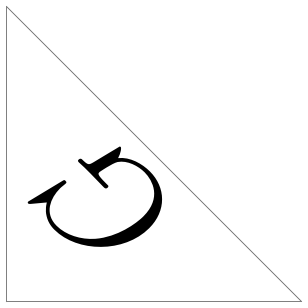}
}
\subcaptionbox{Level $1$ \label{fig:sierpinskiConstruction:l1}}{
\includegraphics[scale=\curveScale]{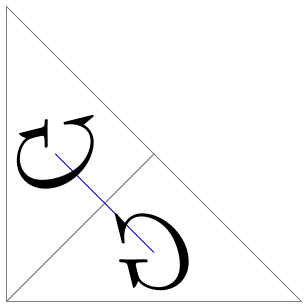}
}
\subcaptionbox{Level $2$ \label{fig:sierpinskiConstruction:l2}}{
\includegraphics[scale=\curveScale]{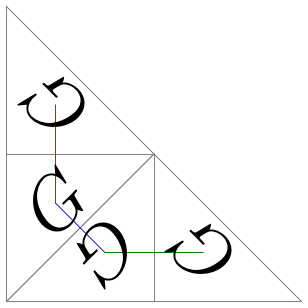}
}
\subcaptionbox{Level $3$ \label{fig:sierpinskiConstruction:l3}}{
\includegraphics[scale=\curveScale]{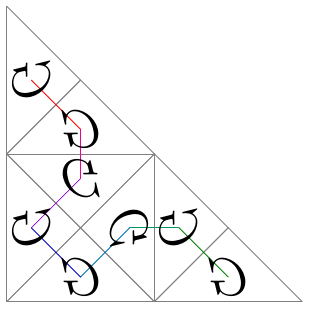}
}
\subcaptionbox{Level $4$ \label{fig:sierpinskiConstruction:l4}}{
\includegraphics[scale=\curveScale]{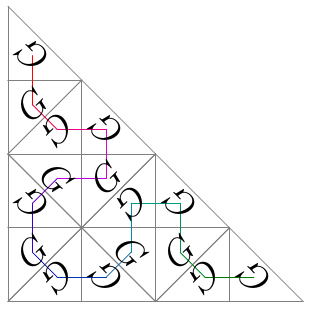}
}
\subcaptionbox{Level $5$ \label{fig:sierpinskiConstruction:l5}}{
\includegraphics[scale=\curveScale]{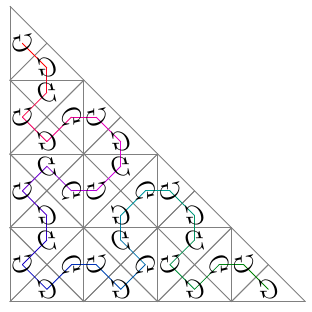}
}
\caption{Construction of the 2D Sierpinski curve.} \label{fig:sierpinskiConstruction}
\end{figure}

Up to now, we have only considered SFCs on $k^d$-trees. The Sierpinski curve, also known as the Sierpinski-Knopp curve, uses simplices (i.e.\ triangles in the 2D case) instead of hypercubes. This means that it can be applied to suitable triangular meshes. In the construction in \Cref{fig:sierpinskiConstruction}, all triangles have the same state $G$, but in different rotated and mirrored versions. To be more precise, not the state is rotated, but their local coordinate system, a concept that will be explained in \Cref{ex:localHilbertModel}.
We will refer to such a model as a \emph{local} model, since the state of a node does not reveal the node's orientation in the global coordinate system. This is a choice of model that is particularly comfortable for the 2D Sierpinski curve: A \emph{global} model like those of the Hilbert and Peano curves above would require eight different states since the right-angled edges of the triangles can point to eight different directions. In \Cref{fig:sierpinskiConstruction}, it can be seen that the Sierpinski curve also satisfies the palindrome property. 

For local models, the neighbor-finding algorithm has to be extended: Consider for example the nodes $v = (2, 1, G)$ and $w = (2, 2, G)$ in the Sierpinski curve. These nodes share a common facet, the hypotenuse of their respective triangles. Their parents $(1, 0, G)$ and $(1, 1, G)$ also share a common facet, but it is not their hypotenuse. This means that when finding neighbors of parents in the recursion step of the neighbor-finding algorithm, the neighbor at a possibly different facet has to be found. This problem can be resolved using a third lookup table which will be introduced in \Cref{sec:modeling:algNeighbors}.

\begin{figure}[htb]
\centering
\subcaptionbox{Level $0$ \label{fig:hilbertConstructionLocal:l0}}{
\includegraphics[scale=\curveScale]{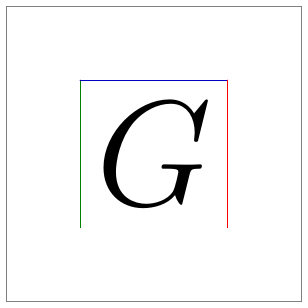}
}
\subcaptionbox{Level $1$ \label{fig:hilbertConstructionLocal:l1}}{
\includegraphics[scale=\curveScale]{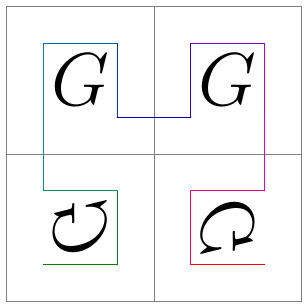}
}
\subcaptionbox{Level $2$ \label{fig:hilbertConstructionLocal:l2}}{
\includegraphics[scale=\curveScale]{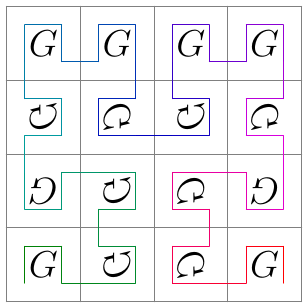}
}
\caption{Local Construction of the 2D Hilbert curve, where the curve is one level finer than the grid.} \label{fig:hilbertConstructionLocal}
\end{figure}

\Cref{fig:hilbertConstructionLocal} shows such a local model for the Hilbert curve. There are four possible orientations of a square, corresponding to the four states $H, A, B, R$ in the global model and to the four group elements from \Cref{remark:hilbert2DGroup}. We used $G$ instead of $H$ since $H$ is a symmetric letter. 

\begin{remark} \label{remark:hilbertLocalNoRegularity}
In the local model of the Hilbert curve, an assumption from \Cref{ex:neighbors:2} is violated: The nodes $v = (2, 0, G)$ and $v = (2, 5, G)$ have the same state and their right neighbors also have the same state, but they have a different orientation. This means that the second lookup table is not well-defined. There are two ways to circumvent this: extending the lookup tables or choosing a global model instead. To avoid more complications, we choose the second approach in this thesis. This problem will be revisited in \Cref{remark:regularityModeling}.
\end{remark}

\begin{figure}[htb]
\centering
\subcaptionbox{Level $0$ \label{fig:peanoConstructionLocal:l0}}{
\includegraphics[scale=\curveScale]{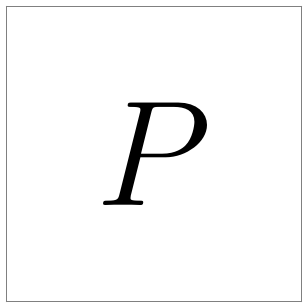}
}
\subcaptionbox{Level $1$ \label{fig:peanoConstructionLocal:l1}}{
\includegraphics[scale=\curveScale]{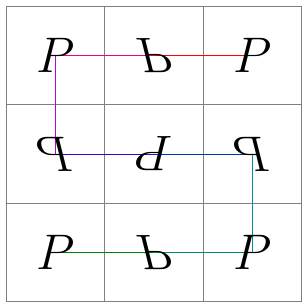}
}
\subcaptionbox{Level $2$ \label{fig:peanoConstructionLocal:l2}}{
\includegraphics[scale=\curveScale]{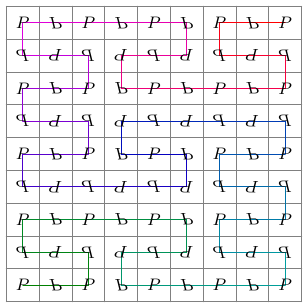}
}
\caption{Local Construction of the 2D Peano curve.} \label{fig:peanoConstructionLocal}
\end{figure}

The Peano curve can also be modeled locally. This is shown in \Cref{fig:peanoConstructionLocal}. In \Cref{remark:hilbertLocalNoRegularity}, we have seen that the local model of the Hilbert curve poses some problems. The palindrome of the Peano curve implies that the orientation of a square and the neighbor direction already uniquely determine the state of a neighbor in that direction. Thus, these problems do not occur for the Peano curve.

\newcommand{\boCurveScale}{1.0}
\begin{figure}[htb]
\centering
\subcaptionbox{Level $0$ \label{fig:betaOmegaConstruction:l0}}{
\includegraphics[scale=\boCurveScale]{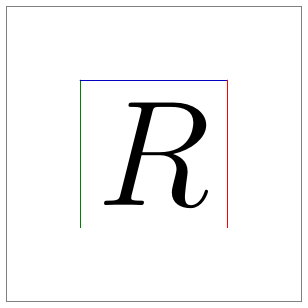}
}
\subcaptionbox{Level $1$ \label{fig:betaOmegaConstruction:l1}}{
\includegraphics[scale=\boCurveScale]{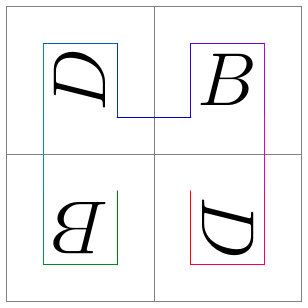}
}
\subcaptionbox{Level $2$ \label{fig:betaOmegaConstruction:l2}}{
\includegraphics[scale=\boCurveScale]{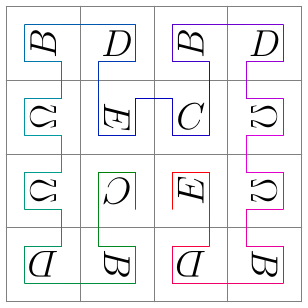}
}
\subcaptionbox{Level $3$ \label{fig:betaOmegaConstruction:l3}}{
\includegraphics[scale=\boCurveScale]{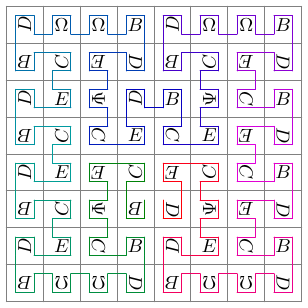}
}
\caption{Construction of the 2D $\beta\Omega$-curve, where the curve is one level finer than the grid.} \label{fig:betaOmegaConstruction}
\end{figure}

The $\beta\Omega$-curve shown in \Cref{fig:betaOmegaConstruction} exposes some new properties. The construction intends to create a closed curve, which comes at some expense: First of all, it needs more states than the curves presented above even though the model from \Cref{fig:betaOmegaConstruction} already uses a partially local approach. More importantly, this is an example of a curve where not all of the functions $s \mapsto \cstate(s, j)$ are invertible. For example, \Cref{fig:betaOmegaConstruction} shows that the parent of a node with state $B$ and index $1$ at level 2 may have state $B$ or state $D$. Another related phenomenon is that the root state $R$ only occurs at the root node.

\begin{figure}[htb]
\centering
\subcaptionbox{Level $0$ \label{fig:gosperConstruction:l0}}{
\includegraphics[scale=\curveScale]{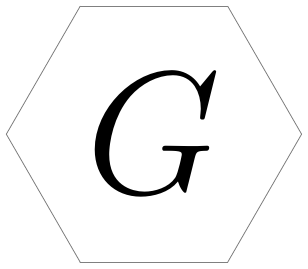}
}
\subcaptionbox{Level $1$ \label{fig:gosperConstruction:l1}}{
\includegraphics[scale=\curveScale]{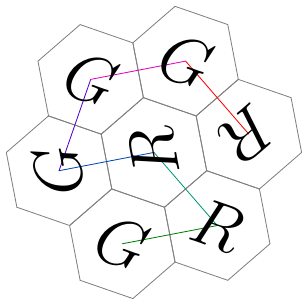}
}
\subcaptionbox{Level $2$ \label{fig:gosperConstruction:l2}}{
\includegraphics[scale=\curveScale]{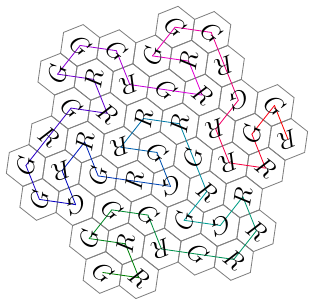}
}
\caption{Construction of the 2D Gosper curve.} \label{fig:gosperConstruction}
\end{figure}

\begin{figure}[htb]
\centering
\includegraphics[scale=1.0]{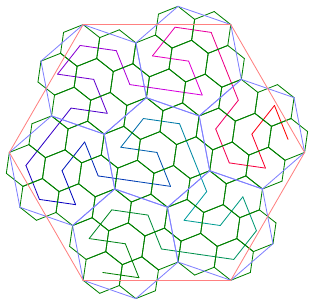}
\caption{Levels 0, 1, 2 of the 2D Gosper curve superimposed.} \label{fig:gosperMultiLevel}
\end{figure}

\Cref{fig:gosperConstruction} shows a semi-local model of the Gosper curve, also called Gosper Flowsnake, using two states. Its tree nodes correspond to hexagons. However, the space filled by the Gosper curve is not a hexagon but a fractal called Gosper island \cite{bader2012}. This is possible because in the Gosper curve, the children of a hexagon do \emph{not} form a partition of the hexagon --- instead, they cover a region neither including nor being included in the parent hexagon. The Gosper curve is thus not well-suited for being used on adaptive grids.  The level-2 curve together with the boundaries of hexagons at levels 0, 1, 2 is shown in \Cref{fig:gosperMultiLevel}. In this model of the Gosper curve, the functions $s \mapsto \cstate(s, j)$ are not all invertible.

It seems geometrically clear, that using a suitable model, the algorithms presented here also work for the Gosper curve. However, the proof of correctness given here is partially restricted to curves where all children are contained in the parent polygon and thus does not apply to the Gosper curve. Moreover, the semi-local model of the Gosper curve given here suffers from the problem described in \Cref{remark:hilbertLocalNoRegularity}.

\begin{figure}[htb]
\centering
\subcaptionbox{Level $1$ \label{fig:hilbert3DConstruction:l1}}{
\includegraphics[scale=0.85]{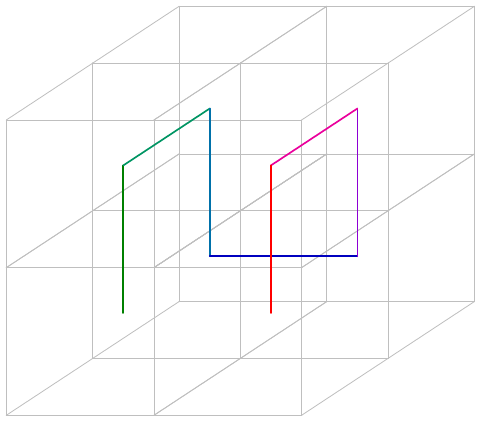}
}
\subcaptionbox{Level $2$ \label{fig:hilbert3DConstruction:l2}}{
\includegraphics[scale=0.85]{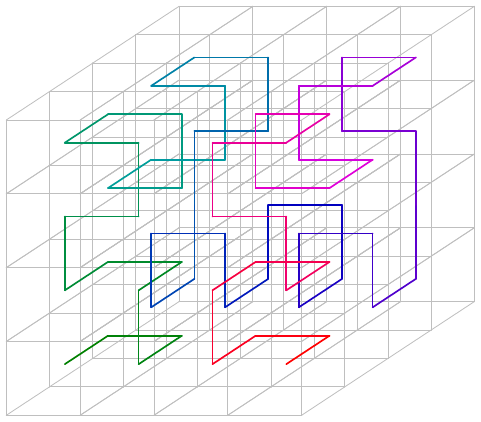}
}
\caption{Construction of the 3D Hilbert curve.} \label{fig:hilbert3DConstruction}
\end{figure}

Up to now, we have only considered two-dimensional SFCs. For many applications, SFCs in dimensions $d \geq 3$ are needed. Curves like Hilbert, Peano, Morton and Sierpinski allow higher-dimensional variants. The Hilbert and the Peano curves have many variants in each dimension $d \geq 2$ and for the Hilbert curve, there is no \quot{canonical} variant even in 3D. \Cref{fig:hilbert3DConstruction} shows two levels of a 3D Hilbert curve. The 3D Hilbert curve yields a sequential order on a $2^3$-tree, also called octree. The book by Bader \cite{bader2012} presents many more SFCs. We suppose that the methods presented here work for all of these SFCs except possibly for the H-index, a modification of the Sierpinski curve that works on squares and is not presented here.

\cleardoublepage

\section{Modeling Space-Filling Curves} \label{sec:modeling}

In this section, we show a general way of modeling common space-filling curves geometrically. This model essentially expresses the vertex-labeling method (see Bartholdi and Goldsman \cite{bartholdi2001}) using matrices. While the original vertex-labeling method only used local models of SFCs, we also allow multiple states in our model so that it applies to more SFCs. We then show how to abstract the geometric information that is contained in this model to an integer-based model that can be efficiently processed using algorithms to find neighbors of tree cells. An implementation of this transformation process is available in the \sfcpp\ library. We intend to make every aspect of this modeling approach algorithmically verifiable. However, at the time of writing, only some aspects are implemented. %

\subsection{Trees, Matrices and States} \label{sec:modeling:treesMatricesStates}

In the following, we will derive a modeling framework for SFCs following the example of the Hilbert curve in two dimensions as examined in \Cref{sec:overview}. When a level of the Hilbert curve is refined to obtain the next level, each square is subdivided into four smaller squares. The squares in the Hilbert curve form a tree: Each square contains four subsquares in the next level, these can be viewed as children in a tree. The children can be ordered in the order the curve passes through them in the corresponding level. We can define the notion of such a tree abstractly:

\begin{definition} \label{def:btree}
Let $b \in \bbN$ with $b \geq 2$. We define the set $\childIndices \equalDef \{0, \hdots, b-1\}$ of indices. A $b$-index-tree is a tuple $(\vertices, \child, \parent, \level, \rootv)$ that satisfies the following conditions:
\begin{itemize}
\item $\vertices$ is a set and $\rootv \in \vertices$. Elements of $\vertices$ are called (tree) nodes (or vertices). $\rootv$ is called the root node.
\item $\child: \vertices \times \childIndices \to \vertices \setminus \{\rootv\}$ is a bijective function. This implies that $\vertices$ is infinite. $\child(v, j)$ is called the $j$-th child of $v$.
\item $\parent: \vertices \setminus \{\rootv\} \to \vertices$ is a function satisfying $\parent(\child(v, j)) = v$ for all $v \in \vertices, j \in \childIndices$. $\parent(v)$ is called the parent of $v$.
\item $\level: \vertices \to \natZero$ is a function satisfying $\level(\rootv) = 0$ and $\level(v) = \level(\parent(v)) + 1$ for all $v \in \vertices \setminus \{\rootv\}$. This immediately yields $\level(\child(v, j)) = \level(v) + 1$ and $\level(v) = 0 \Rightarrow v = \rootv$. The level function $\level$ thus allows proofs by induction on the level.
\end{itemize}

For $v \in \vertices \setminus \{\rootv\}$, we then define $\idx(v)$ to be the unique index satisfying $\child(\parent(v), \idx(v)) = v$. The root node $\rootv$ and the functions $\parent, \level, \idx$ are all uniquely determined by $\child$ if they exist. Because of this, we will also write $(\vertices, \child)$ instead of $(\vertices, \child, \parent, \level, \rootv)$ for brevity.
\end{definition}

In the case of the 2D Hilbert curve, the branching factor $b$ of the tree is $b = 4 = 2^2$. In the more general setting of a $k^d$-tree, we would have $b = k^d$.

As shown in \Cref{sec:overview}, a square arising in the construction of the Hilbert curve can be described by its level in the tree and its position along the curve. Moreover, the tree operations on such nodes can be efficiently realized because they correspond to simple manipulations in the base-$b$ representation of the position. The following definition makes these operations explicit and yields our first example of a $b$-index-tree:

\begin{definition} \label{def:levelPositionTree}
The \defEmph{level-position $b$-index-tree} $T = (\vertices, \child, \parent, \level, \rootv)$ is defined by
\begin{IEEEeqnarray*}{+rCl+x*}
\vertices & \equalDef & \{(l, j) \in \natZero^2 \mid j < b^l\} \\ %
\rootv & \equalDef & (0, 0) \\
\parent((l, j)) & \equalDef & (l-1, j \intdiv b)  \quad (\text{if } l > 0) \\
\child((l, j), i) & \equalDef & (l+1, jb + i) \\
\level((l, j)) & \equalDef & l~.
\end{IEEEeqnarray*}
Here, $j \intdiv b \equalDef \lfloor j / b \rfloor$ is the integer division. It then follows that $\idx((l, j)) = j \bmod b$ if $l > 0$. 
\end{definition}

The first three levels of a level-position $b$-index-tree are shown in \Cref{fig:levelPositionTree}. A level-position $b$-index-tree contains no information about the underlying SFC except for the branching factor $b$. Our next step is to derive a geometric model and include this as an additional information into the tree. This geometric $b$-index-tree will simplify geometric definitions which can then be transferred to a computationally efficient tree like the level-position $b$-index-tree using an isomorphism. Isomorphisms will be covered in \Cref{sec:modeling:isomorphisms}.

As we have seen in \Cref{sec:overview:otherSFC}, not every SFC is constructed using squares. We will use polytopes for modeling. Polytopes are a generalization of convex polygons in arbitrary dimension. The following definition introduces some basic concepts of convex geometry together with customary conventions for matrices. For a detailed introduction, see Ziegler \cite{ziegler2012}. We follow the convention from Ziegler \cite{ziegler2012} and use the term \quot{polytope} only for convex polytopes. \textbf{From now on, we assume that $d \geq 2$ denotes the dimension of the space we are working in.}

\begin{definition} \label{def:polytope}
Let $k \in \natOne$ and $q_1, \hdots, q_k \in \bbR^d$. 
\begin{enumerate}[(a)]
\item The set
\begin{IEEEeqnarray*}{+rCl+x*}
\convHull{\{q_1, \hdots, q_k\}} \equalDef \left\{\sum_{i=1}^k t_i q_i ~\middle|~ t_i \in [0, 1], \sum_{i=1}^k t_i = 1\right\} \subseteq \bbR^d
\end{IEEEeqnarray*}
is called \defEmph{convex hull} of $q_1, \hdots, q_k$. It is the smallest convex set containing $q_1, \hdots, q_k$. 
For the matrix
\begin{IEEEeqnarray*}{+rCl+x*}
Q & \equalDef & \begin{pmatrix}
| & & | \\
q_1 & \hdots & q_k \\
| & & |
\end{pmatrix}
\end{IEEEeqnarray*}
containing $q_1, \hdots, q_k$ as columns, let $\convHull{Q} \equalDef \convHull{\{q_1, \hdots, q_k\}}$. Furthermore, we set $\convHull{\emptyset} \equalDef \emptyset$.

\item A set $P \subseteq \bbR^d$ is called \defEmph{polytope} if $P = \convHull{S}$ for some finite set $S \subseteq \bbR^d$. Especially, $\emptyset = \convHull{\emptyset}$ is also a polytope. The dimension $\dim(P)$ is defined as the dimension of the affine subspace $\aff(P)$ of $\bbR^d$ spanned by the polytope $P$.

\item Let $P$ be a polytope. A subset $F \subseteq P$ is called a \defEmph{face} of $P$ if there exist $c \in \bbR^d$ and $c_0 \in \bbR$ such that $c^\top x \leq c_0$ for all $x \in P$ and $F = P \cap \{x \in \bbR^d \mid c^\top x = c_0\}$. Again, we define $\dim(F) \equalDef \dim(\aff(F))$.

Let $F$ be a face of $P$. If $\dim(F) = \dim(P) - 1$, $F$ is called a \defEmph{facet} of $P$. 
If $F = \{v\}$ for some $v \in \bbR^d$, i.e.\ $\dim(F) = 0$, $v$ is called a \defEmph{vertex} of $P$.
We denote by $\polytopeVert(P)$ the set of all vertices of $P$.
\end{enumerate}
\end{definition}

We can now define the initial level-0 square $[0, 1]^2$ of the Hilbert curve construction via $[0, 1]^2 = \convHull{\rootPoints}$, where $\rootPoints$ contains the four vertices of the square:
\begin{IEEEeqnarray*}{+rCl+x*}
\rootPoints & \equalDef & \begin{pmatrix}
0 & 1 & 0 & 1 \\
0 & 0 & 1 & 1
\end{pmatrix}~. \IEEEyesnumber \label{eq:hilbert2DRootPoints}
\end{IEEEeqnarray*}
There are different choices for $\rootPoints$ generating the same polytope, since permuting the vertices or inserting more points of the square as columns does not affect $\convHull{\rootPoints}$.

To include our geometrical knowledge about the Hilbert curve into the tree nodes, we want to find such a matrix $Q$ for every node. The columns of $Q$ should contain the vertices of the associated square. Then, the associated square is given by $\convHull{Q}$. For example, a corresponding matrix $L$ of the lower left subsquare of the matrix $\rootPoints$ from Equation \eqref{eq:hilbert2DRootPoints} can be obtained using matrix multiplication:
\begin{IEEEeqnarray*}{+rCl+x*}
L & = & \rootPoints M = \begin{pmatrix}
| & & | \\
l_1 & \hdots & l_k \\
| & & |
\end{pmatrix}, \text{ where } M \equalDef \begin{pmatrix}
1 & 1/2 & 1/2 & 1/4 \\
0 & 1/2 & 0 & 1/4 \\
0 & 0 & 1/2 & 1/4 \\
0 & 0 & 0 & 1/4
\end{pmatrix} \text{ and } l_j = \sum_{i=1}^4 M_{ij} q_i.
\end{IEEEeqnarray*}
Similarly, $\rootPoints \cdot M \cdot M$ is the lower left subsquare of the lower left subsquare of $\rootPoints$ and so forth. 
It is important to note that all columns of $M$ sum to one. As we will see in \Cref{lemma:specialMatrixProperties}, this makes the operation $Q \mapsto Q \cdot M$ commute with translations. First, we will introduce some helpful terminology:

\begin{definition} \label{def:transitionMatrix} \label{def:offsetMatrix}
Let $\oneVec{n} \equalDef (1, \hdots, 1)^\top \in \bbR^n$.
\begin{enumerate}[(a)]
\item A matrix $M \in \bbR^{m \times n}$ is called \defEmph{transition matrix} if $\oneVec{m}^\top M = \oneVec{n}^\top$, that is, the entries of each column sum to one.
\item A matrix $B \in \bbR^{d \times m}$ is called \defEmph{offset matrix} if there exists $b \in \bbR^d$ with $B = b\oneVec{m}^\top$, that is, all columns of $B$ are identical.
\end{enumerate}
\end{definition}

\begin{lemma} \label{lemma:specialMatrixProperties}
Let $B = b\oneVec{m}^\top \in \bbR^{d \times m}$ be an offset matrix.
\begin{enumerate}[(a)]
\item For any matrix $A \in \bbR^{d \times d}$, $AB = (Ab)\oneVec{m}^\top$ is an offset matrix.
\item For any transition matrix $M \in \bbR^{m \times n}$, $BM = b\oneVec{m}^\top M = b\oneVec{n}^\top$ is an offset matrix with the same columns as $B$. Especially, if $m = n$, we have $BM = B$.
\item The product of two transition matrices is a transition matrix.
\end{enumerate}

\begin{proof}
It remains to prove (c): Let $M \in \bbR^{m \times n}, N \in \bbR^{n \times p}$ be transition matrices, then $\oneVec{m}^\top MN = \oneVec{n}^\top N = \oneVec{p}^\top$.
\end{proof}
\end{lemma}

We can (by definition of $\convHull{Q}$) choose all entries of $M$ to be non-negative if and only if $\convHull{L} = \convHull{QM} \subseteq \convHull{Q}$, since elements in $\convHull{Q}$ are exactly those vectors that are a convex combination of columns of $Q$. %
However, we need to allow negative entries if we want to be able to model the Gosper curve, see \Cref{fig:gosperConstruction}. In some cases, using negative values might also simplify modeling.

Once we have found transition matrices for the four subsquares of a square in the Hilbert curve construction, we have to decide how to order them in order to obtain the Hilbert curve. As explained in \Cref{sec:overview} the order of the subsquares is determined by the state $s \in \states \equalDef \{H, A, B, R\}$ of a square. Furthermore, the function $\cstate$ allows us to track the state of an element when descending in a tree. The following definition introduces such a model abstractly:

\begin{definition} \label{def:bStateSystem}
Let $\states$ be a finite set, $\rootState \in \states$ and let $\cstate: \states \times \childIndices \to \states$. The tuple $\bStateSystem = (\states, \cstate, \rootState)$ is called \defEmph{$b$-state-system}, if every state $s \in \calS$ is reachable from the root state $\rootState$, i.\,e. there exist $n \in \natZero, s_0, \hdots, s_n \in \states$ and $j_1, \hdots, j_n \in \childIndices$ such that $s_0 = \rootState, s_n = s$ and $s_{k} = \cstate(s_{k-1}, j_k)$ for $k \in \{1, \hdots, n\}$.
\end{definition}

\begin{remark}
The reachability condition in \Cref{def:bStateSystem} can be easily satisfied by restricting the set of states to all reachable states. It is useful to simplify some definitions but will not be used throughout most of this thesis.
\end{remark}

Together with appropriate transition matrices and a matrix for the root square, a $b$-state-system can be used to specify a SFC:

\begin{definition} \label{def:bSpecification}
Let $\bStateSystem$ be a $b$-state-system as in \Cref{def:bStateSystem}. A tuple $(\bStateSystem, \transitionMat, \rootPoints)$ is called \defEmph{$b$-specification} if there exist $n_s \in \natOne$ for $s \in \states$ (specifying the number of vertices of a polytope of state $s$) such that
\begin{enumerate}[(a)]
\item $\rootPoints \in \bbR^{d \times n_{\rootState}}$ and
\item $\transitionMat: \states \times \childIndices \to \bigcup_{n, m \in \natOne} \bbR^{n \times m}$ is a function such that for all $s \in \states, j \in \childIndices$, $\transitionMat^{s,j} \in \bbR^{n_s \times n_{\cstate(s, j)}}$ is a transition matrix.
\end{enumerate}
\end{definition}

\begin{figure}[htb]
\centering
\subcaptionbox{Level $0$ \label{fig:hilbertCorners:l0}}{
\includegraphics[scale=\curveScale]{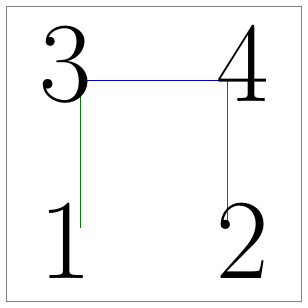}
}
\subcaptionbox{Level $1$ \label{fig:hilbertCorners:l1}}{
\includegraphics[scale=\curveScale]{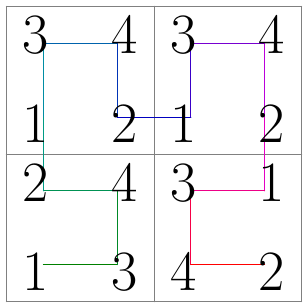}
}
\subcaptionbox{Level $2$ \label{fig:hilbertCorners:l2}}{
\includegraphics[scale=\curveScale]{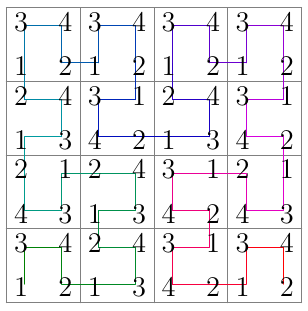}
}
\caption{Vertex column indices in the local model of the 2D Hilbert curve.} \label{fig:hilbertCorners}
\end{figure}

\begin{example} \label{ex:localHilbertModel}
In this example, we define a local $b$-specification of the Hilbert curve, where $b = 4$. This specification corresponds to the local construction of the Hilbert curve shown in \Cref{fig:hilbertConstructionLocal}. Our model only has one state $G$, i.e.\ $\states = \{G\}$. The child state function is thus $\cstate: \states \times \childIndices \to \states, (G, j) \mapsto G$. The root state is $\rootState = G$. The root point matrix $\rootPoints$ is the same as in Equation \eqref{eq:hilbert2DRootPoints}. The only interesting part of this specification are the transition matrices. In the global model, the point matrix belonging to the lower left subsquare of $\rootPoints$ is
\begin{IEEEeqnarray*}{+rCl+x*}
L & = & \begin{pmatrix}
0 & 1/2 & 0 & 1/2 \\
0 & 0 & 1/2 & 1/2
\end{pmatrix}~.
\end{IEEEeqnarray*}
Because $L$ has the same state as $\rootPoints$ in the local model, we have to mirror its coordinate system to modify the arrangement of its subsquares. In this case, we can do this by exchanging the order of the second and third vertex and setting
\begin{IEEEeqnarray*}{+rCl+x*}
L & = & \begin{pmatrix}
0 & 0 & 1/2 & 1/2 \\
0 & 1/2 & 0 & 1/2
\end{pmatrix}~.
\end{IEEEeqnarray*}
\Cref{fig:hilbertCorners} shows for each square the order of its vertices in the matrix from the local specification. We can achieve this reordering using correspondingly permuted transition matrices:
\begin{IEEEeqnarray*}{+rClrCl+x*}
M^{G, 1} & = & \begin{pmatrix}
1/2 & 1/4 & 0 & 0 \\
0 & 1/4 & 0 & 0 \\
1/2 & 1/4 & 1 & 1/2 \\
0 & 1/4 & 0 & 1/2
\end{pmatrix} & \qquad M^{G, 2} & = & \begin{pmatrix}
1/4 & 0 & 0 & 0 \\
1/4 & 1/2 & 0 & 0 \\
1/4 & 0 & 1/2 & 0 \\
1/4 & 1/2 & 1/2 & 1
\end{pmatrix}  \\
M^{G,0} & = & \begin{pmatrix}
1 & 1/2 & 1/2 & 1/4 \\
0 & 0 & 1/2 & 1/4 \\
0 & 1/2 & 0 & 1/4 \\
0 & 0 & 0 & 1/4
\end{pmatrix} & M^{G, 3} & = & \begin{pmatrix}
0 & 0 & 1/4 & 1/2 \\
1/2 & 1 & 1/4 & 1 \\
0 & 0 & 1/4 & 0 \\
1/2 & 0 & 1/4 & 0
\end{pmatrix}~.
\end{IEEEeqnarray*}

As stated in \Cref{remark:hilbertLocalNoRegularity}, the local specification of the Hilbert curve cannot be directly used for the algorithms presented here. In the following, we thus have to revert to global models of the Hilbert curve.
\end{example}

Now that we have seen how to specify a space-filling curve, we want to turn this specification into a tree. Since this specification includes states, the tree should also include states. We will thus first augment the definition of a $b$-index-tree to include states:

\begin{definition} \label{def:recbtree}
Let $(\vertices, \child)$ be a $b$-index-tree as in \Cref{def:btree}, let $\states$ be a finite set and let $\state: \vertices \to \states$ be a function. Then, $T_s = (\vertices, \child, \states, \state)$ is called a state-$b$-index-tree. For $v \in \vertices$, $\state(v)$ is called the \defEmph{state} of $v$.
\end{definition}

\begin{remark} \label{remark:statesFromStateSystem}
A state-$b$-index-tree can be obtained from a $b$-index-tree and a $b$-state-system by setting $\state(\rootv) \equalDef \rootState$ and then inductively defining
\begin{IEEEeqnarray*}{+rCl+x*}
\state(\child(v, j)) & \equalDef & \cstate(\state(v), j)~. & \qedhere
\end{IEEEeqnarray*}
\end{remark}

From a $b$-specification, we can now build a tree where each tree node comprises the following components: 
\begin{itemize}
\item the level of the node,
\item its position in the curve inside the given level,
\item its state, and
\item a point matrix specifying its corresponding polytope.
\end{itemize}

The corresponding model is introduced in the following definition.

\begin{definition} \label{def:geometricBTree}
Let $(\bStateSystem, \transitionMat, \rootPoints)$ be a $b$-specification as in \Cref{def:bSpecification}.
Out of a basic set 
\begin{IEEEeqnarray*}{+rCl+x*}
\tilde{\vertices} \equalDef \left\{(l, j, s, Q) \mid l, j \in \natZero, s \in \states, Q \in \bbR^{d \times n_s}\right\}
\end{IEEEeqnarray*}
with root node $\rootv \equalDef (0, 0, \rootState, \rootPoints)$ and child function
\begin{IEEEeqnarray*}{+rCl+x*}
\tilde{\child}: \tilde{\vertices} \times \childIndices \to \tilde{\vertices}, ((l, j, s, Q), i) & \mapsto & (l+1, jb + i, \cstate(s, i), Q \cdot M^{s, i})~,
\end{IEEEeqnarray*}
we construct our node set recursively:
\begin{IEEEeqnarray*}{+rCl+x*}
\vertices_0 & \equalDef & \{r\} \\
\vertices_{l+1} & \equalDef & \{\tilde{\child}(v, j) \mid v \in \vertices_l, j \in \childIndices\} \\
\vertices & \equalDef & \bigcup_{l=0}^\infty \vertices_l~.
\end{IEEEeqnarray*}
Then, we set our child function on $\vertices$ to $\child \equalDef \tilde{\child}|_{\vertices \times \childIndices}^\vertices$. The restriction of the codomain to $\vertices$ is valid by construction of $\vertices$. For $v = (l, j, s, Q) \in \vertices$, define
\begin{IEEEeqnarray*}{+rCl+x*}
\state(v) & \equalDef & s~, \\
Q(v) & \equalDef & Q~.
\end{IEEEeqnarray*}
Note that the function $v \mapsto Q(v)$ does not belong to the required functions for a state-$b$-index-tree but will be frequently used for geometric purposes later. With these definitions, $(\vertices, \child, \states, \state)$ is a state-$b$-index-tree with root node $R$ and
\begin{IEEEeqnarray*}{+rCl+x*}
\level((l, j, s, Q)) & = & l \\
\idx((l, j, s, Q)) & = & j \bmod b  \quad (\text{if } l > 0)~.
\end{IEEEeqnarray*}

We call this tree the \defEmph{geometric $b$-index-tree} for the $b$-specification $(\bStateSystem, \transitionMat, \rootPoints)$.
\end{definition}

The geometric $b$-index-tree will be used to define neighborship in space-filling curves geometrically. It can also be implemented for rendering space-filling curves, which has been done to generate most of the figures in this thesis.

Next, we want to introduce two trees that are convenient for computations because they contain information about the state of a node, but not its point matrix. The operations of the first tree are easier and more efficient to implement if the functions $s \mapsto \cstate(s, j)$ are invertible for each $j \in \childIndices$. Otherwise, the second tree may be used.

\begin{definition} \label{def:tree:algebraicWithoutHistory}
Let $\bStateSystem$ be a $b$-state-system. Similar to \Cref{def:geometricBTree}, we can define a state-$b$-index-tree $T = (\vertices, \child, \states, \state)$ satisfying
\begin{IEEEeqnarray*}{+rCl+x*}
\vertices & \subseteq & \natZero \times \natZero \times \states \\
\rootv & = & (0, 0, \rootState) \\
\child((l, j, s), i) & = & (l+1, jb + i, \cstate(s, i)) \\
\level((l, j, s)) & = & l \\
\idx((l, j, s)) & = & j \bmod b \quad (\text{if } l > 0) \\
\state((l, j, s)) & = & s
\end{IEEEeqnarray*}
for every $(l, j, s) \in \vertices$. We call $T$ the \defEmph{algebraic $b$-index-tree without history} for $\bStateSystem$.

If $s \mapsto \cstate(s, j)$ is in\-ver\-ti\-ble for every $j \in \childIndices$ with inverse function $s \mapsto \pstate(s, j)$, we also have
\begin{IEEEeqnarray*}{+rCl+x*}
\parent((l, j, s)) & = & (l-1, j \intdiv b, \pstate(s, j \bmod b))
\end{IEEEeqnarray*}
for all $(l, j, s) \in \vertices \setminus \{\rootv\}$. In this case, assuming constant-time integer arithmetic operations, all functions of $T$ can be realized in $\compLeq{1}$ time. If $b$ is a power of two, the operations on the position only operate on a constant number of bits.
\end{definition}

For the Morton, Hilbert, Peano and Sierpinski curves, the functions $s \mapsto \cstate(s, j)$ are invertible. For the semi-local models of the $\beta\Omega$-curve and the Gosper curve from \Cref{sec:overview:otherSFC}, this is not true. In this case, we also need to store the states of ancestors in a node:

\begin{definition} \label{def:tree:algebraicWithHistory}
Let $\bStateSystem$ be a $b$-state-system. Similar to \Cref{def:geometricBTree}, we can define a state-$b$-index-tree $T = (\vertices, \child, \states, \state)$ satisfying
\begin{IEEEeqnarray*}{+rCl+x*}
\vertices & \subseteq & \natZero \times \natZero \times \bigcup_{n=1}^\infty \states^n \\
\rootv & = & (0, 0, \rootState) \\
\child((l, j, (s_p, s), i) & = & (l+1, jb + i, (s_p, s, \cstate(s, i))) \\
\parent((l, j, (s_p, s))) & = & (l-1, j \intdiv b, s_p) \text{ if $l > 0$} \\
\level((l, j, (s_p, s))) & = & l \\
\idx((l, j, (s_p, s))) & = & j \bmod b \\
\state((l, j, (s_p, s))) & = & s
\end{IEEEeqnarray*}
for every $(l, j, (s_p, s)) \in \vertices$ where $s_p$ is a (possibly empty) tuple of states. We call $T$ the \defEmph{algebraic $b$-index-tree with history} for $\bStateSystem$.

At first, it might seem that not all of these operations are realizable in $\compLeq{1}$ since they involve creating tuples of arbitrary length. However, we will show in \Cref{sec:implementation:otherCode} how to implement all operations in $\compLeq{1}$ using partial trees for the state tuples.
\end{definition}

\subsection{Algebraic Neighbors} \label{sec:modeling:algNeighbors}

Now, we want to define what it means for a node $w$ to be a neighbor of $v$. First of all, we need a finite set $\facets$ whose elements encode the facets of a node, i.e.\ the directions where a neighbor can be looked for. For the 2D Hilbert curve, we might for example choose $\facets = \{0, 1, 2, 3\}$, where 0 encodes the left, 1 the right, 2 the lower and 3 the upper side of a square. If a node $v$ has a neighbor $w$ in direction $f \in \facets$, we distinguish two cases:
\begin{itemize}
\item Case 1: $\parent(v) = \parent(w)$. In this case, we assume that the index $\idx(w)$ of $w$ is uniquely determined by $\idx(v)$, $\state(\parent(v))$ and $f$. We can model this relationship by a function $\nfunc$:
\begin{IEEEeqnarray*}{+rCl+x*}
\idx(w) & = & \nfunc(\idx(v), \state(\parent(v)), f)~.
\end{IEEEeqnarray*}

\item Case 2: $\parent(v) \neq \parent(w)$. In this case, we assume that $\nfunc(\idx(v), \state(\parent(v)), f) = \ndef$, where $\ndef$ is a value representing an undefined result. In addition, we assume that $\parent(w)$ is a neighbor of $\parent(v)$ in direction $f_p \in \facets$, where $f_p$ can be modeled by another functional relationship:
\begin{IEEEeqnarray*}{+rCl+x*}
f_p & = & \pface(\idx(v), \state(\parent(v)), f)~.
\end{IEEEeqnarray*}
For most global models, we can assume that $f_p = f$, i.e.\ the direction of the neigbors stays the same. In local models, this is generally not the case due to the change of coordinate system. Finally, we assume that $\idx(w)$ is given by a third functional relationship:
\begin{IEEEeqnarray*}{+rCl+x*}
\idx(w) & = & \ofunc(\idx(v), \state(\parent(v)), \state(\parent(w)), f)~.
\end{IEEEeqnarray*} 
This is the case if the states of the parents provide enough information to know which of their children can be neighbors in which direction.
\end{itemize}

\begin{definition} \label{def:ndef}
We use $\ndef$ as a not otherwise used object representing an undefined or invalid result. For any set $M$, let $M_\ndef \equalDef M \cup \{\ndef\}$. %
\end{definition}

We can now take the three functions $\nfunc, \ofunc$ and $\pface$ that were derived from geometric intuition above and use them to define neighborship:

\begin{definition} \label{def:neighborbtree}
A tuple $T = (T_s, \facets, \nfunc, \ofunc, \pface)$ is called \defEmph{neighbor-$b$-index-tree} if
\begin{itemize}
\item $T_s = (\vertices, \child, \states, \state)$ is a state-$b$-index-tree as in \Cref{def:recbtree}.
\item $\facets$ is a finite set, called the set of facets.
\item $\nfunc: \childIndices \times \states \times \facets \to \childIndices_\ndef$ is a function.
\item $\ofunc: \childIndices \times \states \times \states \times \facets \to \childIndices_\ndef$ is a function.
\item $\pface: \childIndices \times \states \times \facets \to \facets_\ndef$ is a function.
\end{itemize}

Let $f \in \facets$, $v, w \in \vertices \setminus \{\rootv\}$ and  $j \equalDef \nfunc(\idx(v), \state(\parent(v)), f)$. We define neighborship inductively as follows: 
\begin{itemize}
\item $w$ is called a depth-1 $f$-neighbor of $v$ if $j \neq \ndef$ and $w = \child(\parent(v), j)$.
\item Let $\depth \geq 2$. $w$ is called a depth-$\depth$ $f$-neighbor of $v$ if
\begin{itemize}
\item $j = \ndef$,
\item $f_p \equalDef \pface(\idx(v), \state(\parent(v)), f) \neq \ndef$,
\item $\parent(w)$ is a depth-$(\depth-1)$ $f_p$-neighbor of $\parent(v)$, and
\item $\idx(w) = \ofunc(\idx(v), \state(\parent(v)), \state(\parent(w)), f)$.
\end{itemize}
\item $w$ is called an $f$-neighbor of $v$ if there exists $\depth \geq 1$ such that $w$ is a depth-$\depth$ $f$-neighbor of $v$.
\end{itemize}
\end{definition}

Although this definition makes no assumptions about the functions $\nfunc, \ofunc$ and $\pface$, we can still derive some intuitive properties of neighborship:

\begin{lemma} \label{lemma:neighborProperties}
Let $v \in \vertices$ and $f \in \facets$. If there is a depth-$\depth$ $f$-neighbor $w$ of $v$, it is its only $f$-neighbor, $\depth$ is unique, $\level(w) = \level(v)$ and $\depth \leq \level(v)$.

\begin{proof}
We prove this by induction on $\level(v)$. If $\level(v) = 0$, there is by definition no neighbor of $v$.

Now assume that the claim is true up to level $l \geq 0$ and let $\level(v) = l+1$. Furthermore, let $j \equalDef \nfunc(\idx(v), \state(\parent(v)), f)$. Suppose that $w$ is a depth-$\depth$ $f$-neighbor of $v$ and $u$ is a depth-$\depth'$ $f$-neighbor of $v$. We distinguish two cases:
\begin{enumerate}[(1)]
\item If $j \neq \ndef$, it follows that $\depth = \depth' = 1 \leq \level(v)$. But then, $u = \child(\parent(v), j) = w$. The levels are identical because $\level(w) = \level(\child(\parent(v), j)) = \level(\parent(v)) + 1 = \level(v)$.
\item If $j = \ndef$, then $k, k' > 1$. Thus, $\parent(w)$ must be a depth-$(\depth - 1)$ $\pface(\idx(v), \state(\parent(v)), f)$-neighbor of $\parent(v)$ and $\parent(u)$ must be a depth-$(\depth' - 1)$ $\pface(\idx(v), \state(\parent(v)), f)$-neighbor of $\parent(v)$. Since $\level(\parent(v)) = \level(v) - 1 = l$, we can apply the induction hypothesis to obtain $\parent(u) = \parent(w)$ and $\depth - 1 = \depth' - 1 \leq \level(\parent(v)) = \level(\parent(w)) = \level(\parent(u))$. Thus, $\depth = \depth' \leq \level(\parent(v)) + 1 = \level(v)$. By definition, 
\begin{IEEEeqnarray*}{+rCl+x*}
\idx(u) & = & \ofunc(\idx(v), \state(\parent(v)), \state(\parent(u)), f) = \ofunc(\idx(v), \state(\parent(v)), \state(\parent(w)), f) = \idx(w)
\end{IEEEeqnarray*}
and thus
\begin{IEEEeqnarray*}{+rCl+x*}
u & = & \child(\parent(u), \idx(u)) = \child(\parent(w), \idx(w)) = w~. & \qedhere
\end{IEEEeqnarray*}
\end{enumerate}
\end{proof}
\end{lemma}

With the preceding lemma, the depth of a neighbor is well-defined. This will lead to a characterization of the neighbor-finding algorithm's runtime later. 
\begin{definition} \label{def:neighborDepth}
Let $T$ be a neighbor-structured $b$-index-tree as in \Cref{def:neighborbtree}. The $f$-neighbor-depth of $v \in \vertices$, $\neighborDepth(v, f)$, is defined as
\begin{itemize}
\item $\depth$, if there is a depth-$\depth$ $f$-neighbor of $v$; and
\item $\level(v)+1$, if there is no $f$-neighbor of $v$.
\end{itemize}
\end{definition}

\subsection{Geometric Neighbors and Regularity Conditions} \label{sec:modeling:geometricNeighbors}

This section presents sufficient conditions for a $b$-specification such that under these conditions,
\begin{itemize}
\item the geometric $b$-index-tree from \Cref{def:geometricBTree} is extended to a neighbor-$b$-index-tree,
\item a geometric definition of $f$-neighborship is given for such a tree, and
\item the equivalence of this geometric definition and \Cref{def:neighborbtree} is proven.
\end{itemize}

Readers that understand the geometric intuitions behind \Cref{def:neighborbtree} and are mainly interested in algorithms may skip this rather technical section and directly go to \Cref{sec:modeling:isomorphisms} or \Cref{sec:algorithms}.

To define conditions for regularity of geometric $b$-trees, we first need to define what it means for two nodes to \quot{look essentially the same}. We define this by defining an equivalence relation on their point matrices, where two point matrices are defined to be equivalent if their points are affinely transformed versions of each other.

\begin{definition} \label{def:matrixEquivalence}
Let 
\begin{IEEEeqnarray*}{+rCl+x*}
\affAut(\bbR^d) \equalDef \{\tau: \bbR^d \to \bbR^d, x \mapsto Ax + b \mid A \in \bbR^{d \times d} \text{ invertible}, b \in \bbR^d\}
\end{IEEEeqnarray*}
be the group of all invertible affine transformations $\tau: \bbR^d \to \bbR^d$. Furthermore, let $\matrixCol{Q}{i}$ denote the vector containing the $i$-th column of the matrix $Q$.

For $Q, Q' \in \bbR^{d \times m}$ and $\tau \in \affAut(\bbR^d)$, we write
\begin{IEEEeqnarray*}{+rCl+x*}
\tau: Q \sim Q' & \equivDef & \forall i \in \{1, \hdots, m\}: \matrixCol{Q'}{i} = \tau(\matrixCol{Q}{i})~.
\end{IEEEeqnarray*}
If $\tau(x) = Ax + b$ for all $x \in \bbR^d$, this is equivalent to $Q' = AQ + b\oneVec{m}^\top$. We write $Q \sim Q'$ if there exists $\tau \in \affAut(\bbR^d)$ such that $\tau: Q \sim Q'$.

It is easy to show that
\begin{itemize}
\item $\id_{\bbR^d}: Q \sim Q$,
\item $\tau: Q \sim Q'$ implies $\tau^{-1}: Q' \sim Q$, and
\item ($\pi: Q \sim Q'$ and $\tau: Q' \sim Q''$) implies $(\tau \circ \pi): Q \sim Q''$
\end{itemize}
for all $Q, Q', Q'' \in \bbR^{d \times m}$ and $\pi, \tau \in \affAut(\bbR^d)$. This shows that $\sim$ is an equivalence relation.

For $\tau \in \affAut(\bbR^d)$, $Q_1, Q_1' \in \bbR^{d \times m}$ and $Q_2, Q_2' \in \bbR^{d \times n}$, we define
\begin{IEEEeqnarray*}{+rCl+x*}
\tau: (Q_1, Q_2) \sim (Q_1', Q_2') & \equivDef & (\tau: Q_1 \sim Q_1' \text{ and } \tau: Q_2 \sim Q_2')~.
\end{IEEEeqnarray*}
This means that $Q_1$ and $Q_2$ are related to $Q_1'$ and $Q_2'$ by the \emph{same} affine automorphism $\tau$. Again, we write $(Q_1, Q_2) \sim (Q_1', Q_2')$ if there exists $\tau \in \affAut(\bbR^d)$ such that $\tau: (Q_1, Q_2) \sim (Q_1', Q_2')$. The relation $\sim$ on pairs of matrices is also an equivalence relation.
\end{definition}

\begin{remark}
Given two matrices $Q, Q' \in \bbR^{d \times m}$, it can be algorithmically checked\footnote{This assumes arbitrary precision arithmetic, which is also required for exact computations on polytopes. This assumption is realistic if only rational numbers are used for modeling, which is usually the case.} whether $Q \sim Q'$: Let $u_1, \hdots, u_m$ be the columns of $Q$ and let $u_1', \hdots, u_m'$ be the columns of $Q'$. Set $U \equalDef \{u_1, \hdots, u_m\}$ and $U' \equalDef \{u_1', \hdots, u_m'\}$. Let $B = \{u_{i_1}, \hdots, u_{i_k}\} \subseteq U$ be an affine basis of $\aff(U)$. Set $B' \equalDef \{u_{i_1}', \hdots, u_{i_k}'\}$. Then, there is a unique affine map $\pi: \aff(U) = \aff(B) \to \aff(B')$ satisfying $\pi(u_{i_n}) = u_{i_n}'$ for all $n \in \{1, \hdots, k\}$.

\begin{itemize}
\item If $\tau: Q \sim Q'$, it follows that $\pi = \tau|_{\aff(B)}^{\aff(B')}$ and thus, $B'$ is affinely independent and $\pi(u_i) = u_i'$ for all $i \in \{1, \hdots, m\}$.
\item If $B'$ is affinely independent and $\pi(u_i) = u_i'$ for all $i \in \{1, \hdots, m\}$, $\dim(\aff(U)) = \dim(\aff(U'))$ and $\pi$ can be extended to an affine isomorphism $\tau \in \affAut(\bbR^d)$ with $\tau: Q \sim Q'$.
\end{itemize}

Thus, the condition that $B'$ is affinely independent and $\pi(u_i) = u_i'$ for all $i \in \{1, \hdots, m\}$ is equivalent to $Q \sim Q'$ and can be algorithmically verified. Since $(Q_1, Q_2) \sim (Q_1', Q_2')$ if and only if the concatenated matrices $(Q_1|Q_2)$ and $(Q_1'|Q_2')$ are equivalent, equivalence of matrix pairs can also be algorithmically verified. %
\end{remark}

\begin{remark}
Let $m, n \in \natOne$. For $Q \in \bbR^{d \times m}$, let $[Q]$ be the equivalence class of $Q$ with respect to $\sim$. The set $\calM$ of transition matrices in $\bbR^{m \times n}$ is a semigroup by \Cref{lemma:specialMatrixProperties} (c). We can define a right group action of $\calM$ on the quotient $\bbR^{d \times m}/{\sim}$ by
\begin{IEEEeqnarray*}{+rCl+x*}
[Q] \cdot M & \equalDef & [Q \cdot M]~.
\end{IEEEeqnarray*}
To show that this is well-defined, assume that $\tau: Q \sim Q'$, where $\tau(x) = Ax + b$ for all $x \in \bbR^d$. Then, $Q' = AQ + b\oneVec{m}^\top$ and thus
\begin{IEEEeqnarray*}{+rCl+x*}
Q'M & = & (AQ + b\oneVec{m}^\top)M = AQM + b(\oneVec{m}^\top M) = A(QM) + b\oneVec{n}^\top~,
\end{IEEEeqnarray*}
which means $\tau: QM \sim Q'M$. %
This is the idea behind the next lemma.
\end{remark}

To work with matrix equivalences in a geometric $b$-tree, the following identities are essential:

\begin{lemma} \label{lemma:matrixEquivalences}
Let $\tau \in \affAut(\bbR^d)$. For matrices of suitable dimensions, the following implications hold:
\begin{itemize}
\item If $\tau: A \sim B$, then $\tau: (A, A) \sim (B, B)$.
\item If $\tau: (A, A') \sim (B, B')$ and $M, M'$ are transition matrices, then \[\tau: (AM, A'M') \sim (BM, B'M')~.\]
\item If $\tau: (A, A') \sim (B, B')$, then $\tau: (A', A) \sim (B', B)$.
\end{itemize}

Especially, in a geometric $b$-index-tree $T$, the following implications hold:
\begin{itemize}
\item If $\tau: Q(v) \sim Q(w)$, then $\tau: (Q(v), Q(v)) \sim (Q(w), Q(w))$.
\item If $\tau: (Q(v), Q(v')) \sim (Q(w), Q(w'))$ and $j, j' \in \childIndices$, then 
\begin{IEEEeqnarray*}{+rCl+x*}
\tau: (Q(v'), Q(v)) & \sim & (Q(w'), Q(w))~, \\
\tau: (Q(\child(v, j)), Q(\child(v', j'))) & \sim & (Q(\child(w, j)), Q(\child(w', j')))~, \\
\tau: (Q(v), Q(\child(v', j'))) & \sim & (Q(w), Q(\child(w', j')))~.
\end{IEEEeqnarray*}
\end{itemize}
\begin{proof}
These identites follow easily from \Cref{def:matrixEquivalence} and \Cref{lemma:specialMatrixProperties}.
\end{proof}
\end{lemma}

To prove geometric results, we need to establish some well-known facts about polytopes.

\begin{proposition} \label{prop:polytopeProperties}
Let $S \subseteq \bbR^d$ be finite and let $F \subseteq P$ be a face of the polytope $P \equalDef \convHull{S}$. Then,
\begin{enumerate}[(a)]
\item $\convHull{\polytopeVert(P)} = P$,
\item $\polytopeVert(P) \subseteq S$,
\item $F$ is a polytope and $\polytopeVert(F) = F \cap \polytopeVert(P)$,
\item If $F$ and $F'$ are facets of $P$ and $\dim(F \cap F') = \dim(P) - 1$, then $F = F'$.
\item If $\tau \in \affAut(\bbR^d)$, then $\tau(P)$ is a polytope with $\dim(\tau(P)) = \dim(P)$, $\tau(F)$ is a face of $\tau(P)$, $\dim(\tau(F)) = \dim(F)$, $\polytopeVert(\tau(P)) = \tau(\polytopeVert(P))$  and $\convHull{\tau(S)} = \tau(\convHull{S})$.
\item If $P$ and $P'$ are polytopes, then $P \cap P'$ is also a polytope.
\end{enumerate}

\begin{proof}
Follows partially from Ziegler \cite{ziegler2012} or elementary from \Cref{def:polytope}.
\end{proof}
\end{proposition}

In the following, we want to investigate the structure of a geometric $b$-index-tree $T$. Our goal is to specify the set $\facets$ and the functions $\nfunc, \ofunc$ and $\pface$ from \Cref{def:neighborbtree}. To this end, we want to specify a map $(v, f) \mapsto \vertexFace{v}{f}$ that takes a node $v \in \vertices$ and an encoding $f \in \facets$ and returns a facet $\vertexFace{v}{f}$ of the polytope $\convHull{Q(v)}$. Having said that, not all such maps allow us to reasonably specify functions $\nfunc, \ofunc$ and $\pface$. We have to ensure that for two nodes $v, w$ having the same state, the facets $\vertexFace{v}{f}$ and $\vertexFace{w}{f}$ are the \quot{same} facet in the sense that they are the same when viewed from the local coordinate system of $v$ and $w$, respectively. To make that precise, we look at the indices of the columns of $Q(v)$ and $Q(w)$ that are contained in the facet $\vertexFace{v}{f}$ and $\vertexFace{w}{f}$, respectively. The set of these indices should be the same for $v$ and $w$.

\begin{definition} \label{def:polytopeFacetFunctions}
For $Q \in \bbR^{d \times m}$, let $\facetsOf{Q}$ be the set of all facets of $\convHull{Q}$. For $f \in \facetsOf{Q}$, define $\facetIndices{f}{Q} = \{j \in \{1, \hdots, m\} \mid \matrixCol{Q}{j} \in f\}$, the set of indices of matrix columns contained in the facet $f$. Define $\polytopeFacetIndices{Q} \equalDef \{\facetIndices{f}{Q} \mid f \in \facetsOf{Q}\}$, which means that
\begin{IEEEeqnarray*}{+rCl+x*}
\facetIndices{\cdot}{Q}: \facetsOf{Q} \to \polytopeFacetIndices{Q}
\end{IEEEeqnarray*}
is surjective. It is even bijective since it follows from \Cref{prop:polytopeProperties} (a) -- (c) that $\facetFromIndices{\cdot}{Q}: \polytopeFacetIndices{Q} \to \facetsOf{Q}, J \mapsto \convHull{\{\matrixCol{Q}{j} \mid j \in J\}}$ is its inverse. %
\end{definition}

To be able to define the facet $f$ of a node $v$, we first show that equivalent matrices have the same facet indices and then require nodes of the same state to have equivalent matrices.

\begin{proposition} \label{prop:facetIndexIndependence}
Let $Q, Q' \in \bbR^{d \times m}$. If $Q \sim Q'$, then \[\polytopeFacetIndices{Q} = \polytopeFacetIndices{Q'}~.\]

\begin{proof}
Since $\sim$ is symmetric, it is sufficient to show \quot{$\subseteq$}. Let $\calI \in \polytopeFacetIndices{Q}$. By definition, there exists a facet $F$ of $\convHull{Q}$ such that $\calI = \{i \in \{1, \hdots, m\} \mid \matrixCol{Q}{i} \in \polytopeVert(F)\}$. Choose $\tau \in \affAut(\bbR^d)$ with $\tau: Q \sim Q'$. For all $i \in \{1, \hdots, m\}$, \Cref{prop:polytopeProperties} yields
\begin{IEEEeqnarray*}{+rCl+x*}
\matrixCol{Q}{i} \in \polytopeVert(F) & \Leftrightarrow & \matrixCol{Q'}{i} = \tau(\matrixCol{Q}{i}) \in \tau(\polytopeVert(F)) = \polytopeVert(\tau(F))~,
\end{IEEEeqnarray*}
where $\tau(F)$ is a facet of $\convHull{Q'} = \tau(\convHull{Q})$. This shows $\calI \in \polytopeFacetIndices{Q'}$.
\end{proof}
\end{proposition}

To turn a geometric $b$-index-tree $T$ into a neighbor-$b$-index-tree in a meaningful way, we need to impose some regularity conditions on the structure of $T$. We establish these regularity conditions in two steps, where the first one is needed to define the $f$-facet of a node, which is in turn useful to define the second step conveniently.

\begin{definition} \label{def:preregular}
A geometric $b$-index-tree $T = (\vertices, \child, \states, \state)$ is called \defEmph{pre-regular} if the following conditions are satisfied:
\begin{enumerate}[(P1)]
\item \label{item:P1} For all $v, v' \in \vertices$, $\state(v) = \state(v')$ implies $Q(v) \sim Q(v')$, 
\item \label{item:P2} For all $v \in V$, $\dim(\convHull{Q(v)}) = d$.
\end{enumerate}
\end{definition}

\begin{remark}
The geometric $b$-index-trees corresponding to SFCs presented in \Cref{sec:overview} obviously satisfy these criteria. They even appear to satisfy a much stronger version of \hyperref[item:P1]{(P1)}: For \emph{any} two nodes $v, v' \in \vertices$ irrespective of their states, there exists $\tau \in \affAut(\bbR^d)$ such that $\tau: Q(v) \sim Q(v')$, where $\tau(x) = Ax + b$ with an invertible matrix $A \in \bbR^{d \times d}$ that is a scalar multiple of an orthogonal matrix.
\end{remark}

\textbf{For the rest of this section, we assume $T = (\vertices, \child, \states, \state)$ to be a pre-regular geometric $b$-index-tree.}

\begin{definition} \label{def:geometricBTree:facets}
Let $\facets_s$ be a finite set for every $s \in \states$ and $\facets \equalDef \bigcup_{s \in \states} \facets_s$ (not necessarily disjoint). The intuition behind this is that the set $\facets_s$ is a set representing the facets of a node in state $s$. For every $s \in \calS$, choose $v_s \in \vertices$ satisfying $\state(v_s) = s$ (this is possible because of the reachability condition in \Cref{def:bStateSystem}). Let $\Phi_s: \facets_s \to \polytopeFacetIndices{Q(v_s)}$ be a bijective function for each $s \in \states$. Note that \hyperref[item:P1]{(P1)} together with \Cref{prop:facetIndexIndependence} implies that $\polytopeFacetIndices{Q(v_s)}$ is independent of the choice of $v_s$. We call $(\facets, \Phi)$ a \defEmph{facet specification} for $T$.

For $v \in \vertices, f \in \facets$, we can then define the facet $f$ of $v$, $\vertexFace{v}{f}$, by
\begin{IEEEeqnarray*}{+rCl+x*}
\vertexFace{v}{f} & \equalDef & \begin{cases}
\facetFromIndices{\Phi_{\state(v)}(f)}{Q(v)} = \convHull{\{\matrixCol{Q(v)}{i} \mid i \in \Phi_{\state(v)}(f)\}} &, \text{ if } f \in \facets_{\state(v)} \\
\emptyset &, \text{ otherwise}.
\end{cases}
\end{IEEEeqnarray*}
Since $\Phi_{\state(v)}$ is bijective and by \Cref{def:polytopeFacetFunctions}, $\facetFromIndices{\cdot}{Q(v)}$ is bijective, the function $\calF_{\state(v)} \to \facetsOf{Q(v)}, f \mapsto \vertexFace{v}{f}$ is also bijective. This means that $\vertexFace{v}{f} = \vertexFace{v}{f'}$ implies $f = f'$ if $\vertexFace{v}{f} \neq \emptyset$.
\end{definition}

Note that definition of $\vertexFace{v}{f}$ depends on the choice of $(\facets_s)_{s \in \states}$ and $(\Phi_s)_{s \in \states}$, i.e.\ on the choice of the facet specification $(\calF, \Phi)$. We will assume for the rest of this section that one such choice is fixed.

\begin{lemma} \label{lemma:facetTransformation}
Let $\tau: Q(v) \sim Q(w)$ and $\state(v) = \state(w)$. Then, $\vertexFace{w}{f} = \tau(\vertexFace{v}{f})$.

\begin{proof}
Let $s \equalDef \state(v) = \state(w)$. We may assume that $f \in \facets_s$, otherwise the claim is trivial. With $\calI \equalDef \Phi_{s}(f)$, we obtain
\begin{IEEEeqnarray*}{+rCl+x*}
\vertexFace{w}{f} & = & \convHull{\{\matrixCol{Q(w)}{i} \mid i \in \calI\}} = \convHull{\{\tau(\matrixCol{Q(v)}{i}) \mid i \in \calI\}} \\
& = & \convHull{\tau(\{\matrixCol{Q(v)}{i} \mid i \in \calI\})} \stackrel{\ref{prop:polytopeProperties}}{=} \tau(\convHull{\{\matrixCol{Q(v)}{i} \mid i \in \calI\}}) = \tau(\vertexFace{v}{f})~. & \qedhere
\end{IEEEeqnarray*}
\end{proof}
\end{lemma}

\begin{definition} \label{def:section}
For $v, v' \in \vertices$, let $\polytopeSection{v}{v'} \equalDef \convHull{Q(v)} \cap \convHull{Q(v')}$. By \Cref{prop:polytopeProperties} (f), $\polytopeSection{v}{w}$ is a polytope. Furthermore, if $\tau: (Q(v), Q(v')) \sim (Q(w), Q(w'))$, \Cref{prop:polytopeProperties} yields
\begin{IEEEeqnarray*}{+rCl+x*}
\polytopeSection{w}{w'} & = & \convHull{Q(w)} \cap \convHull{Q(w')} = \convHull{\tau(Q(v))} \cap \convHull{\tau(Q(v'))} \\
& = & \tau(\convHull{Q(v)} \cap \convHull{Q(v')}) = \tau(\polytopeSection{v}{v'})~.
\end{IEEEeqnarray*}
\end{definition}

\begin{definition}[Geometric neighbors] \label{def:geometric:neighbor}
Let $v \in \vertices$ and $f \in \facets_{\state(v)}$. A node $v' \in \vertices$ is called \defEmph{geometric $f$-neighbor} of $v \in \vertices$ if $\level(v) = \level(v')$ and there is $f' \in \facets_{\state(v')}$ such that $\polytopeSection{v}{v'} = \vertexFace{v}{f} = \vertexFace{v'}{f'}$. Note that this implies $\dim(\polytopeSection{v}{v'}) = d-1$, hence $v \neq v'$ and thus $\level(v) = \level(v') \neq 0$, since only $\rootv$ has level $0$ and $\dim(\polytopeSection{\rootv}{\rootv}) = d \neq d-1$ by \hyperref[item:P2]{(P2)}.
\end{definition}

\begin{definition} \label{def:geometric:regular}
A pre-regular geometric $b$-index-tree $T$ is called \defEmph{regular} if the following conditions hold:
\begin{enumerate}[(R1)]
\item \label{item:R1} If $v_2, v_2'$ are geometric $f$-neighbors of $v_1$ and $v_1'$ respectively satisfying $\state(v_i) = \state(v_i')$ for $i \in \{1, 2\}$, then $(Q(v_1), Q(v_2)) \sim (Q(v_1'), Q(v_2'))$.
\item \label{item:R2} For all $v \in \vertices$ and $j \in \childIndices$, $\convHull{Q(\child(v, j))} \subseteq \convHull{Q(v)}$.
\item \label{item:R3} For all $v, v' \in \vertices$ with $\level(v) = \level(v')$, the following conditions are satisfied:
\begin{enumerate}[(i)]
\item If $\dim(\polytopeSection{v}{v'}) = d$, then $v = v'$.
\item If $\dim(\polytopeSection{v}{v'}) = d-1$, there exists $f \in \facets$ such that $v'$ is a geometric $f$-neighbor of $v$.
\end{enumerate}
\end{enumerate}
Note that regularity is a property independent of the choice of the facet specification $(\facets, \Phi)$, but a formulation using sets of matrix column indices would be less convenient. %
\end{definition}

\begin{remark} \label{remark:regularityModeling}
Condition \hyperref[item:R2]{(R2)} is somewhat restrictive as it rules out the Gosper curve. However, it is satisfied by the most popular curves and it is also important for using a curve in an adaptive setting. Furthermore, it is intuitively obvious that the algorithms given below work for the Gosper curve as well.

Condition \hyperref[item:R1]{(R1)} is also restrictive because it is not satisfied by the local model of the Hilbert curve and the semi-local model of the Gosper curve, cf.\ \Cref{remark:hilbertLocalNoRegularity}. This is a problem because this means that the function $\ofunc$ will not be well-defined. A simple remedy is to choose a global model instead, where the state contains more information about the orientation of a polytope. Another possibility would be to add another parameter to $\ofunc$: Let $w$ be a geometric $f$-neighbor of $v$ and $v$ be a geometric $f'$-neighbor of $w$. Instead of setting
\begin{IEEEeqnarray*}{+rCl+x*}
\ofunc(\idx(v), \state(\parent(v)), \state(\parent(w)), f) & \equalDef & \idx(w)~,
\end{IEEEeqnarray*}
we could set 
\begin{IEEEeqnarray*}{+rCl+x*}
\ofunc(\idx(v), \state(\parent(v)), \state(\parent(w)), f, f') & \equalDef & \idx(w)~,
\end{IEEEeqnarray*}
i.e.\ include $f'$ as an additional parameter. This has two main drawbacks:
\begin{itemize}
\item The lookup table for $\ofunc$ gets bigger (though it might be still smaller than the lookup table in a global model without additional parameter).
\item The facet $f'$ has to be known, which in general requires additional lookup tables and thus complicates the model.
\end{itemize}
Hence, we will not further investigate this approach.

Condition \hyperref[item:R2]{(R2)} yields the relation
\begin{IEEEeqnarray*}{+rCl+x*}
\bigcup_{j \in \childIndices} \convHull{Q(\child(v, j))} \subseteq \convHull{Q(v)}~.
\end{IEEEeqnarray*}
We might also demand equality instead, which is again satisfied by all curves from \Cref{sec:overview} except the Gosper curve. This would yield information about the existence of neighbors in some cases. Here, we stay with the weaker regularity condition because it will be sufficient for our purposes.

In condition \hyperref[item:R3]{(R3)}, we could require $\polytopeSection{v}{v'}$ to be a face of $v$ and $v'$ independent of its dimension, but we will not need this requirement and it might be more difficult to verify algorithmically.
\end{remark}

To prove useful results about our geometric approach, we need some more preparations.

\begin{lemma} \label{lemma:vertexFacetIntersection}
Let $v \in \vertices$ and $f, f' \in \facets$ such that $\dim(\vertexFace{v}{f} \cap \vertexFace{v}{f'}) \geq d-1$. Then, $f = f'$.

\begin{proof}
Under the assumptions above, let $F \equalDef \vertexFace{v}{f}$ and $F' \equalDef \vertexFace{v}{f'}$. From \Cref{def:geometricBTree:facets}, we know that $\dim(F), \dim(F') \leq d-1$. The assumption $\dim(F \cap F') \geq d-1$ yields $\dim(F) = \dim(F') = d-1$. Since $F$ and $F'$ are facets of $\convHull{Q(v)}$ and since $\dim(\convHull{Q(v)} = d$ by \hyperref[item:P2]{(P2)}, %
\Cref{prop:polytopeProperties} (d) yields $F = F'$, i.e.\ $\vertexFace{v}{f} = \vertexFace{v}{f'}$. As noted in \Cref{def:geometricBTree:facets}, this implies $f = f'$. 
\end{proof}
\end{lemma}

\begin{lemma} \label{lemma:neighborAffinity}
Let $\tau: (Q(v), Q(v')) \sim (Q(w), Q(w'))$ with $\state(v) = \state(w)$ and $\state(v') = \state(w')$.
\begin{enumerate}[(a)]
\item If $f, f' \in \facets$ such that $\vertexFace{v}{f} \supseteq \vertexFace{v'}{f'}$, then $\vertexFace{w}{f} \supseteq \vertexFace{w'}{f'}$.
\item If $v'$ is a geometric $f$-neighbor of $v$ and $\level(w) = \level(w')$, then $w'$ is a geometric $f$-neighbor of $w$.
\end{enumerate}

\begin{proof}
\leavevmode
\begin{enumerate}[(a)]
\item By \Cref{lemma:facetTransformation}, we obtain $\vertexFace{w}{f} = \tau(\vertexFace{v}{f}) \supseteq \tau(\vertexFace{v'}{f'}) = \vertexFace{w'}{f'}$.
\item By \Cref{lemma:facetTransformation} and \Cref{def:section}, we obtain $\polytopeSection{w}{w'} = \tau(\polytopeSection{v}{v'})$, $\vertexFace{w}{f} = \tau(\vertexFace{v}{f})$ and $\vertexFace{w'}{f'} = \tau(\vertexFace{v'}{f'})$. Hence, $\polytopeSection{w}{w'} = \vertexFace{w}{f} = \vertexFace{w'}{f'}$, which means that $w'$ is a geometric $f$-neighbor of $w$. \qedhere
\end{enumerate}
\end{proof}
\end{lemma}

\textbf{For the rest of this section, we assume $T$ to be a regular geometric $b$-index-tree.}

\begin{lemma} \label{lemma:geometricNeighborUniqueness}
For any $v \in \vertices$ and $f \in \facets$, there is at most one geometric $f$-neighbor of $v$.

\begin{proof}
Assume that $w \in \vertices$ and $w' \in \vertices$ are geometric $f$-neighbors of $v$. Let $F \equalDef \polytopeSection{v}{w} = \vertexFace{v}{f} = \polytopeSection{v}{w'}$. Choose a point $p$ in the interior of $F$. By \hyperref[item:P2]{(P2)}, we have $\dim(\convHull{Q(v)}) = \dim(\convHull{Q(w)}) = \dim(\convHull{Q(w')}) = d$. Choosing $\varepsilon > 0$ sufficiently small, we can assume that ball $B_\varepsilon(p)$ with center $p$ and radius $\varepsilon$ intersects $\convHull{Q(v)}, \convHull{Q(w)}$ and $\convHull{Q(w')}$ in half-balls $H_v, H_w$ and $H_{w'}$, respectively. Since $H_v \cap H_w, H_v \cap H_{w'} \subseteq F$ by assumption, it follows that $H_w = H_{w'}$. Therefore, $\dim(\polytopeSection{w}{w'}) \geq \dim(\aff(H_w)) = d$. But then, $w = w'$ by \hyperref[item:R3]{(R3)}.
\end{proof}
\end{lemma}

With these preparations, we can now extend $T$ to a neighbor-$b$-index-tree and then show in the following proposition that this extension is well-defined:

\begin{definition} \label{def:neighborFunctions}
Let $j \in \childIndices, s \in \states, f \in \facets$.
\begin{enumerate}[(a)]
\item If there exists $v \in \vertices$ with $\state(v) = s$ and $j' \in \childIndices$ such that $\child(v, j')$ is a geometric $f$-neighbor of $\child(v, j)$, we set
\begin{IEEEeqnarray*}{+rCl+x*}
\nfunc(j, s, f) & \equalDef & j'~.
\end{IEEEeqnarray*}

\item If there exist $v, v' \in \vertices$, $j' \in \childIndices$ and $f' \in \facets$ such that $\state(v) = s$, $v'$ is a geometric $f'$-neighbor of $v$ and $\child(v', j')$ is a geometric $f$-neighbor of $\child(v, j)$, we set
\begin{IEEEeqnarray*}{+rCl+x*}
\pface(j, s, f) & \equalDef & f' \\
\ofunc(j, s, \state(v'), f) & \equalDef & j'~.
\end{IEEEeqnarray*} 
\end{enumerate}
In all cases not covered, we set $\nfunc(j, s, f)$, $\pface(j, s, f)$ or $\ofunc(j, s, s', f)$ to $\ndef$, respectively.
\end{definition}

\begin{proposition}
The functions in \Cref{def:neighborFunctions} are well-defined.

\begin{proof}
We assume to be given a choice of variables in \Cref{def:neighborFunctions} and show that any other choice leads to the same definition:
\begin{enumerate}[(a)]
\item Let $w \in \vertices$ with $\state(w) = s$ and $j'' \in \childIndices$ such that $\child(w, j'')$ is a geometric $f$-neighbor of $\child(w, j)$. We show that $j' = j''$:

\hyperref[item:P1]{(P1)} yields $Q(v) \sim Q(w)$ and thus, by \Cref{lemma:matrixEquivalences},
\[(Q(\child(v, j)), Q(\child(v, j'))) \sim (Q(\child(w, j)), Q(\child(w, j')))~.\]
By \Cref{lemma:neighborAffinity}, $\child(w, j')$ is a geometric $f$-neighbor of $\child(w, j)$ and thus $j' = j''$ by the uniqueness of geometric neighbors (\Cref{lemma:geometricNeighborUniqueness}).
\item Let $w, w' \in \vertices$, $j'' \in \childIndices$ and $f'' \in \facets$ such that $\state(w) = s$, $w'$ is a geometric $f''$-neighbor of $w$ and $\child(w', j'')$ is a geometric $f$-neighbor of $\child(w, j)$. We show that $f' = f''$ and that $S(v') = S(w')$ implies $j' = j''$:
\begin{enumerate}[(i)]
\item We obtain
\begin{equation}
\vertexFace{v}{f'} = \polytopeSection{v}{v'} \stackrel{\hyperref[item:R2]{\text{(R2)}}}{\supseteq} \polytopeSection{\child(v, j)}{\child(v', j')} = \vertexFace{\child(v, j)}{f}  \label{eq:facetInclusion}
\end{equation}
and similarly $\vertexFace{w}{f''} \supseteq \vertexFace{\child(w, j)}{f}$. Since $\state(v) = \state(w)$, \hyperref[item:P1]{(P1)} yields $Q(v) \sim Q(w)$ and thus $(Q(v), Q(\child(v, j))) \sim (Q(w), Q(\child(w, j)))$ by \Cref{lemma:matrixEquivalences}.
By \Cref{lemma:neighborAffinity}, Equation \eqref{eq:facetInclusion} implies $\vertexFace{w}{f'} \supseteq \vertexFace{\child(w, j)}{f}$. Consequently,
\begin{IEEEeqnarray*}{+rCl+x*}
d-1 & = & \dim(\vertexFace{\child(w, j)}{f}) \leq \dim(\vertexFace{w}{f'} \cap \vertexFace{w}{f''})
\end{IEEEeqnarray*}
and thus $f' = f''$ by \Cref{lemma:vertexFacetIntersection}.

\item Now assume that $\state(v') = \state(w')$. Then, $(Q(v), Q(v')) \sim (Q(w), Q(w'))$ by \hyperref[item:R1]{(R1)} and thus $(Q(\child(v, j), Q(\child(v', j'))) \sim (Q(\child(w, j)), Q(\child(w', j')))$ by \Cref{lemma:matrixEquivalences}. This means that $\child(w', j')$ is a geometric $f$-neighbor of $\child(w, j)$ by \Cref{lemma:neighborAffinity}. Since $\child(w, j)$ can have only one geometric $f$-neighbor by \Cref{lemma:geometricNeighborUniqueness}, it follows that $\child(w', j') = \child(w', j'')$ and thus $j' = j''$. \qedhere
\end{enumerate}
\end{enumerate}
\end{proof}
\end{proposition}

\begin{definition} \label{def:extendedGeometricBTree}
Given a geometric $b$-index-tree $T$ and a facet specification $(\facets, \Phi)$ as in \Cref{def:geometricBTree:facets}, we can define the \defEmph{extended geometric $b$-index-tree} $\hat{T} \equalDef (T, \facets, \nfunc, \ofunc, \pface)$, where $\nfunc, \ofunc$ and $\pface$ are defined as in \Cref{def:neighborFunctions}.
\end{definition}

We can now relate the geometric neighborship-definition from \Cref{def:geometric:neighbor} to the algebraic neighborship-definition from \Cref{def:neighborbtree}.

\begin{Theorem}
Let $\hat{T}$ be an extended geometric $b$-tree for the facet specification $(\facets, \Phi)$. For any $v, v' \in \vertices$ and $f \in \facets$, $v'$ is a geometric $f$-neighbor of $v$ in $\hat{T}$ if and only if it is an $f$-neighbor of $v$ in $\hat{T}$. 

\begin{proof}
We will prove both implications by induction on $\level(v)$. The induction basis is trivial because both neighborship definitions do not allow $\level(v) = 0$, which would mean that $v = \rootv$.
\begin{itemize}
\item[\quot{$\Rightarrow$}:] Let $v' \in \vertices$ be a geometric $f$-neighbor of $v \in \vertices$ with $\level(v) \geq 1$. Since $v \neq \rootv \neq v'$, let $v_p \equalDef \parent(v), v_p' \equalDef \parent(v')$.
\begin{itemize}
\item Case 1: $v_p = v_p'$. The definition of $\nfunc$ implies that $\nfunc(\idx(v), \state(v), f) = \idx(v')$. By definition, $v'$ is an $f$-neighbor of $v$.
\item Case 2: $v_p \neq v_p'$. Let $H \equalDef \polytopeSection{v_p}{v_p'}$. \hyperref[item:R2]{\text{(R2)}} implies $\convHull{Q(v)} \subseteq \convHull{Q(v_p)}$ and $\convHull{Q(v')} \subseteq \convHull{Q(v_p')}$ and therefore
\begin{IEEEeqnarray*}{+rCl+x*}
\dim(H) & = & \dim(\aff(H)) = \dim(\aff(\polytopeSection{v_p}{v_p'})) \\
& \geq & \dim(\aff(\polytopeSection{v}{v'})) = \dim(\polytopeSection{v}{v'}) = d-1~.
\end{IEEEeqnarray*}
By \hyperref[item:R3]{(R3)}, there exists $f_p \in \facets_{\state(v_p)}$ such that $v_p'$ is a geometric $f_p$-neighbor of $v_p$.
Invoking the induction hypothesis, we obtain that $v_p'$ is an $f_p$-neighbor of $v_p$. By definition, we have $f_p = \pface(\idx(v), \state(v_p), f)$ and $\idx(v') = \ofunc(\idx(v), \state(v_p), \state(v_p'), f)$. This means that $v'$ is an $f$-neighbor of $v$.
\end{itemize}

\item[\quot{$\Leftarrow$}:] Let $v' \in \vertices$ be an $f$-neighbor of $v \in \vertices$ with $\level(v) \geq 1$. Since $v \neq \rootv \neq v'$, let $v_p \equalDef \parent(v), v_p' \equalDef \parent(v')$.
\begin{itemize}
\item Case 1: $v_p = v_p'$. By definition, $\idx(v') = \nfunc(\idx(v), \state(v_p), f)$. By definition of $\nfunc$, there exists $w_p \in \vertices$ with $\state(w_p) = \state(v_p)$ such that $w' \equalDef \child(w_p, \idx(v'))$ is a geometric $f$-neighbor of $w \equalDef \child(w_p, \idx(v))$. \hyperref[item:P1]{(P1)} implies $Q(w_p) \sim Q(v_p)$ and thus $(Q(w), Q(w')) \sim (Q(v), Q(v'))$ by \Cref{lemma:matrixEquivalences}. By \Cref{lemma:neighborAffinity}, $v'$ is a geometric $f$-neighbor of $v$. 

\item Case 2: $v_p \neq v_p'$. By definition, $v_p'$ is a $\pface(\idx(v), \state(v_p), f)$-neighbor of $v_p$. By the induction hypothesis, $v_p'$ is then also a geometric $\pface(\idx(v), \state(v_p), f)$-neighbor of $v_p$. Furthermore, we have $\idx(v') = \ofunc(\idx(v), \state(v_p), \state(v_p'), f)$. By definition of $\ofunc$, there exist $w_p, w_p' \in \vertices$ with $\idx(w_p) = \idx(v_p)$, $\state(w_p) = \state(v_p)$ and $\state(w_p') = \state(v_p')$ such that $w_p'$ is a geometric $\pface(\idx(v), \state(v_p), f)$-neighbor of $w_p$ and $w' \equalDef \child(w_p', \idx(v'))$ is a geometric $f$-neighbor of $w \equalDef \child(w_p, \idx(v))$. \hyperref[item:R1]{(R1)} yields $(Q(w_p), Q(w_p')) \sim (Q(v_p), Q(v_p'))$, which leads to $(Q(w), Q(w')) \sim (Q(v), Q(v'))$ using \Cref{lemma:matrixEquivalences}. 
By \Cref{lemma:neighborAffinity}, $v'$ is a geometric $f$-neighbor of $v$. \qedhere
\end{itemize}
\end{itemize}
\end{proof}
\end{Theorem}

\subsection{Algorithmic Verification} \label{sec:modeling:verification}

In this section, we want to give an algorithmically verifiable necessary and sufficient criterion for the regularity of a geometric $b$-tree $T$. It works by finding a representative finite collection of nodes such that checks on this collection can verify regularity on the whole tree. Such a collection can also be used to compute the functions $\nfunc, \ofunc, \pface$. As in the last section, the representations are introduced in two stages, where the first stage can be used to check pre-regularity, which is then required for the definition of neighborship that is in turn used for the second stage.

\begin{definition}
A sequence $u$ of representants $u_s \in \vertices$ for $s \in \states$ is called \defEmph{pre-representation} for $T$ if $u_{\state(\rootv)} = \rootv$ and $\state(u_s) = s$ for all $s \in \states$.
\end{definition}

The reachability condition in \Cref{def:bStateSystem} ensures that nodes $u_s$ with $\state(u_s) = s$ for every $s \in \states$ exist. A pre-representation for $T$ can be found algorithmically as follows:
\begin{itemize}
\item Initialize $u_{\state(\rootv)}$ to $\rootv$ and all other representants to $\ndef$.
\item Iteratively check the children of found representants to find a representant for each state.
\end{itemize}

\begin{definition}
A pre-representation $u$ for $T$ is called \defEmph{pre-regular} if the following conditions are satisfied:
\begin{enumerate}[(P1')]
\item For all $s \in \states$ and $j \in \childIndices$, $Q(\child(u_s, j)) \sim Q(u_{\state(\child(u_s, j))})$.
\item For all $s \in \states$, $\dim(\convHull{Q(u_s)}) = d$.
\end{enumerate}
\end{definition}

\begin{proposition} \label{prop:preRegularEquivalence}
Let $u$ be a pre-representation for $T$. Then, $T$ is pre-regular if and only if $u$ is pre-regular.

\begin{proof}
The direction \quot{$\Rightarrow$} follows immediately from the definition. We will show \quot{$\Leftarrow$} here. Thus, let $u$ be pre-regular. To show that \hyperref[item:P1]{(P1)} is satisfied, we show that $Q(x) \sim Q(u_{\state(x)})$ for all $x \in \vertices$ by induction on $\level(x)$. If $\level(x) = 0$, $x = \rootv$ and the assertion follows directly from $u_{\state(r)} = r$. Now assume that $\level(x) = l+1 \geq 1$ and the claim is true up to level $l$. Set $x_p \equalDef \parent(x)$. Then, $\level(x_p) = l$ and thus $Q(x_p) \sim Q(u_{\state(x_p)})$ by the induction hypothesis. Setting $y \equalDef \child(u_{\state(x_p)}, \idx(x))$, we obtain $\state(y) = \cstate(\state(x_p), \idx(x)) = \state(x)$ and thus
\begin{IEEEeqnarray*}{+rCl+x*}
Q(x) & = & Q(\child(x_p, \idx(x))) \stackrel{\ref{lemma:matrixEquivalences}}{\sim} Q(y) \stackrel{\text{(P1')}}{\sim} Q(u_{\state(y)}) = Q(u_{\state(x)})~,
\end{IEEEeqnarray*}
which completes the induction.

To show \hyperref[item:P2]{(P2)}, let $x \in \vertices$. By \hyperref[item:P1]{(P1)}, there exists $\tau \in \affAut(\bbR^d)$, such that $\convHull{Q(x)} = \tau(\convHull{Q(u_{\state(x)})}$. But then,
\begin{IEEEeqnarray*}{+rCl+x*}
\dim(\convHull{Q(x)}) & = & \dim(\tau(\convHull{Q(u_{\state(x)})})) \stackrel{\ref{prop:polytopeProperties}}{=} \dim(\convHull{Q(u_{\state(x)})}) \stackrel{\text{(P2')}}{=} d~. & \qedhere
\end{IEEEeqnarray*}
\end{proof}
\end{proposition}

\textbf{For the rest of this section, we assume $T$ to be a pre-regular geometric $b$-index-tree with a neighbor structure given by a fixed facet representation $(\facets, \Phi)$.}

\begin{definition}
Let $u$ be a pre-representation for $T$. A tuple $(u, v, w)$ with representants $(v_{s, s', f}, w_{s, s', f}) \in ((\vertices \setminus \{r\}) \times (\vertices \setminus \{r\})) \cup \{(\ndef, \ndef)\}$ for all $s, s' \in \states$ and $f \in \facets$ is called \defEmph{representation} for $T$ if the following conditions are satisfied:
\begin{enumerate}[(a)]
\item For any $s \in \states, j, j' \in \childIndices$ and $f' \in \facets$ such that $y \equalDef \child(u_s, j')$ is a geometric $f'$-neighbor of $x \equalDef \child(u_s, j)$, we have $(v_{\state(x), \state(y), f'}, w_{\state(x), \state(y), f'}) \neq (\ndef, \ndef)$.
\item For every $s, s' \in \states, f \in \facets$ with $(v_{s, s', f}, w_{s, s', f}) \neq (\ndef, \ndef)$, the following conditions are satisfied:
\begin{enumerate}[(1)]
\item $\state(v_{s, s', f}) = s$,
\item $\state(w_{s, s', f}) = s'$,
\item $w_{s, s', f}$ is a geometric $f$-neighbor of $v_{s, s', f}$,
\item For any $j, j' \in \childIndices$ and $f' \in \facets$ such that $y \equalDef \child(w_{s, s', f}, j')$ is a geometric $f'$-neighbor of $x \equalDef \child(v_{s, s', f}, j)$, we have $(v_{\state(x), \state(y), f'}, w_{\state(x), \state(y), f'}) \neq (\ndef, \ndef)$.
\end{enumerate}
\end{enumerate}
\end{definition}

A representation for $T$ can be found algorithmically as follows:
\begin{itemize}
\item Find a pre-representation $u$ for $T$ as described above.
\item Initialize all representants $v_{s, s', f}$ and $w_{s, s', f}$ to $\ndef$.
\item Repeatedly check conditions (a) and (b)(4) for all found representants to find new representants until these conditions are satisfied.
\end{itemize}

\begin{definition}
A pre-regular representation $(u, v, w)$ for $T$ is called \defEmph{regular} if the following conditions are satisfied:
\begin{enumerate}[(R1')]
\item $(Q(x), Q(y)) \sim (Q(v_{\state(x), \state(y), f'}), Q(w_{\state(x), \state(y), f'}))$ 
\begin{itemize}
\item for all $x \equalDef \child(u_s, j'), y \equalDef \child(u_s, j'')$, where $y$ is a geometric $f'$-neighbor of $x$, and
\item for all $x \equalDef \child(v_{s, s', f}, j'), y \equalDef (w_{s, s', f}, j'')$, where $y$ is a geometric $f'$-neighbor of $x$.
\end{itemize}
\item For all $s \in \states$ and $j \in \childIndices$, $\convHull{Q(\child(u_s, j))} \subseteq \convHull{Q(u_s)}$.
\item For all $x, y \in \vertices$ of the form
\begin{enumerate}[(1)]
\item $x = \child(u_s, j'), y = \child(u_s, j'')$, or
\item $x = \child(v_{s, s', f}, j'), y = (w_{s, s', f}, j'')$,
\end{enumerate}
the following conditions are satisfied:
\begin{itemize}
\item If $\dim(\polytopeSection{x}{y}) = d$, then $x = y$.
\item If $\dim(\polytopeSection{x}{y}) = d-1$, then there exists $f' \in \facets$ such that $y$ is a geometric $f'$-neighbor of $x$.
\end{itemize}
\end{enumerate}
\end{definition}

\begin{proposition} \label{prop:regularEquivalence}
Let $(u, v, w)$ be a representation for $T$. Then, $T$ is regular if and only if $(u, v, w)$ is regular.

\begin{proof}
Again, the implication \quot{$\Rightarrow$} follows directly from the definition and we only show \quot{$\Leftarrow$}. Let $(u, v, w)$ be regular.

To show that \hyperref[item:R2]{(R2)} holds, let $x \in \vertices$ and $j \in \childIndices$. By \hyperref[item:P1]{(P1)}, there exists $\tau \in \affAut(\bbR^d)$ such that $\tau: Q(x) \sim Q(u_{\state(x)})$. \Cref{lemma:matrixEquivalences} then yields $\tau: Q(\child(x, j)) \sim Q(\child(u_{\state(x)}, j))$. Hence,
\begin{IEEEeqnarray*}{+rCl+x*}
\convHull{Q(\child(x, j))} & \stackrel{\ref{prop:polytopeProperties}}{=} & \tau(\convHull{Q(\child(u_{\state(x)}, j))}) \stackrel{\text{(R2')}}{\subseteq} \tau(\convHull{Q(u_{\state(x)})}) \\
& \stackrel{\ref{prop:polytopeProperties}}{=} & \convHull{Q(x)}~.
\end{IEEEeqnarray*}

To show that \hyperref[item:R1]{(R1)} and \hyperref[item:R3]{(R3)} hold, we prove the following statements for all $x, y \in \vertices$ with $\level(x) = \level(y)$ by induction on $\level(x)$:
\begin{enumerate}[(i)]
\item If $y$ is a geometric $f$-neighbor of $x$ for some $f \in \facets$, then $(Q(x), Q(y)) \sim (Q(v_{\state(x), \state(y), f}), Q(w_{\state(x), \state(y), f}))$.
\item If $\dim(\polytopeSection{x}{y}) = d$, then $x = y$.
\item If $\dim(\polytopeSection{x}{y}) = d-1$, then there exists $f' \in \facets$ such that $y$ is a geometric $f'$-neighbor of $x$.
\end{enumerate}
The induction basis is trivial: If $\level(x) = \level(y) = 0$, then $x = y = \rootv$, which means that $y$ is \emph{not} a geometric $f$-neighbor of $x$ and furthermore $\dim(\polytopeSection{x}{y}) = d$ by \hyperref[item:P2]{(P2)}. Now, let the statement be true for levels $\leq l$ and let $\level(x) = \level(y) = l+1$. Define $x_p \equalDef \parent(x), y_p \equalDef \parent(y), H \equalDef \polytopeSection{x}{y}$ and $H_p \equalDef \polytopeSection{x_p}{y_p}$. Since \hyperref[item:R2]{(R2)} has already been shown, we can conclude $H_p \supseteq H$ and thus $\dim(H_p) \geq \dim(H)$. If $\dim(H) < d-1$, nothing has to be shown. We therefore assume $\dim(H) \geq d-1$ and investigate different cases:
\begin{itemize}
\item Case 1: $\dim(H_p) = d$. By the induction hypothesis, (ii) holds for $x_p$ and $y_p$, which means $x_p = y_p$. \hyperref[item:P1]{(P1)} yields $Q(x_p) \sim Q(u_{\state(x_p)})$. Set $x_p' \equalDef u_{\state(x_p)}$, $x' \equalDef \child(x_p', \idx(x))$ and $y' \equalDef \child(x_p', \idx(y))$. With an appropriate $\tau \in \affAut(\bbR^d)$, this yields
\begin{IEEEeqnarray*}{+rCl+x*}
\dim(\polytopeSection{x'}{y'}) & = & \dim(\tau(\polytopeSection{\child(x_p, \idx(x))}{\child(x_p, \idx(y))})) \\
& = & \dim(\polytopeSection{x}{y}) = \dim(H)~. \IEEEyesnumber \label{eq:regularEquivalence:1}
\end{IEEEeqnarray*}
\begin{itemize}
\item Case 1.1: $\dim(H) = d$. Then, (R3') yields $x' = y'$, i.e.\ $\idx(x) = \idx(y)$ and thus $x = y$.
\item Case 1.2: $\dim(H) = d-1$. Then, by (R3'), there exists $f \in \facets$ such that $y'$ is a geometric $f$-neighbor of $x'$. Since $(Q(x), Q(y)) \sim (Q(x'), Q(y'))$, $y$ is a geometric $f$-neighbor of $x$. This shows (iii). To show (i), assume that $y$ is a geometric $\tilde{f}$-neighbor of $x$. Then, $\vertexFace{x}{\tilde{f}} = \polytopeSection{x}{y} = \vertexFace{x}{f}$ and thus $\tilde{f} = f$. Statement (i) then follows from
\begin{IEEEeqnarray*}{+rCl+x*}
(Q(x), Q(y)) & \sim & (Q(x'), Q(y')) \stackrel{\text{(R1')}}{\sim} (Q(v_{\state(x), \state(y), f}), Q(w_{\state(x), \state(y), f})~.
\end{IEEEeqnarray*}
\end{itemize}

\item Case 2: $\dim(H) = \dim(H_p) = d-1$. By the induction hypothesis, (iii) holds for $x_p$ and $y_p$. Hence, a facet $f_p \in \facets$ exists such that $y_p$ is a geometric $f_p$-neighbor of $x_p$. Let $x_p' \equalDef v_{\state(x_p), \state(y_p), f_p}$, $y_p' \equalDef w_{\state(x_p), \state(y_p), f_p}$, $x' \equalDef \child(x_p', \idx(x))$ and $y' \equalDef \child(y_p', \idx(y))$. By statement (i) of the induction hypothesis, $(Q(x_p), Q(y_p)) \sim (Q(x_p'), Q(y_p'))$. The argument from Equation \eqref{eq:regularEquivalence:1} yields $\dim(\polytopeSection{x'}{y'}) = d-1$. By (R3'), there exists $f \in \facets$ such that $y'$ is a geometric $f$-neighbor of $x'$. The statements (i) and (iii) then follow as in Case 1.2. \qedhere
\end{itemize}
\end{proof}
\end{proposition}

By examining children of representants in a regular representation, the values of the functions $\nfunc, \ofunc, \pface$ can be computed to automatically generate lookup tables for these functions. This process is implemented in the \sfcpp\ library.

\subsection{Tree Isomorphisms} \label{sec:modeling:isomorphisms}

In this chapter, we introduce isomorphisms between trees. We then show isomorphisms between various trees that have already been defined and add the neighbor structure from \Cref{sec:modeling:geometricNeighbors} to algebraic $b$-index-trees. Furthermore, we show that neighbor-finding on isomorphic trees is equivalent. Isomorphisms will also be used later to
\begin{itemize}
\item formulate conditions for a certain state invariance of neighborship, cf.\ \Cref{sec:algorithms:symmetry},
\item establish a framework to describe the computation of additional information for a node, for example its state, its point matrix or its coordinates. This is shown in \Cref{sec:algorithms:other}.
\end{itemize}

\begin{definition}[Isomorphism] \label{def:isomorphism}
\leavevmode
\begin{enumerate}[(1)]
\item Let $T = (\vertices, \child)$ and $T' = (\vertices', \child')$ be $b$-index-trees. A map $\vertexIso: \vertices \to \vertices'$ is called \defEmph{$b$-index-tree isomorphism} if it is bijective and 
\begin{IEEEeqnarray*}{+rCl+x*}
\vertexIso(\child(v, j)) & = & \child'(\vertexIso(v), j)
\end{IEEEeqnarray*}
for all $v \in \vertices, j \in \childIndices$. We write $\vertexIso: T \cong T'$.

\item Let $T = (\vertices, \child, \states, \state)$ and $T' = (\vertices', \child', \states', \state')$ be state-$b$-index-trees. A pair $(\vertexIso, \stateIso)$ is called \defEmph{state-$b$-index-tree isomorphism} if $\vertexIso: (\vertices, \child) \cong (\vertices', \child')$ and $\stateIso: \states \to \states'$ is a bijective function satisfying
\begin{IEEEeqnarray*}{+rCl+x*}
\stateIso(\state(v)) & = & \state'(\vertexIso(v))
\end{IEEEeqnarray*}
for all $v \in \vertices$. We write $(\vertexIso, \stateIso): T \cong T'$.

\item Let $T = (\vertices, \child, \states, \state, \facets, \nfunc, \ofunc, \pface)$ and $T' = (\vertices', \child', \states', \state', \facets', \nfunc', \ofunc', {\pface}')$ be neigh\-bor-$b$-index-trees. A tuple $(\vertexIso, \stateIso, \facetIso)$ is called \defEmph{neighbor-$b$-index-tree isomorphism} if $(\vertexIso, \stateIso): (\vertices, \child, \states, \state) \cong (\vertices', \child', \states', \state')$ and $\facetIso: \facets \to \facets'$ is a bijective function satisfying
\begin{IEEEeqnarray*}{+rCl+x*}
\nfunc(j, s, f) & = & \nfunc'(j, \stateIso(s), \facetIso(f))~, \\
\ofunc(j, s, \tilde{s}, f) & = & \ofunc'(j, \stateIso(s), \stateIso(\tilde{s}), \facetIso(f))~, \\
\facetIso(\pface(j, s, f)) & = & {\pface}'(j, \stateIso(s), \facetIso(f))
\end{IEEEeqnarray*}
for all $j \in \childIndices, s, \tilde{s} \in \states, f \in \facets$. We write $(\vertexIso, \stateIso, \facetIso): T \cong T'$.
\end{enumerate}

In any of these cases, if an isomorphism between $T$ and $T'$ exists, we write $T \cong T'$ and say that $T$ and $T'$ are isomorphic. This is an equivalence relation. %
\end{definition}

The next two propositions show that isomorphisms have convenient properties:

\begin{proposition} \label{lemma:isomorphismProperties}
Let $\vertexIso: (\vertices, \child, \parent, \level, \rootv) \cong (\vertices', \child', \parent', \level', \rootv')$. Then,
\begin{IEEEeqnarray*}{+rCl+x*}
\vertexIso(\rootv) & = & \rootv' \\
\parent'(\vertexIso(v)) & = & \vertexIso(\parent(v)) \quad (\text{if }v \neq \rootv) \\
\level'(\vertexIso(v)) & = & \level(v) \\
\idx'(\vertexIso(v)) & = & \idx(v) \quad (\text{if }v \neq \rootv)
\end{IEEEeqnarray*}
for all $v \in \vertices$.

Between two $b$-index-trees $T, T'$, there is exactly one isomorphism $\vertexIso_{T \to T'}: T \cong T'$.

\begin{proof}
Since $\vertexIso$ is bijective, there exists $v \in \vertices$ with $\vertexIso(v) = \rootv'$. Assume that $v \neq \rootv$, then there exists $w \in \vertices$ and $j \in \childIndices$ such that $\child(w, j) = v$. But this means that 
\begin{IEEEeqnarray*}{+rCl+x*}
\rootv' & = & \vertexIso(v) = \vertexIso(\child(w, j)) = \child'(\vertexIso(w), j)~,
\end{IEEEeqnarray*}
which is a contradiction. Thus, $\vertexIso(\rootv) = \rootv'$. This shows that the other equations are well-defined for $v \neq \rootv$. Let $v \in \vertices \setminus \{\rootv\}$. Then,
\begin{IEEEeqnarray*}{+rCl+x*}
\parent'(\vertexIso(v)) & = & \parent'(\vertexIso(\child(\parent(v), \idx(v)))) = \parent'(\child'(\vertexIso(\parent(v)), \idx(v))) = \vertexIso(\parent(v)) \\
\idx'(\vertexIso(v)) & = & \idx'(\vertexIso(\child(\parent(v), \idx(v)))) = \idx'(\child'(\vertexIso(\parent(v)), \idx(v)) = \idx(v)~.
\end{IEEEeqnarray*}
The level equation follows by induction on the level of $v$.

The equations $\vertexIso(\rootv) = \rootv'$ and $\vertexIso(\child(v, j)) = \child'(\vertexIso(v), j)$ uniquely define a function $\vertexIso$ and this function is an isomorphism. This can also be shown by induction on the level. 
\end{proof}
\end{proposition}

\begin{proposition} \label{lemma:isomorphismNeighbors}
Let $T, T'$ be neighbor-$b$-index-trees and $(\vertexIso, \stateIso, \facetIso): T \cong T'$ as in \Cref{def:isomorphism}. Let $v, w \in \vertices$ and $f \in \facets$. Then, $w$ is a depth-$\depth$ $f$-neighbor of $v$ in $T$ if and only if $\vertexIso(w)$ is a depth-$\depth$ $\facetIso(f)$-neighbor of $\vertexIso(w)$ in $T'$.

\begin{proof}
\leavevmode
\begin{itemize}
\item[\quot{$\Rightarrow$}:] We prove this statement by induction on $\level(v)$. Let $w$ be a depth-$\depth$ $f$-neighbor of $v$ in $T$. Then, $v, w \neq \rootv$ and thus also $\vertexIso(v), \vertexIso(w) \neq \rootv'$.
\begin{itemize}
\item Case 1: $\parent(v) = \parent(w)$. Then, $\idx(w) = \nfunc(\idx(v), \state(\parent(v)), f)$ and $\depth = 1$. By \Cref{lemma:isomorphismProperties}, we have $\parent'(\vertexIso(v)) = \vertexIso(\parent(v)) = \vertexIso(\parent(w)) = \parent'(\vertexIso(w))$. Furthermore,
\begin{IEEEeqnarray*}{+rCl+x*}
\nfunc'(\idx'(\vertexIso(v)), \state'(\parent'(\vertexIso(v))), \facetIso(f)) & = & \nfunc'(\idx(v), \state'(\vertexIso(\parent(v))), \facetIso(f)) \\
& = & \nfunc'(\idx(v), \stateIso(\state(\parent(v))), \facetIso(f)) \\
& = & \nfunc(\idx(v), \state(\parent(v)), f) \\
& = & \idx(w) = \idx'(\vertexIso(w))~.
\end{IEEEeqnarray*}
This shows that $\vertexIso(w)$ is a depth-$1$ $\facetIso(f)$-neighbor of $\vertexIso(v)$ in $T'$.

\item Case 2: $\parent(v) \neq \parent(w)$. Then, $\parent'(\vertexIso(v)) = \vertexIso(\parent(v)) \neq \vertexIso(\parent(w)) = \parent'(\vertexIso(w))$. By assumption, $f_p \equalDef \pface(\idx(v), \state(\parent(v)), f) \neq \ndef$ and $w_p \equalDef \parent(w)$ is a depth-$(\depth-1)$ $f_p$-neighbor of $v_p \equalDef \parent(v)$. By the induction hypothesis, $\vertexIso(w_p)$ is a depth-$(\depth-1)$ $\facetIso(f_p)$-neighbor of $\vertexIso(v_p)$. Here, the definition yields
\begin{IEEEeqnarray*}{+rCl+x*}
\facetIso(f_p) & = & {\pface}'(\idx(v), \stateIso(\state(\parent(v))), \facetIso(f)) \\
& = & {\pface}'(\idx'(\vertexIso(v)), \state'(\parent'(\vertexIso(v))), \facetIso(f))
\end{IEEEeqnarray*}
and
\begin{IEEEeqnarray*}{+rCl+x*}
\idx'(\vertexIso(w)) & = & \idx(w) = \ofunc(\idx(v), \state(\parent(v)), \state(\parent(w)), f) \\
& = & \ofunc'(\idx(v), \stateIso(\state(\parent(v))), \stateIso(\state(\parent(w))), \facetIso(f)) \\
& = & \ofunc'(\idx'(\vertexIso(v)), \state'(\parent'(\vertexIso(v))), \state'(\parent'(\vertexIso(w))), \facetIso(f))
\end{IEEEeqnarray*}
by assumption. Thus, $\vertexIso(w)$ is a depth-$\depth$ $\facetIso(f)$-neighbor of $\vertexIso(v)$ in $T'$.
\end{itemize}

\item[\quot{$\Leftarrow$}:] This direction follows from what we have proven above by applying the inverse isomorphism. \qedhere
\end{itemize}
\end{proof}
\end{proposition}

Now, we can revisit the geometric $b$-index-tree from \Cref{def:geometricBTree} and the algebraic $b$-index-trees with or without history from \Cref{def:tree:algebraicWithHistory} and \Cref{def:tree:algebraicWithoutHistory}.

\begin{definition} \label{def:extendedAlgebraicBIndexTree}
For neighbor-finding on SFCs, we want to turn an algebraic $b$-index-tree $T = (\vertices, \child, \states, \state)$ (with or without history) for a $b$-state-system $\bStateSystem$ into a neighbor-$b$-index-tree. Assume that $\hat{T}' = (\vertices', \child', \states, \state', \facets, \nfunc, \ofunc, \pface)$ is an extended geometric $b$-index-tree for a $b$-specification $(\bStateSystem, \transitionMat, \rootPoints)$ and a facet specification $(\facets, \Phi)$. This has been defined in \Cref{def:extendedGeometricBTree}. We can now define the \defEmph{extended algebraic $b$-index-tree} $\hat{T} \equalDef (\vertices, \child, \states, \state, \facets, \nfunc, \ofunc, \pface)$ (with or without history). Then, $(\vertexIso, \stateIso, \facetIso): \hat{T}' \cong \hat{T}$, where $\vertexIso = \vertexIso_{\hat{T}' \to \hat{T}}$, $\stateIso = \id_\states$ and $\facetIso = \id_\facets$. By \Cref{lemma:isomorphismNeighbors}, neighbor-finding on $\hat{T}$ is equivalent to neighbor-finding on $\hat{T}'$.
\end{definition}

Using an isomorphism, we can also define coordinates on $b$-index-trees, which is especially useful if they represent $k^d$-trees. A coordinate vector of a tree node is a vector $u \in \natZero^d$.

\begin{definition}[Coordinates] \label{def:coordinateTree}
Let $\bStateSystem$ be a $b$-state-system. Let $b = k^d$ and for every $s \in \states$, let $\kappa_s: \childIndices \to \{0, \hdots, k-1\}^d$ be a bijective function that assigns a $d$-dimensional coordinate vector to a child index in a node of state $s$. Using a similar construction to \Cref{def:geometricBTree}, we can define a $b$-index-tree $T_\kappa = (\vertices, \child, \parent, \level, \rootv)$ such that
\begin{IEEEeqnarray*}{+rCl+x*}
\vertices & \subseteq & \natZero \times \natZero^d \times \states~, \\
\rootv & = & (0, (0, \hdots, 0), \rootState) \in \vertices~, \\
\level((l, u, s)) & = & l~, \\
\idx((l, u, s)) & = & \kappa_s^{-1}(u_1 \bmod k, \hdots, u_d \bmod k)~, \\
\child((l, u, s), j) & = & (l+1, k \cdot u + \kappa_s(j), \cstate(s, j)) \\
\parent((l, u, s)) & = & (l-1, (u_1 \intdiv k, \hdots, u_d \intdiv k), s_p) \text{ for some $s_p \in \states$.}
\end{IEEEeqnarray*}

We call $T_\kappa$ the coordinate-$b$-index-tree for $\kappa$.

Now, let $T'$ be another $b$-index-tree. We can define coordinate vectors on $T'$ by using an isomorphism $\vertexIso \equalDef \vertexIso_{T' \to T_\kappa}$ (cf.\ \Cref{lemma:isomorphismProperties}): Define $\hat{\kappa}: \vertices' \to \natZero^d$ by $\hat{\kappa}(v') \equalDef u$, where $(l, u, s) \equalDef \vertexIso(v')$.
\end{definition}

\cleardoublepage

\section{Algorithms} \label{sec:algorithms}

In this section, we first present a new algorithm for finding the $f$-neighbor of a node $v$ in a neighbor-$b$-index-tree if it exists. We then prove its correctness and its average-case and worst-case runtime complexity. For most space-filling curves, we obtain an average-case runtime complexity of $\compLeq{1}$. In \Cref{sec:algorithms:other}, we show how to embed the neighbor-finding algorithm efficiently in a traversal of all nodes at a fixed level. Furthermore, we show how, based on the frameworks of isomorphisms introduced in \Cref{sec:modeling:isomorphisms}, we can elegantly formulate algorithms for the computation of states, point matrices and coordinate vectors. In \Cref{sec:algorithms:symmetry}, we give formal criteria under which executing the algorithm with a wrong state of a node for all facets $f \in \facets$ still yields all neighbors, but in permuted order.

\subsection{General Neighbor-Finding Algorithm} \label{sec:algorithms:generalNeighbors}

\Cref{alg:generalTrees} is a general neighbor-finding algorithm for neighbor-$b$-index-trees. To formulate a bound on its runtime, we use the neighbor-depth $\neighborDepth(v, f)$ from \Cref{def:neighborDepth}.

\begin{algorithm}[htb]
\caption{Neighbor-finding in a neighbor-$b$-index-tree} \label{alg:generalTrees}
\begin{algorithmic}[1]
\Function{Neighbor}{$v \in \vertices, f \in \facets$}
	\If{$\level(v) = 0$} \Comment{If $v$ is the root node,}
		\State \Return $\ndef$ \label{line:generalTrees:rootReturn} \Comment{return an invalid result.}
	\EndIf
	\State $p_v \assign \parent(v)$ \Comment{Let $p_v$ be the parent of $v$,}
	\State $j_v \assign \idx(v)$ \Comment{$j_v$ the index of $v$,}
	\State $s_p \assign \state(p_v)$ \Comment{$s_p$ the state of the parent,}
	\State $j_w \assign \nfunc(j_v, s_p, f)$ \Comment{and $j_w$ the index of the direct neighbor of $v$.}
	\If{$j_w \neq \ndef$} \Comment{If this neighbor index is valid,}
		\State \Return $\child(p_v, j_w)$ \label{line:generalTrees:firstReturn} \Comment{return the neighbor.}
	\EndIf
	\State $f_p \assign \pface(j_v, s_p, f)$ \Comment{Find the corresponding facet of the parent.}
	\If{$f_p = \ndef$} \Comment{If it does not exist,}
		\State \Return $\ndef$ \Comment{return an invalid result.}
	\EndIf
	\State $p_w \assign$ \Call{Neighbor}{$p_v, f_p$} \label{line:generalTrees:recursion} \Comment{Find the parent's neighbor recursively.}
	\If{$p_w = \ndef$} \Comment{If it does not exist,}
		\State \Return $\ndef$ \Comment{return an invalid result.}
	\EndIf
	\State $s_{p_w} \assign \state(p_w)$ \Comment{Let $s_{p_w}$ be the parent's neighbor's state.}
	\State $\tilde{j}_w \assign \ofunc(j_v, s_p, s_{p_w}, f)$ \label{line:generalTrees:opponentCall} \Comment{Let $\tilde{j}_w$ be the index of its child adjacent to $v$.}
	\If{$\tilde{j}_w = \ndef$} \Comment{If it is invalid,}
		\State \Return $\ndef$ \Comment{return an invalid result.}
	\EndIf
	\State \Return $\child(p_w, \tilde{j}_w)$ \label{line:generalTrees:lastReturn} \Comment{Return the adjacent child.} 
\EndFunction
\end{algorithmic}
\end{algorithm}

\Cref{thm:generalTrees:correct} proves major results about the correctness and the runtime of \Cref{alg:generalTrees}. The runtime assumptions in \Cref{thm:generalTrees:correct} (b) are satisfied by the extended algebraic $b$-index-trees with or without history from \Cref{def:extendedAlgebraicBIndexTree}.

\begin{theorem} \label{thm:generalTrees:correct}
Let $T = (\vertices, \child, \states, \state, \facets, \nfunc, \ofunc, \pface)$ be a neighbor-$b$-index-tree and $v \in \vertices, f \in \facets$.
\begin{enumerate}[(a)]
\item If $v$ has an $f$-neighbor $w$, then $\textsc{Neighbor}(v, f) = w$. If $v$ has no $f$-neighbor, then $\textsc{Neighbor}(v, f) = \ndef$.
\item If $\state, \nfunc, \ofunc, \pface$ can be computed in $\compLeq{1}$ and $\vRuntime$ is a nondecreasing function such that $\vRuntime(\level(v))$ is an upper bound for the runtime of $\child(v, j), \parent(v), \idx(v)$, there is a constant $D \in \bbR$ such that
\begin{IEEEeqnarray*}{+rCl+x*}
\neighborRuntimeBound(v, f) & \equalDef & \neighborDepth(v, f) \cdot (3 \cdot \vRuntime(\level(v)) + D)
\end{IEEEeqnarray*}
is an upper bound for the actual runtime $\neighborRuntime(v, f)$ of $\textsc{Neighbor}(v, f)$.
\end{enumerate}

\begin{proof}
Choose $D$ as a constant comprising all computational overhead generated from calls to $\state, \nfunc, \ofunc, \pface$, assignments, if statements etc. excluding those inside a possible recursion in \Cref{line:generalTrees:recursion}.

We prove both parts by induction on $\level(v)$. If $\level(v) = 0$, then $v = \rootv$ and $v$ has no neighbor. The algorithm correctly returns $\ndef$ in \Cref{line:generalTrees:rootReturn} with a time less than $D \leq \neighborRuntimeBound(v, f)$.

Now assume that the claims are true up to level $l$ and that $\level(v) = l+1$. Note that $s_p = \state(\parent(v))$. We have to distinguish several cases: 
\begin{itemize}
\item Case 1: $j_w \neq \ndef$.
\begin{enumerate}[(a)]
\item Let $w \equalDef \child(\parent(v), j_w)$. The node $w$ is a depth-1 $f$-neighbor of $v$, since $\idx(w) = j_w = \nfunc(\idx(v), s_p, f) = \nfunc(\idx(v), \state(\parent(v)), f)$. Thus, the algorithm returns the correct result in \Cref{line:generalTrees:firstReturn}.
\item The runtime in this case consists of one call to $\parent, \idx, \child$ respecively. This yields the term $3 \cdot \vRuntime(\level(v))$, the rest is covered by $D$.
\end{enumerate}

\item Case 2: $j_w = \ndef$ and $f_p = \ndef$.
\begin{enumerate}[(a)]
\item By definition, $v$ has no $f$-neighbor. Thus, the algorithm correctly returns $\ndef$.
\item It follows by definition that $\neighborDepth(v, f) = \level(v) + 1 \geq 1$ and thus
\begin{IEEEeqnarray*}{+rCl+x*}
\neighborRuntime(v, f) & \leq & D + 2\vRuntime(\level(v)) \leq \neighborRuntimeBound(v, f)~.
\end{IEEEeqnarray*}
\end{enumerate}

\item Case 3: $j_w = \ndef$, $f_p \neq \ndef$ and $p_w = \ndef$.
\begin{enumerate}[(a)]
\item By the induction hypothesis, $p_v = \parent(v)$ has no $f$-neighbor. Since $j_w = \ndef$, this means that $v$ has no neighbor and the algorithm correctly returns $\ndef$.
\item The runtime of this call to $\textsc{Neighbor}$ consists of the recursive call, the calls to $\parent, \idx$ and the rest covered by $D$. Using the induction hypothesis and $\neighborDepth(v, f) = \level(v) + 1 = \level(\parent(v)) + 2 = \neighborDepth(\parent(v), f_p) + 1$, we obtain
\begin{IEEEeqnarray*}{+rCl+x*}
\neighborRuntime(v, f) & \leq & 2\vRuntime(\level(v)) + D + \neighborDepth(\parent(v), f_p) \cdot (3\vRuntime(\level(\parent(v))) + D) \\
& \leq & (\neighborDepth(\parent(v), f_p) + 1) (3\vRuntime(\level(v)) + D) = \neighborRuntimeBound(v, f)~.
\end{IEEEeqnarray*}
\end{enumerate}

\item Case 4: $j_w = \ndef$, $f_p \neq \ndef$ and $p_w \neq \ndef$. 
\begin{enumerate}[(a)]
\item In this case, the induction hypothesis tells us that $p_w$ is the $f_p$-neighbor of $p_v$. If there is an $f$-neighbor $w$ of $v$, $\parent(w)$ must be an $f$-neighbor of $p_v = \parent(v)$, so $\parent(w) = p_w$. Furthermore, \[\idx(w) = \ofunc(\idx(v), \state(\parent(v)), \state(\parent(w)), f) = \ofunc(j_v, s_p, s_{p_w}, f) = \tilde{j}_w~.\] This means that $\tilde{j}_w \neq \ndef$ and we have
\begin{IEEEeqnarray*}{+rCl+x*}
\child(p_w, \tilde{j}_w) = c(\parent(w), \idx(w)) = w~.
\end{IEEEeqnarray*}
If $j_w = \ndef$, there cannot be any $f$-neighbor of $v$ and the algorithm correctly returns $\ndef$.

\item If $\tilde{j}_w \neq \ndef$, we have $\neighborDepth(\parent(v), f_p) + 1 = \neighborDepth(v, f)$. Otherwise, $\parent(v)$ has an $f$-neighbor, but $v$ does not, and \Cref{lemma:neighborProperties} yields $\neighborDepth(\parent(v), f_p) + 1 \leq \level(\parent(v)) + 1 < \level(v) + 1 = \neighborDepth(v, f)$. But then,
\begin{IEEEeqnarray*}{+rCl+x*}
\neighborRuntime(v, f) & \leq & 3\vRuntime(\level(v)) + D + \neighborDepth(\parent(v), f_p) \cdot (3\vRuntime(\level(\parent(v))) + D) \\
& \leq & (\neighborDepth(\parent(v), f_p) + 1) (3\vRuntime(\level(v)) + D) \leq \neighborRuntimeBound(v, f)~. & \qedhere
\end{IEEEeqnarray*}
\end{enumerate}
\end{itemize}

\end{proof}
\end{theorem}

\Cref{thm:generalTrees:correct} can be employed to derive more runtime results:

\begin{corollary} \label{cor:generalTrees:runtime}
Under the assumptions of \Cref{thm:generalTrees:correct} (b), the following runtime bounds hold:
\begin{enumerate}[(a)]
\item \Cref{alg:generalTrees} has a worst-case-runtime complexity of $\compLeq{\level(v) \cdot \vRuntime(\level(v))}$.
\item Let $\vertices_l \equalDef \{v \in \vertices \mid \level(v) = l\}$ be the set of all nodes at level $l$. If there exist constants $c \in \reals$ and $q \in (0, 1)$ such that 
\begin{IEEEeqnarray*}{+rCl+x*}
\frac{|\{(v, f) \in \vertices_l \times \facets \mid \neighborDepth(v, f) \geq k\}|}{|\vertices_l| \cdot |\facets|} & \leq & cq^{k-1}
\end{IEEEeqnarray*}
for every $l \in \natZero$ and every $k \in \natOne$, then \Cref{alg:generalTrees} has an average-case runtime of $\compLeq{\vRuntime(\level(v))}$. The average is taken across each set $\vertices_l \times \facets$.
\end{enumerate}

\begin{proof}
\leavevmode
\begin{enumerate}[(a)]
\item By \Cref{lemma:neighborProperties} and \Cref{def:neighborDepth}, we have $\neighborDepth(v, f) \leq \level(v) + 1$. The claim then follows from \Cref{thm:generalTrees:correct}.
\item We want to determine the average neighbor-depth of nodes $v \in \calV_l$ with level $l$. Observe that
\begin{IEEEeqnarray*}{+rCl+x*}
\frac{1}{|\vertices_l| \cdot |\facets|} \sum_{v \in \vertices_l, f \in \facets} \neighborDepth(v, f) & = & \frac{1}{|\vertices_l| \cdot |\facets|} \sum_{v \in \vertices_l, f \in \facets} \sum_{k=1}^{\neighborDepth(v, f)} 1 = \frac{1}{|\vertices_l| \cdot |\facets|} \sum_{k=1}^{l+1} \sum_{\substack{v \in \vertices_l, f \in \facets \\ \neighborDepth(v, f) \geq k}} 1 \\
& = & \sum_{k=1}^{l+1} \frac{|\{(v, f) \in \vertices_l \times \facets \mid \neighborDepth(v, f) \geq k\}|}{|\vertices_l| \cdot |\facets|} \\
& \leq & \sum_{k=1}^{l+1} cq^{k-1} \leq \sum_{k=1}^\infty cq^{k-1} = \frac{c}{1 - q}~,
\end{IEEEeqnarray*}
which is a bound independent of $l$. By \Cref{thm:generalTrees:correct}, this implies an average-case runtime of $\compLeq{\vRuntime(\level(v))}$. \qedhere
\end{enumerate}
\end{proof}
\end{corollary}

\begin{remark} \label{remark:averageNeighborDepth}
The conditions of \Cref{cor:generalTrees:runtime} (b) are satisfied by sufficiently \quot{normal} space-filling curves like those used in this thesis. For example, in the 2D Hilbert curve, a square of level $l_1$ contains $4^{l_2 - l_1}$ subsquares of level $l_2 > l_1$. Of these, for all $f \in \facets$, $2^{l_2 - l_1}$ squares have no $f$-neighbor. Thus, we can choose $q = 2/4 = 1/2$ and $c = 1$. Assuming constant-time arithmetic operations, this yields an average-case runtime in $\compLeq{1}$ for an extended algebraic $b$-index-tree with or without history as defined in \Cref{def:extendedAlgebraicBIndexTree}.
\end{remark}

\subsection{Other Algorithms} \label{sec:algorithms:other}

In this section, we want to show other algorithms related to \Cref{alg:generalTrees}. For example, \Cref{alg:generalTrees} can be embedded into a traversal of all nodes at level $l$ as shown in \Cref{alg:traversal}. During the traversal, the state of a node is computed \quot{on the fly} and does not cause significant overhead. Thus, under the conditions of \Cref{remark:averageNeighborDepth}, a traversal can be performed in $\compLeq{n}$, where $n = b^l$ is the number of nodes (i.e.\ grid cells) at level $l$.

\begin{algorithm}[htb] %
\caption{Traversing a tree} \label{alg:traversal}
\begin{algorithmic}[1]
\Function{Traverse}{$T = (\vertices, \child, \states, \state, \facets, \nfunc, \ofunc, \pface), l \in \natZero$}
	\State \Call{TraverseRecursively}{$T, \rootv, l$}
\EndFunction

\Function{TraverseRecursively}{$T = (\vertices, \child, \states, \state, \facets, \nfunc, \ofunc, \pface), v \in \vertices, l \in \natZero$}
	\If{$\level(v) = l$}
		\ForAll{$f \in \facets$}
			\State $w \assign$ \Call{Neighbor}{$\idx(v), \state(\parent(v)), f$}
			\State Do something with $w$
		\EndFor
	\Else
		\ForAll{$j \in \childIndices$} 
		\State \Call{TraverseRecursively}{$T, \child(v, j), l$} 
		\EndFor
	\EndIf
\EndFunction
\end{algorithmic}
\end{algorithm}

Other useful algorithms like state computation and conversion between curve position and coordinates can be formulated elegantly in the framework of \Cref{sec:modeling:isomorphisms}. They are special cases of \Cref{alg:isomorphism}, an algorithm to evaluate an isomorphism between two $b$-index-trees that can be seen as a generalization of several well-known algorithms. From \Cref{lemma:isomorphismProperties}, we know that there is exactly one isomorphism $\vertexIso_{T \to T'}: T \cong T'$ for each pair $(T, T')$ of $b$-index-trees. To distinguish the functions of different trees, we use the tree as a subscript of these functions. For example, $\child_{T'}$ means the child function of $T'$.

\begin{algorithm}[htb]
\caption{Computing a tree isomorphism} \label{alg:isomorphism}
\begin{algorithmic}[1]
\Function{Isomorphism}{$T, T', v \in \vertices_T$}
	\If{$v = \rootv_T$} \Comment{If $v$ is the root node of $T$,}
		\State \Return $\rootv_{T'}$ \label{line:isomorphism:rootReturn} \Comment{return the root node of $T'$.}
	\EndIf
	\State \Return $\child_{T'}($\Call{Isomorphism}{$T, T', \parent_T(v)$}$, \idx_T(v))$ \Comment{Else proceed recursively.}
\EndFunction
\end{algorithmic}
\end{algorithm}

Obviously, \Cref{alg:isomorphism} runs in time $\compLeq{\level(v)}$ if the functions $\child_{T'}$ and $\parent_T$ are computable in $\compLeq{1}$, i.e.\ with a runtime bound independent of $\level(v)$. In the following, we give examples of how to use this algorithm.

\begin{example}[State computation] \label{ex:algorithms:state}
Let $T'$ be an algebraic $b$-index-tree (with or without history). Given a level $l$ and a position $j$, we want to find the unique node $(l, j, s) \in \vertices_{T'}$. To this end, we set $T$ to the level-position $b$-index-tree from \Cref{def:levelPositionTree}. We then obtain $(l, j, s) = \vertexIso_{T \to T'}((l, j))$. Since $\child_{T'}$ and $\parent_T$ are efficiently computable, \Cref{alg:isomorphism} can be used to compute $s$. Here, $s$ can be a single state or a state history as in \Cref{def:tree:algebraicWithHistory}.
\end{example}

\begin{example}[Coordinate conversion]
In this example, we want to show how to convert between array index and grid coordinates for suitable grids. Assume that $T$ is the level-position $b$-index-tree from \Cref{def:levelPositionTree} and $T'$ is a coordinate-$b$-index-tree as in \Cref{def:coordinateTree}. Using the $b$-index-tree isomorphisms $\vertexIso_{T \to T'}$ and $\vertexIso_{T' \to T}$, we can convert between an array index $j$ and a coordinate representation $(u_1, \hdots, u_d)$. Note that in order to be able to compute $\parent_{T'}$ efficiently, it is necessary for some SFCs to define $T'$ using a state-history as in the algebraic $b$-index-tree with history from \Cref{def:tree:algebraicWithHistory}.
\end{example}

\begin{example}[Point matrix computation]
By choosing $T$ as the level-position $b$-index-tree and $T'$ as a geometric $b$-index-tree, computing $\vertexIso_{T \to T'}(v)$ not only yields the state, but also the point matrix associated with $v$.
\end{example}

\subsection{Exploiting Symmetry} \label{sec:algorithms:symmetry}

In order to apply \Cref{thm:generalTrees:correct}, the function $\state$ must be efficiently computable, which means in many cases that state information must be stored in $v$. For example, this is realized in an extended algebraic $b$-index-tree. Storing states in memory may significantly increase memory usage. Therefore, the user might want to apply the algorithm without knowing the state of a tree node, i.e.\ just given a level $l$ and an index $j$. One possibility is then to compute the corresponding tree node $v$ in time $\compLeq{\level(v)}$ as shown in \Cref{ex:algorithms:state} and then apply the neighbor-finding algorithm. In this section, we show that if an extended algebraic $b$-index-tree $T$ without history (cf.\ \Cref{def:extendedAlgebraicBIndexTree}) satisfies rather common conditions, setting $v = (l, j, s)$ and calling $\textsc{Neighbor}(v, f)$ for all $f \in \facets$ yields the same neighbors for each $s \in \states$, just in permuted order. This essentially corresponds to the existence of a local model for the given SFC, cf.\ \Cref{remark:hilbert2DSymmetry}, i.e.\ it can be applied to global models of Hilbert, Peano and Sierpinski curves.

In this section, let $T = (\vertices, \child, \states, \state, \facets, \nfunc, \ofunc, \pface)$ be an extended algebraic $b$-index-tree without history for the $b$-specification $(\states, \cstate, \rootState, \transitionMat, \rootPoints)$. We first motivate the introduction of a group structure on $\states$ for a certain class of SFCs and then show that such a group structure together with suitable facet correspondences can lead to this symmetry property.

\begin{definition} \label{def:stateGroup}
If the functions $\sigma_j: \states \to \states, s \mapsto \cstate(s, j)$ are bijective for each $j \in \childIndices$, we define the group $\calG_T$ as the subgroup of the permutation group (or symmetric group) on $\calS$ spanned by these $\sigma_j$.
\end{definition}

\begin{example} \label{ex:hilbertPermutationGroup}
For the 2D Hilbert curve, we can use \Cref{table:hilbertChildState} to explicitly obtain the permutations in cycle notation:
\begin{IEEEeqnarray*}{+rCl+x*}
\sigma_0 & = & (HA)(BR) \\
\sigma_1 & = & \id_\states \\
\sigma_2 & = & \id_\states \\
\sigma_3 & = & (HB)(AR)~.
\end{IEEEeqnarray*}
Hence, $\calG_T = \{\id, (HA)(BR), (HB)(AR), (HR)(AB)\}$, which is isomorphic to the Klein four-group $\bbZ_2 \times \bbZ_2$ as noted in \Cref{remark:hilbert2DGroup}. Note that the permutation $(HR)(AB)$ is not equal to any of the $\sigma_j, j \in \childIndices$. This corresponds to the fact that the state $R$ does not occur at Level 1 of the Hilbert curve.
\end{example}

\begin{remark}
In many cases, the group $\calG_T$ can be identified with the set $\states$ of states: For the Hilbert, Peano and Sierpinski curves, the the function $\phi: \calG_T \to \states, \pi \mapsto \pi(\rootState)$ is a bijection. This is due to the fact that as indicated in \Cref{remark:hilbert2DGroup}, every state corresponds to a particular rotation or reflection of the coordinate system. Together, these rotations and reflections form a group. A permutation $\sigma_j$ that describes the state transition from a parent to its $j$-th child also corresponds to such a rotation or reflection.

Under these assumptions, we can define a group multiplication $\cdot: \states \times \states \to \states, (s_1, s_2) \mapsto \phi(\phi^{-1}(s_1) \circ \phi^{-1}(s_2))$ on the set of states. Then, $\phi: \calG_T \to \states$ is a group isomorphism. Moreover, every index $j \in \childIndices$ corresponds to a state $\istate(j) \equalDef \phi^{-1}(\sigma_j)$. With these definitions, we can write
\begin{IEEEeqnarray*}{+rCl+x*}
\cstate(s, j) & = & \sigma_j(s) = \sigma_j(\phi(\phi^{-1}(s))) = \phi^{-1}(\istate(j))(\phi(\phi^{-1}(s))) \\
& = & \phi^{-1}(\istate(j))(\phi^{-1}(s)(\rootState)) = (\phi^{-1}(\istate(j)) \circ \phi^{-1}(s))(\rootState) \\
& = & \phi(\phi^{-1}(\istate(j)) \circ \phi^{-1}(s)) = \istate(j) \cdot s~. \\
\pstate(s, j) & = & \istate(j)^{-1} \cdot \istate(j) \cdot \pstate(s, j) = \istate(j)^{-1} \cdot \cstate(\pstate(s, j), j) = \istate(j)^{-1} \cdot s~. & \qedhere
\end{IEEEeqnarray*}
\end{remark}

Now assume that some multiplication $\cdot: \states \times \states \to \states$ and a function $\istate: \childIndices \to \states$ are given such that $(\states, \cdot)$ is a group, $\cstate(s, j) = \istate(j) \cdot s$ and $\pstate(s, j) = \istate(j)^{-1} \cdot s$ for all $s \in \states, j \in \childIndices$. For any $g \in \states$, define the neighbor-$b$-index-tree $T_g$ by the following conventions:
\begin{IEEEeqnarray*}{+rCl+x*}
\vertices_{T_g} & \equalDef & \{(l, j, sg) \mid (l, j, s) \in \vertices\} \\
\rootv_{T_g} & \equalDef & (0, 0, \rootState g) \\
\child_{T_g}((l, j, s), i) & \equalDef & (l+1, jb + i, \istate(i) \cdot s) \\
\parent_{T_g}((l, j, s)) & \equalDef & (l-1, j \intdiv b, \istate(j \bmod b)^{-1} \cdot s) \\
\level_{T_g}((l, j, s)) & \equalDef & l \\
\idx_{T_g}((l, j, s)) & \equalDef & j \bmod b \\
\states_{T_g} & \equalDef & \states_T \\
\state_{T_g}((l, j, s)) & \equalDef & s \\
\facets_{T_g} & \equalDef & \facets \\
\nfunc_{T_g} & \equalDef & \nfunc_T \\
\ofunc_{T_g} & \equalDef & \ofunc_T \\
\pface_{T_g} & \equalDef & \pface_T~.
\end{IEEEeqnarray*}

Note that while $\vertices_{T_g}$ depends on $g$, the \quot{implementation} of all functions here is independent of $g$. This means that we can replace any occurence of $\vertices_{T_g}$ in their domain by $\bigcup_{g' \in \states} \vertices_{T_{g'}}$ without changing efficiency aspects. If \Cref{alg:generalTrees} is implemented using these extended domains, it works for all trees $T_g$ simultaneously. This is because it never uses $\vertices_{T_g}$ and $\rootv_{T_g}$, the only objects where the implementation depends on $g$. How can we exploit this? %

Let
\begin{IEEEeqnarray*}{+rCl+x*}
\vertexIso_g: \vertices_T \to \vertices_{T_g}, (l, j, s) & \mapsto & (l, j, sg) \\
\stateIso_g: \states_T = \states \to \states_{T_g} = \states, s & \mapsto & sg~.
\end{IEEEeqnarray*}
Then $(\vertexIso_g, \stateIso_g): T \simeq T_g$, meaning that $T$ and $T_g$ are isomorphic as state-$b$-index-trees in the sense of \Cref{def:isomorphism}. Our last assumption is that for every $g \in \states$, there exists a map $\facetIso_g: \facets \to \facets$ such that $(\vertexIso_g, \stateIso_g, \facetIso_g): T \simeq T_g$, meaning that $T$ and $T_g$ are isomorphic as neighbor-$b$-index-trees in the sense of \Cref{def:isomorphism}. This corresponds to a permutation of facets induced by a rotation or reflection associated with the state transition $s \mapsto sg$.

Now assume that a level $l$ and a position $j$ is given but the state $s \in \states$ of the corresponding tree node, i.e.\ the state satisfying $v_s \equalDef (l, j, s) \in \vertices_T$, is unknown. We want to investigate the results of calling $\textsc{Neighbor}((l, j, s'), f)$ where $s' \in \calS$, i.e.\ just pretending that the state is $s'$. Defining the (unknown) state $g \equalDef s^{-1}s'$, we obtain
\begin{IEEEeqnarray*}{+rCl+x*}
\vertexIso_g(v_s) & = & (l, j, sg) = (l, j, ss^{-1}s') = (l, j, s')~.
\end{IEEEeqnarray*}
By \Cref{lemma:isomorphismNeighbors}, a node $w \in \vertices_T$ is an $f$-neighbor of $v_s$ if and only if $\vertexIso_g(w)$ is a $\facetIso_g(f)$-neighbor of $\vertexIso_g(v_s)$. Thus,
\begin{IEEEeqnarray*}{+rCl+x*}
\{\vertexIso_g(\textsc{Neighbor}((l, j, s), f)) \mid f \in \facets\} & = & \{\textsc{Neighbor}((l, j, s'), \facetIso_g(f)) \mid f \in \facets\} \\
& = & \{\textsc{Neighbor}((l, j, s'), f) \mid f \in \facets\}~,
\end{IEEEeqnarray*}
where $\vertexIso_g(\ndef) \equalDef \ndef$. This means that using the state $s'$ instead of $s$, all the neighbors are still found, but in (possibly) unknown order and with a different state. For the runtime complexity, \Cref{thm:generalTrees:correct} and \Cref{cor:generalTrees:runtime} can be applied to the tree $T_g$ instead of $T$.\footnote{The value of $g$ depends on $l$ and $j$, which makes the conclusion of an average-case runtime for this method more difficult. However, there are only finitely many possible values for $g$ and they all have the same average-case runtime complexity of $\compLeq{\vRuntime(\level(v))}$ under the conditions of \Cref{cor:generalTrees:runtime} (b). The average-case runtime for computing neighbors of $(l, j, s')$ for all $j \in \{0, \hdots, b^l - 1\}$ is therefore bounded by the sum of the average-case complexities for all $g \in \states$, which is again $\compLeq{\vRuntime(\level(v))}$.}

\cleardoublepage

\section{Optimizations} \label{sec:optimizations}

In this section, we want to introduce different kinds of optimizations that can be used to make \Cref{alg:generalTrees} (and other algorithms) faster in practice. Some of these optimizations work for all curves and some only work for special curves.

\subsection{Curve-Indepentent Optimizations}

\begin{optimization}[Avoiding recursion] \label{opt:noRecursion}
The recursion in \Cref{alg:generalTrees} is a single recursion, so it can be replaced by two loops. The first loop walks up the tree until it finds a direct neighbor using $\nfunc$ or until it reaches the root node. In the former case, the second loop walks down the tree, tracking the neighbors using $\ofunc$ until it reaches the level of the original node. In general, the second loop will need the states $s$ and facets $f$ computed in the first loop. These can be stored in arrays during the first loop. \Cref{opt:palindrome} investigates cases where the arrays and the second loop can be omitted. This optimization can also be applied to \Cref{alg:isomorphism}.
\end{optimization}

\begin{optimization}[Loop unrolling] \label{opt:loopUnrolling}
In the iterative algorithm described in \Cref{opt:noRecursion}, it can be effective to unroll the first iteration of each loop. This is the case since by \Cref{remark:averageNeighborDepth}, most curves have a high chance of finding the neighbor in the first loop. This probability gets even higher when applying \Cref{opt:multiLevelTables}. The unrolled iteration can then be optimized specifically.
\end{optimization}

\begin{optimization}[Multi-level tables] \label{opt:multiLevelTables}
\Cref{alg:generalTrees} traverses the tree one level per recursion. If enough space is available for the lookup tables, these lookup tables can be modified so that $k \geq 2$ levels can be handled per recursion. The algorithm then effectively behaves almost as if $k$ levels of the $b$-index-tree each had been \quot{flattened} to a single level of a $b^k$-index-tree. For example, a call $\nfunc(j, s, f)$ might allow $j \in \{0, \hdots, b^k - 1\}$ instead of $j \in \childIndices = \{0, \hdots, b-1\}$, as well as the calls to $\ofunc$, $\pface$, $\cstate$ and (if available) $\pstate$. Such tables may not be applicable to the case $\level(v) < k$, in which case one may revert to using a smaller lookup table. This case can also appear inside a loop or recursion and it is relevant if and only if $\level(v)$ is not a multiple of $k$. \Cref{ex:hilbertDepth2} shows the computation of a neighbor in the 2D Hilbert curve with depth-2 tables.

In a case where the functions $s \mapsto \cstate(s, j)$ are all invertible, it is possible to replace the state of an ancestor by the state of a child to increase efficiency: For example, assume that $v = (l, j, s)$ is a node of an algbraic $b$-index-tree representing the 2D Hilbert curve. Instead of calling $\nfunc(j \bmod b^k, \state(\parent^k(v)), f)$ to find a potential neighbor of $v$ inside its ancestor $\parent^k(v) = (\parent \circ \hdots \circ \parent)(v)$, the definition of $\nfunc$ may be changed so that $\nfunc(j \bmod b^k, \state(v), f)$ yields the desired result. 

Of course, multi-level tables require more cache memory to be stored and thus can impact the cache performance of a grid traversal.
\end{optimization}

\begin{figure}[htb]
\centering
\subcaptionbox{Positions \label{fig:hilbert2DLevel3:pos}}{
\includegraphics[scale=1.2]{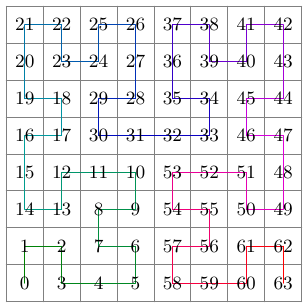}
}
\subcaptionbox{States \label{fig:hilbert2DLevel3:states}}{
\includegraphics[scale=1.2]{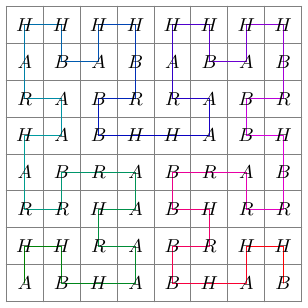}
}
\caption{Positions and states of nodes at level 3 of the 2D Hilbert curve.} \label{fig:hilbert2DLevel3}
\end{figure}

\begin{example} \label{ex:hilbertDepth2}
Consider the computation of the $f$-neighbor of the node $v = (3, 28, R)$ in the 2D Hilbert curve as shown in \Cref{fig:hilbert2DLevel3}, where $f$ encodes the right facet of $v$. With lookup tables of depth 2, we can skip two levels at once: 
\begin{enumerate}[(1)]
\item Compute the index $i = 28 \bmod b^2 = 12$ of $v$ inside its grandparent $p_v \equalDef P(P(v)) = (3 - 2, 28 \intdiv b^2, \widehat{\pstate}(B, 12)) = (1, 1, H)$. Here, the hat in $\widehat{\pstate}$ indicates that a modified lookup table of depth 2 is used.
\item Compute $j_w = \widehat{\nfunc}(12, H, f) = \ndef$. This means that there is no $f$-neighbor of $v$ inside $p_v$. (In a global model of the Hilbert curve, we can assume $f_p = f$ in \Cref{alg:generalTrees}.)
\item Call the algorithm recursively with $p_v$ as a new parameter.
\begin{itemize}
\item Since $\level(p_v) = 1$, lookup tables of depth 2 cannot be used (at least directly). Thus, compute the direct parent $p_{p_v} \equalDef \parent(p_v) = (0, 0, H)$ of $p_v$ and the index $i_p \equalDef \idx(p_v) = 1$.
\item Compute $j_{p_w} \equalDef \nfunc(1, H, f) = 2$.
\item Compute the $f$-neighbor $p_w \equalDef \child(p_{p_v}, i_p) = (1, 2, H)$ of $p_v$ and return it.
\end{itemize}

\item Compute
\begin{IEEEeqnarray*}{+rCl+x*}
\tilde{j}_w & \equalDef & \widehat{\ofunc}(i, \state(p_v), \state(p_w), f) = \widehat{\ofunc}(12, H, H, f) = 3~.
\end{IEEEeqnarray*}

\item Compute the $f$-neighbor $w \equalDef \widehat{\child}(p_w, \tilde{j}_w) = (1 + 2, 2 \cdot b^2 + 3, \widehat{\cstate}(H, 3)) = (3, 35, R)$ of $v$ and return it.
\end{enumerate}
\end{example}

\begin{optimization}[Precomputations] \label{opt:precomputations}
In the case of a regular grid, the level of all nodes handled is known in advance of the computation. This can be used for precomputations. Especially, if arrays are needed for the second loop in \Cref{opt:noRecursion}, they can be preallocated.
\end{optimization}

\begin{optimization}[Eliminating useless checks]
For all SFCs presented here, the cases $f_p = \ndef$ and $\tilde{j}_w = \ndef$ cannot occur. This means that the corresponding if-clauses can be omitted in implementations. By choosing stronger regularity conditions, it would be possible to exclude these two cases for all regular geometric $b$-index-trees, cf.\ \Cref{remark:regularityModeling}.
\end{optimization}

\subsection{Curve-Dependent Optimizations}

\begin{optimization}[State groups] \label{opt:stateGroups}
If the states $\states$ form a group as explained in \Cref{remark:hilbert2DGroup} and \Cref{sec:algorithms:symmetry}, it may be possible to exploit this group structure for more efficient computation of the functions $\cstate$ and $\pstate$. For example, if the group is isomorphic to $\bbZ_2^n$ for some $n \in \natZero$, the states can be stored as non-negative integers using bitwise XOR as the group operation.
\end{optimization}

\begin{optimization}[Bit operations]
If the branching factor $b$ of a given SFC is a power of two, many arithmetic operations can be replaced by bit operations. This can happen manually or automatically by the compiler. For example, division and multiplication by $b$ translates to bit shifts and computing the remainder modulo $b$ corresponds to applying a bit mask.
\end{optimization}

A very effective optimization can be performed for curves that satisfy the palindrome property. Bader \cite{bader2012} uses the term \quot{palindrome property} to describe curves where \quot{at a common (hyper-)face of two subdomains [\ldots], the element orders imposed by the space-filling curve on the two faces are exactly inverse to each other}.\footnote{Bader uses the term \quot{face} for what we defined as \quot{facet}.} Using the terminology from \Cref{sec:modeling}, we can turn this idea into a rigorous definition:

\begin{definition} \label{def:palindromeProperty}
A neighbor-$b$-index-tree satisfies the \defEmph{palindrome property} if the following condition is satifsied: %
For all $f \in \facets, v \in \vertices$ such that $v$ that has no depth-1 $f$-neighbor and there exists a $\pface(\state(v), \idx(v), f)$-neighbor $p_w$ of $\parent(v)$, we have
\[\ofunc(\idx(v), \state(\parent(v)), \state(p_w), f) = b - 1 - \idx(v)~.\]
\end{definition}

\begin{remark} \label{remark:palindromeProperty}
The Peano curves in arbitrary dimension satisfy the palindrome property while the Hilbert curves do not \cite{bader2012}. Using the \sfcpp\ library, it can be checked for arbitrary curve specifications whether $\ofunc(j, s, s', f) \in \{\ndef, b-1-j\}$ for all $j \in \childIndices, s, s' \in \states, f \in \facets$. For example, according to our implementation, this property is satisfied by all implemented generalized Sierpinski models up to $d = 16$, although it is not clear whether all of these models specify regular geometric $b$-index-trees.\footnote{In our experiments, it seems that they specify regular geometric $b$-index-trees but in dimensions $d > 3$, their polytopes (simplices) can get arbitrarily large sidelength ratios.}
\end{remark}

\begin{optimization}[Exploiting the palindrome property] \label{opt:palindrome}
In an extended algebraic $b$-index-tree without history where the palindrome property is satisfied, we have $\Omega(j, s, s', f) = b - 1 - j$ for all parameter combinations $(j, s, s', f)$ occuring in the neighbor-finding algorithm. If a node $v = (l, j, s)$ has a depth-$\depth$ neighbor $w = (l, j', s')$, and their positions have base-$b$-representations $j = (a_{l-1}\hdots a_0)_b$ and $j' = (a_{l-1}'\hdots a_0')_b$, this means that
\begin{IEEEeqnarray*}{+rCl+x*}
a_q' & = & \begin{cases}
a_q &, \depth \leq q \leq l-1 \\
\nfunc(\idx(P^{\depth-1}(v)), \state(P^\depth(v)), f') &, q = \depth-1 \\
b - 1 - a_q &, 0 \leq q \leq \depth-2
\end{cases}
\end{IEEEeqnarray*}
for a suitable facet $f' \in \facets$. We can write this as an equation:
\begin{IEEEeqnarray*}{+rCl+x*}
j' & = & j - (j \bmod b^\depth) + a_{\depth-1}' \cdot b^{\depth-1} + (b^{\depth-1} - 1 - (j \bmod b^{\depth-1}))~. \IEEEyesnumber \label{eq:palindromeNeighbor}
\end{IEEEeqnarray*}
Once $\depth$ and $a_{\depth-1}'$ have been found, the position $j'$ of $w$ can be obtained using this formula. It remains to determine the state $s'$ of $w$ if this information is desired. Since the palindrome property poses a condition on the arrangement of children in an $f$-neighbor of $v$, the state of $v$ and the facet $f$ are likely to be sufficient to determine the state of the $f$-neighbor of $v$. For the Peano and Sierpinski curves, this is the case. Therefore, the second loop mentioned in \Cref{opt:noRecursion} can be omitted for these curves, as well as the arrays needed for the second loop and the lookup table $\ofunc$. If $b = 2^n$ is a power of two, Equation \eqref{eq:palindromeNeighbor} flips the least significant $n(\depth-1)$ bits in the binary representation of $j$, resets the next $n$ bits to $a_{\depth-1}'$ and leaves the rest of the bits untouched. If $b = 2$ as for the Sierpinski curve, the fact that $a_{\depth-1}' \neq a_{\depth-1}$ yields $a_{\depth-1}' = 1 - a_{\depth-1}$, which means that the $\depth$ least significant bits are flipped and the other bits stay the same.
\end{optimization}

\begin{optimization}[Eliminating redundancy]
For many SFCs, the lookup tables here contain more parameters than necessary or some variables are not needed. Removing them speeds up the algorithm and reduces memory usage. Examples:
\begin{itemize}
\item In local models, there is only one state and it does not have to be tracked.
\item In many global models, the lookup table for $\pface$ is not needed since $\pface(j, s, f) = f$ in all occuring cases.
\item When searching for an upper or lower neighbor in the Peano curve model from \Cref{fig:peanoConstruction}, the states $P$ and $R$ behave equivalently to the states $Q$ and $S$, respectively. This is the case because $P$ and $Q$ both are traversed from bottom to top, while $R$ and $S$ are both traversed from top to bottom. Due to the special properties of the Peano curve, all lookup tables for $f \in \{$up, down$\}$ are invariant under exchanging $P$ and $Q$ or $R$ and $S$. Looking for an upper and lower neighbor, only one boolean is needed to track whether the state is in the equivalence class $\{P, Q\}$ or $\{R, S\}$.
\end{itemize}
\end{optimization}

\cleardoublepage

\section{Implementation} \label{sec:implementation}

In this section, we will address different implementation issues: First, we will present code that has been implemented for extracting information from general SFC models. This code can also be used for visualization of these models. In \Cref{sec:implementation:otherCode}, we will discuss implemented algorithms and data structures for various particular SFCs. Runtime measurements for many of these algorithms are provided in \Cref{sec:implementation:experimentalResults}. Most of the code explained here has been written for this thesis and is provided in the \sfcpp\ library.\footnote{\href{https://github.com/dholzmueller/sfcpp}{https://github.com/dholzmueller/sfcpp}}

\subsection{Implementation for General Models} \label{sec:implementation:general}

In the \sfcpp\ library, SFCs can be defined via the \texttt{CurveSpecification} class, which almost directly realizes the definition of a $b$-specification (cf.\ \Cref{def:bSpecification}). Since states are used as array indices, the set $\calS$ of states has to be of the form $\{0, \hdots, n - 1\}$ for some $n \in \natOne$. In addition, the root state $\rootState$ is automatically assumed to be $0$. Since defining a $b$-specification can be quite tedious, the \texttt{KDCurveSpecification} class provides a more convenient way to define SFCs on $k^d$-trees. It allows to define an order of subcubes for each state and can infer the matrices $\transitionMat^{s, j}, s \in \calS, j \in \childIndices$ from these orders.\footnote{This order is analogous to the coordinate functions $\kappa_s$ from \Cref{def:coordinateTree}.} If possible, a local \texttt{CurveSpecification} can also be generated instead of a global one. Already implemented models comprise arbitrary-dimensional Morton, Hilbert, Peano and Sierpinski curves as well as the $\beta\Omega$ and Gosper curves in 2D.

Given a \texttt{CurveSpecification}, a \texttt{CurveInformation} object can be computed that, assuming that the corresponding geometric $b$-index-tree is regular, provides lookup tables for the functions $\nfunc, \ofunc$ and $\pface$, which provide the indices of neighbors inside the same parent, across different parents and the face that different parents share, respectively. %
The computation of the \texttt{CurveInformation} is implemented as follows:
\begin{enumerate}[(1)]
\item Traverse the corresponding geometric $b$-index-tree $T$ to find representants $u_s = (l_s, j_s, s, Q_s) \in \vertices$ of each (reachable) state. In the terminology of \Cref{sec:modeling:verification}, this corresponds to finding a pre-representation.
\item For all states $s \in \states$, find all faces of $\convHull{Q_s}$, specified by the set of indices of columns of $Q_s$ that constitute the vertices of the respective face. To do this, a custom adaptation of the QuickHull algorithm \cite{barber1996} is employed that finds all faces and also works for non-simplicial facets. For cubes, computation times are acceptable up to a dimension of about eight.
\item By using a suitable ordering on the found facet index sets, enumerate the facets of each node canonically. The ordering of the facets is chosen such that for cubes generated out of a global model from a \texttt{KDCurveSpecification}, the index $f \in \facets = \{0, \hdots, |\facets| - 1\}$ corresponds to the facet with normal vector $\vec{n} = (-1)^{f + 1} e_{1 + f \intdiv 2}$, where $e_k$ is the $k$-th unit vector.
\item Using this information, examine the children of all nodes $u_s$ to compute $\nfunc$. 
\item By recursively finding neighboring pairs of nodes with certain states and common facets, compute $\ofunc$ and by examining their children, compute $\pface$. During this step, violations of the regularity condition \hyperref[item:R1]{(R1)} may be detected and reported, for example for the local model of the 2D Hilbert curve. In the terminology of \Cref{sec:modeling:verification}, this corresponds to finding a representation.
\item Check a property similar to the palindrome property by examining the values of $\ofunc$, see \Cref{remark:palindromeProperty}.
\end{enumerate}
The unoptimized brute-force search currently used in steps (4) and (5) may cause the computation to take very long for dimensions $d \geq 5$. This might be subject to future improvements. The current implementation does not use arbitrary-precision arithmetic. The use of floating point numbers appears to be sufficient for practical purposes.

The \texttt{CurveRenderer} class allows to render (visualize) SFCs in various ways using the \LaTeX\ package TikZ. The rendering is done by traversing the specified geometric $b$-index-tree and using the face information from step (2) to create custom objects such as lines connecting subsequent nodes or one-dimensional faces of each polytope. This class has been used to generate most of the figures in this thesis.

It is also possible to compute the group $\calG_T$ defined in \Cref{def:stateGroup} from a specification.

\subsection{Other Code} \label{sec:implementation:otherCode}

Besides the code for general models, there is also code for particular SFCs available. Before we come to neighbor-finding code, we want to examine some examples.

\definecolor{dkgreen}{rgb}{0,0.6,0}
\definecolor{gray}{rgb}{0.5,0.5,0.5}
\definecolor{mauve}{rgb}{0.58,0,0.82}

\lstset{frame=tb,
  language=Java,
  aboveskip=3mm,
  belowskip=3mm,
  showstringspaces=false,
  columns=flexible,
  basicstyle={\small\ttfamily},
  numbers=none,
  numberstyle=\tiny\color{gray},
  keywordstyle=\color{blue},
  commentstyle=\color{dkgreen},
  stringstyle=\color{mauve},
  breaklines=true,
  breakatwhitespace=true,
  tabsize=3
}

\lstset{language=Java}

\begin{lstlisting}[float=htb, caption = A Java implementation of a partial state-tree., label=lst:stateImpl:1]
public class StateHistory {
	// example values for the semi-local model of the 2D Gosper curve
	private static int[][] childStateTable = {
        {0, 1, 1, 0, 0, 0, 1},
        {0, 1, 1, 1, 0, 0, 1}};
	
	private StateHistory parent;
	private int state;
	
	public StateHistory(StateHistory parent, int state) {
		this.parent = parent;
		this.state = state;
	}
	
	public StateHistory getChild(int j) {
		return new StateHistory(this, childStateTable[state][j]);
	}
	
	public StateHistory getParent() {
		return parent;
	}
	
	public int getState() {
		return state;
	}
}
\end{lstlisting}

\paragraph{Implementing a state history} \Cref{lst:stateImpl:1} and \Cref{lst:stateImpl:2} show a Java implementation of an algebraic $b$-index-tree with history (cf.\ \Cref{def:tree:algebraicWithHistory}) that realizes every operation in $\compLeq{1}$. A \texttt{StateHistory} class is implemented that stores the state of a node and the states of all of its ancestors. To realize all operations in time $\compLeq{1}$, it is necessary that multiple nodes can share a common part of the state history. As an example, this implementation uses a lookup table for the semi-local model of the Gosper curve shown in \Cref{fig:gosperConstruction}.

\begin{lstlisting}[float=!ht, caption = A Java implementation of an algebraic $b$-index-tree with history., label = lst:stateImpl:2]
public class Node {
	private static int b = 4;
	private int level;
	private long position;
	private StateHistory history;
	
	public Node(int level, long position, StateHistory history) {
		this.level = level;
		this.position = position;
		this.history = history;
	}
	
	public Node getChild(int index) {
		return new Node(level + 1, position * b + index, history.getChild(index));
	}
	
	public Node getParent() {
		return new Node(level - 1, position / b, history.getParent());
	}
	
	public static Node getRoot() {
		return new Node(0, 0, new StateHistory(null, 0));
	}
	
	public long getIndex() {
		return position % b;
	}
	
	public int getLevel() {
		return level;
	}
	
	public int getState() {
		return history.getState();
	}
}
\end{lstlisting}

\lstset{language=C++} 

\begin{lstlisting}[float=hbt, caption = Efficient Hilbert 2D state computation in C++., label=lst:hilbertState2D]
uint64_t getState(uint64_t level, uint64_t position) {
	uint64_t lowerMask = 0x5555555555555555ul; // binary: 01010101...
	uint64_t flipMask = lowerMask >> (64 - 2 * level);
	uint64_t a = position & lowerMask;
	uint64_t b = (position >> 1) & lowerMask;
	uint64_t aband = a & b;
	uint64_t abnor = flipMask ^ (a | b);
	uint64_t n_3 = __builtin_popcount(aband);
	uint64_t n_0 = __builtin_popcount(abnor);
	return 2 * (n_3 % 2) + (n_0 % 2);
};
\end{lstlisting}

\textbf{Efficient state computation} For the 2D Hilbert curve, there is a neat trick to compute the state of a node very efficiently: In \Cref{ex:hilbertPermutationGroup}, we showed that the group $\calG_T$ of the 2D Hilbert curve is isomorphic to the Klein four-group $\bbZ_2 \times \bbZ_2$ with addition as the group operation. We can identify the root state $H = \istate(1) = \istate(2)$ with $(0, 0)$, the state $A = \istate(0)$ with $(0, 1)$, the state $B = \istate(3)$ with $(1, 0)$ and the state $R$ with $(1, 1)$. To compute the state of a level-position node $(l, j)$, where $(a_{l-1}\hdots a_0)_4$ is the base-four-representation of $j$, we have to compute $s \equalDef \rootState + \sum_{i=0}^{l-1} \istate(a_i) = \sum_{i=0}^{l-1} \istate(a_i)$, where the summation order is irrelevant since the Klein four-group is abelian. With the notation $n_k \equalDef |\{i \in \{0, \hdots, l-1\} \mid a_i = k\}|$ for $k \in \{0, \hdots, 3\}$, the above identification yields $s = (n_3 \bmod 2, n_0 \bmod 2)$. By taking the AND of pairs of bits in the binary representation of $j$, and counting the number of resulting ones, $n_3$ can be computed. The computation of $n_0$ is similar but uses NOR instead of AND. The resulting state $s = (s_1, s_2)$ can be identified with a number $k_s = 2s_1 + s_2$. A C++ implementation of this algorithm for levels $\leq 31$, i.e.\ $j < 2^{64}$, is shown in \Cref{lst:hilbertState2D}. Since the popcount operation (which counts the number of ones in the bit representation of an integer) is natively supported by modern processors, this algorithm is very fast. Under the assumption of constant-time arithmetic integer operations, its runtime is in $\compLeq{1}$.

Unfortunately, this approach cannot be extended to the 3D Hilbert curve since it appears that its state group $\calG_T$ is isomorphic to the alternating group $A_4$, which is not abelian.

\paragraph{Neighbor-finding} The \sfcpp\ library contains several optimized implementations of \Cref{alg:generalTrees} for different SFCs:
\begin{itemize}
\item An implementation of the neighbor-finding algorithm, state computation and coordinate conversion for arbitrary-dimensional Peano curves is provided. The dimension $d$ can be passed as a template parameter. The neighbor-finding algorithm uses all optimizations from \Cref{sec:optimizations} except the bit operations, since $b = 3^d$ is not a power of two for the Peano curve. The implementation is based on a local model of the Peano curve, which means that it does not have to track the state of a tree node. This implementation has been developed for a term paper\footnote{David Holzmüller: Raumfüllende Kurven, 2016.} and has been further optimized for this thesis.
\item For the 2D and 3D Hilbert curves as shown in \Cref{sec:overview}, the neighbor-finding algorithm is implemented. Most optimizations from \Cref{sec:optimizations} are used. Since the Hilbert curves do not satisfy the palindrome property, the corresponding optimization cannot be applied. \Cref{opt:stateGroups} can only be applied in the 2D case since the state group of the 3D Hilbert curve does not allow a XOR-based implementation. Multi-level tables are not implemented but might result in a considerable speedup. The single-level lookup tables were automatically generated from the implemented Hilbert curve model as described in \Cref{sec:implementation:general}. For the 2D Hilbert curve, the efficient state computation algorithm from \Cref{lst:hilbertState2D} is also included.
\item For a local model of the 2D Sierpinski curve, the neighbor-finding algorithm is implemented. A suitable enumeration of the facets allows the algorithm to replace lookup tables by simple arithmetic operations. It is unclear whether multi-level tables would bring an advantage. All other optimizations from \Cref{sec:optimizations} are implemented.
\item For the 2D Morton curve, the $\compLeq{1}$-algorithm by Schrack \cite{schrack1992} is implemented.
\end{itemize}

\subsection{Experimental Results} \label{sec:implementation:experimentalResults}

\newcommand{\plotScale}{0.8}

To measure the performance of the implemented neighbor-finding algorithms, these algorithms were executed repeatedly with fixed level $l$ and state $s = \rootState$ and uniformly distributed random position $j$ and facet $f$. The state was fixed because as explained in \Cref{sec:algorithms:symmetry}, for all curves where neighbor-finding has been implemented, calling the algorithm with any state for all facets yields the same neighbors. The parameters of one call were also chosen to be dependent on those of the previous call to avoid pipelining of multiple calls by the compiler. Since the random number generation requires a large portion of the execution time, the time needed to execute $n \equalDef 5 \cdot 10^6$ random number generations was measured and subtracted from the time measured for $n$ executions of random number generation together with neighbor-finding. For each algorithm and each level, 15 such time measurements were conducted. The median of these 15 measurements (divided by $n$) was then taken as an estimate of the average runtime of a single execution.

\begin{figure}[htb]
\centering
\includegraphics[scale=\plotScale]{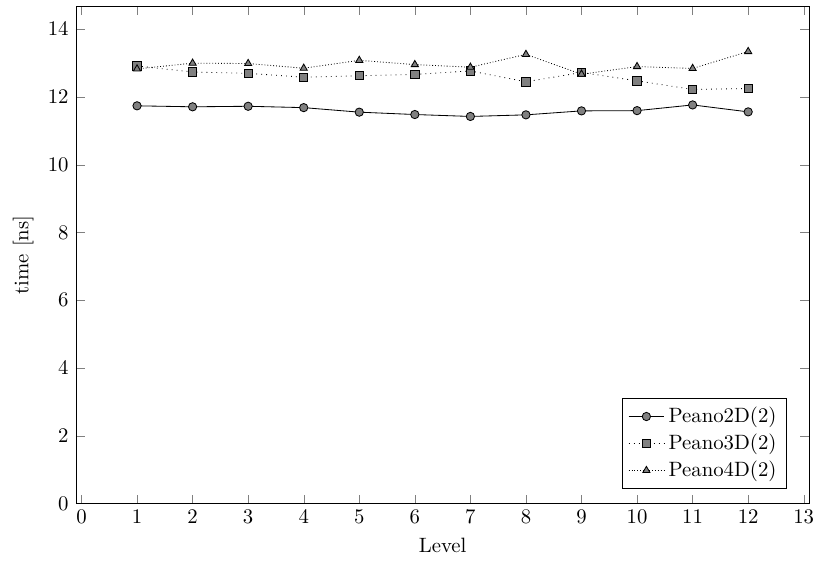}
\caption{Measured average runtimes of finding a neighbor in random direction of a random grid cell in Peano curves of different dimensions.} \label{fig:peanoDimRuntimes}
\end{figure}

\begin{figure}[!htb]
\centering
\includegraphics[scale=\plotScale]{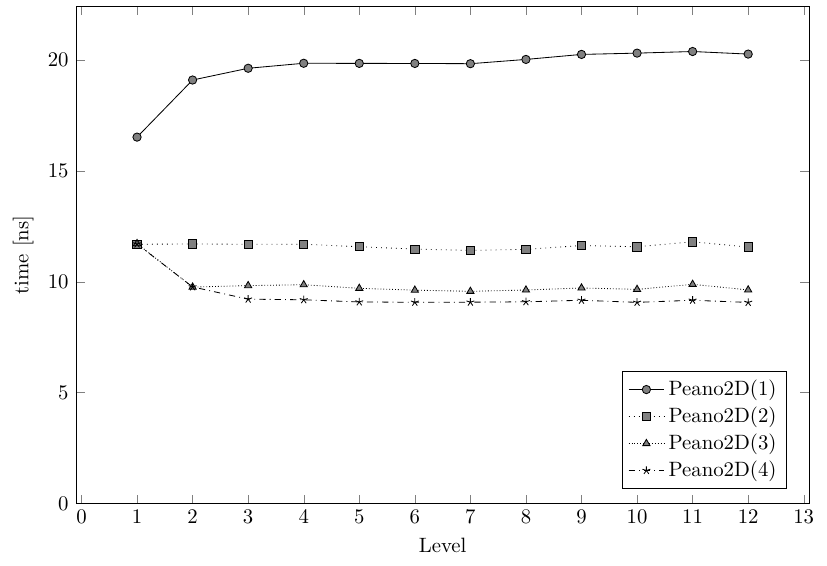}
\caption{Measured average runtimes of finding a neighbor in random direction of a random grid cell in the 2D Peano curve with different lookup table depths.} \label{fig:peanoDepthRuntimes}
\end{figure}

\begin{figure}[!hbt]
\centering
\includegraphics[scale=\plotScale]{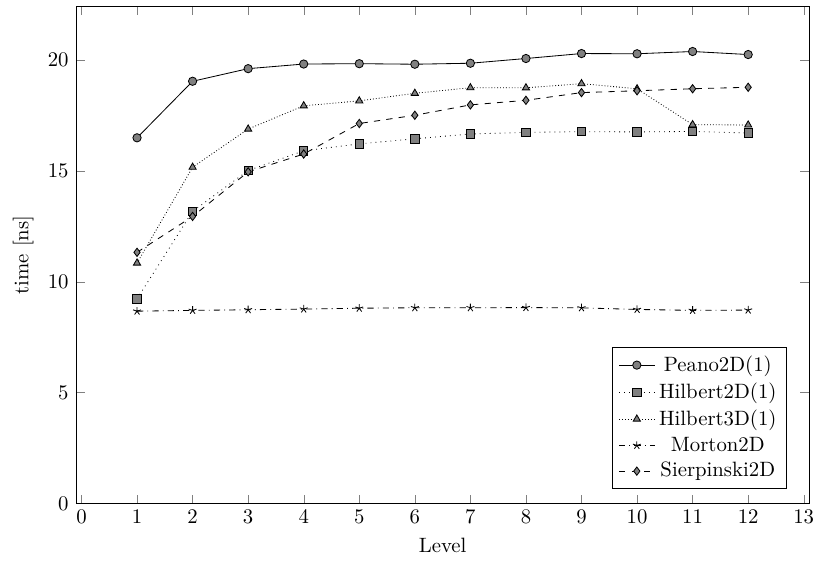}
\caption{Measured average runtimes of finding a neighbor in random direction of a random grid cell in different 2D curves. For the Morton curve, the algorithm by Schrack \cite{schrack1992} was used.} \label{fig:2dRuntimes}
\end{figure}

\Cref{fig:peanoDimRuntimes}, \Cref{fig:peanoDepthRuntimes} and \Cref{fig:2dRuntimes} show the measured average-case runtime of different implementations. The number in parentheses denotes the number of levels included in the lookup tables as explained in \Cref{opt:multiLevelTables}. If no number is provided in parentheses, the algorithm does not use lookup tables. The measurements were conducted on a system with an Intel Core i7-5600U CPU, 12 GB RAM and Ubuntu 16.04. The library was compiled using gcc version 5.4.0 with optimization flag -O2. %

Despite some fluctuations in the measured runtimes, it is clearly visible that all new neighbor-finding algorithms perform similarly well and, with multi-level lookup tables as in \ref{fig:peanoDepthRuntimes}, can even compete with the neighbor-finding algorithm for the Morton curve. It has to be noted, though, that due to its simpler structure, the algorithm for the Morton curve may have a clear advantage when compiler pipelining is allowed.

\begin{figure}[!hbt]
\centering
\includegraphics[scale=\plotScale]{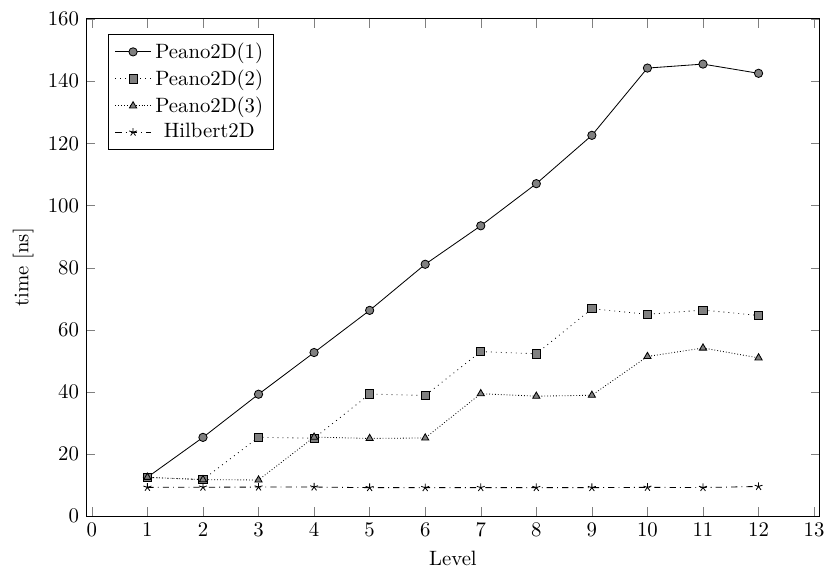}
\caption{Measured average runtimes of computing the state of a random grid cell in different curves with different lookup table depths.} \label{fig:stateRuntimes} 
\end{figure}

\Cref{fig:stateRuntimes} shows measured runtimes of state-computation algorithms for the 2D Peano and Hilbert curves. The measurements were also conducted as described above. The algorithm for the Peano curve is an optimized version of \Cref{alg:isomorphism}, while for the 2D Hilbert curve, the special algorithm from \Cref{lst:hilbertState2D} was used. The effect of the lookup table depth on the Peano algorithm is clearly visible.

\cleardoublepage

\section{Conclusion} \label{sec:conclusion}

In this thesis, we have derived a modeling framework which can be used to model most practical SFCs. These models were used to define trees on which a general neighbor-finding algorithm was formulated. This algorithm can, under certain assumptions, compute positions of neighbor cells in a regular grid ordered by a SFC with an average-case runtime complexity of $\compLeq{1}$. Measurements on various implementations were performed to demonstrate that with some optimizations, the runtime of this algorithm is comparable to the runtime of an existing $\compLeq{1}$ algorithm for the Morton order by Schrack \cite{schrack1992}. It was discussed how related computational issues can be solved and how properties of specified models can be verified. An implementation was introduced that can visualize and compute lookup tables for SFC models.

The application of space-filling curves in computer science is a relatively young research topic. Consequently, there are many questions yet to be explored. In particular, this thesis leaves some challenges still to be addressed. For example, future research might investigate neighbor-finding on adaptive grids. For theoretical modeling, one might define a finite set $\calL \subseteq \vertices$ of leaves. A suitable characterization for such a set might be that for every sequence $(v_n)_{n \in \natZero}$ of nodes with $v_0 = \rootv$ and $v_n = \parent(v_{n+1})$, there is exactly one $n \in \natZero$ with $v_n \in \calL$. One might then define that $v' \in \calL$ is a geometric $f$-neighbor of $v$ if $\dim(\convHull{Q(v')} \cap \vertexFace{v}{f}) = d-1$. For an implementation, the introduction of an additional function similar to $\ofunc$ might be necessary. The average-case runtime complexity of the algorithm might change if the difference between the maximum and minimum level is unbounded. With the definition of neighborship in adaptive grids used by Aizawa and Tanaka \cite{aizawa2009}, neighbor-finding is even easier since they define neighbors to be of equal or bigger size.

It might also be interesting to examine whether the principle of the efficient Hilbert 2D state computation algorithm from \Cref{sec:implementation:otherCode} can be transferred to global models of Sierpinski curves.

In the context of this thesis, some aspects of analyzing SFC models (i.e.\ $b$-spe\-ci\-fi\-ca\-tio\-ns) have already been implemented. Future work could extend this analysis to verify several properties of SFC models. Some properties and approaches for their verification have already been presented in this thesis. Other aspects, for example whether and how a model specifies a space-filling curve as the limit of finite approximations, have been left to future research. A suitable analysis tool might then also be extended to implement automatic code-generation for arbitrary regular SFC models. 

\newpage
\bibliographystyle{plain}
\nocite{*}
\bibliography{sfc-thesis-references}

\end{document}